\PassOptionsToPackage{dvipsnames}{xcolor}
\documentclass[twocolumn,numberedappendix, tighten]{aastex701}

\newcommand{\Msun}{\mathrm{M}_\odot}
\newcommand{\Lhat}{\hat{L}}
\newcommand{\kyr}{\,\mathrm{kyr}}
\newcommand{\AU}{\,\mathrm{AU}}
\newcommand{\kpc}{\,\mathrm{kpc}}
\newcommand{\gcm}{\,\mathrm{g\,cm^{-3}}}
\newcommand{\rdisk}{r_\mathrm{disk}}

\newcommand{\appropto}{\mathrel{\vcenter{
  \offinterlineskip\halign{\hfil$##$\cr
    \propto\cr\noalign{\kern2pt}\sim\cr\noalign{\kern-2pt}}}}}

\usepackage[dvipsnames]{xcolor}
\usepackage[utf8]{inputenc}
\usepackage[english]{babel}
\usepackage{amsmath, amssymb}
\usepackage{graphicx}

\definecolor{linkcolor}{rgb}{0.0,0.3,0.5}

\definecolor{darkgreen}{rgb}{0.13, 0.55, 0.13}

\DeclareGraphicsExtensions{.bmp,.png,.jpg,.pdf}
\usepackage{orcidlink}

\begin{document}
\title{FORGE'd in Fragments: Pop III star formation in a Milky Way-like progenitor halo}

\shorttitle{Pop III star formation in a Milky Way progenitor}
\shortauthors{Pustovoit et al.}

\correspondingauthor{Vasilii Pustovoit}

\author[0009-0007-3199-8511]{Vasilii Pustovoit}
\affiliation{Canadian Institute for Theoretical Astrophysics, University of Toronto, 60 St. George St., Toronto, ON M5S 3H8, Canada}
\affiliation{Department of Physics, University of Toronto, 60 St. George St., Toronto, ON M5S 3H8, Canada}
\email{vpustovoit@cita.utoronto.ca}

\author[0000-0003-1053-1262]{Shivan Khullar}
\affiliation{Center for Computational Astrophysics, Flatiron Institute, 162 5th Ave, New York, NY 10010, USA}
\email{skhullar@flatironinstitute.org}

\author[0000-0002-8659-3729]{Norman Murray}
\affiliation{Canadian Institute for Theoretical Astrophysics, University of Toronto, 60 St. George St., Toronto, ON M5S 3H8, Canada}
\email{murray@cita.utoronto.ca}

\author[0009-0002-4295-9239]{Yasmine J. Meziani}
\affiliation{TAPIR, California Institute of Technology, Mailcode 350-17, Pasadena, CA 91125, USA}
\email{ymeziani@caltech.edu}

\author[0000-0002-4900-6628]{Claude-Andr\'e Faucher-Gigu\`ere}
\affiliation{Department of Physics and Astronomy and Center for Interdisciplinary Exploration and Research in Astrophysics (CIERA), Northwestern University, 1800 Sherman Ave., Evanston, IL 60201, USA}
\email{cgiguere@northwestern.edu}

\author[0000-0003-3729-1684]{Philip F. Hopkins}
\affiliation{TAPIR, California Institute of Technology, Mailcode 350-17, Pasadena, CA 91125, USA}
\email{phopkins@caltech.edu}

\author[0000-0002-1655-5604]{Michael Y. Grudi\'c}
\affiliation{Center for Computational Astrophysics, Flatiron Institute, 162 5th Ave, New York, NY 10010, USA}
\email{mgrudic@flatironinstitute.org}

\author[0000-0003-2358-9949]{J. Richard Bond}
\affiliation{Canadian Institute for Theoretical Astrophysics, University of Toronto, 60 St. George St., Toronto, ON M5S 3H8, Canada}
\email{bond@cita.utoronto.ca}

\begin{abstract}
     We present a radiation-magnetohydrodynamic (RMHD) cosmological zoom-in simulation of Pop III star formation, drawing initial conditions from a Milky Way-like progenitor 
	halo at redshift $z \sim 21$. We follow the collapse of primordial gas over 10 orders of magnitude, through to the formation of an accretion disk that becomes gravitationally unstable $\Delta t \sim -3\kyr$ before the first protostar's formation. 
	Prominent spiral arms then develop and fragment vigorously, producing $\sim 60$ sink formation events over the first $8\kyr$ post-collapse evolution,
	with a total stellar mass reaching $> 150\,\Msun$ by $15\kyr$, the end of our simulation. 
	The disk remains magnetically supercritical throughout: the mass-to-flux ratio falls from $\mu_\Phi \sim 10^2$--$10^3$ to $\sim 15$ by $15\kyr$, 
	as the inflow brings in magnetic flux. The magnetic energy in the disk remains sub-dominant throughout, but grows by $\sim 3.5$ orders of magnitude after the first star forms. 
	The disk starts sub-virial ($2 E_\mathrm{kin}/|E_\mathrm{pot}| \approx 0.6$) and reaches virial equilibrium within $\sim 2\kyr$. The total stellar mass of the population grows as $M_* \appropto t^{2}$, 
	confirming the prediction of accelerating accretion in self-gravitating turbulent collapse, at rates of $\dot{M} \sim 10^{-3}$--$10^{-2}\,\Msun\,\mathrm{yr}^{-1}$.
	The star formation efficiency reaches $\approx 90\%$ within the disk (after
	$\approx 11\kyr$; $\approx 80\%$ at $9.5\kyr$)
	and $\sim 50\%$ within a $0.1$ pc sphere. We find a top heavy IMF ($dN/dm \propto m^{-\alpha}$) with a high-mass slope $\alpha \approx 1$.
	The cosmological environment sets the boundary conditions of the disk: an
	envelope delivers $\sim 3\times 10^{-3}\,\Msun\,\mathrm{yr^{-1}}$ to the outer edge of a disk with a
	radius of $\sim 10^{3}$\,AU; the disk spreads and, in our simulation, its fragmentation
	sites move outward, from $\sim 500$ to $\sim 1350$\,AU, until it has been
	depleted. 
	
\end{abstract}

\section{Introduction}
\label{sec:intro}

The first generation of stars, Population III (Pop III), is thought to have formed from
metal-free primordial gas in dark-matter minihalos at redshifts
$z \sim 20$--$30$ \citep{abel02, bromm02}. 
Because molecular hydrogen (H$_2$) was the only efficient gas coolant
at these epochs, the primordial gas could cool to a temperature floor of
$T \sim 200$ K at densities $n \sim 10^4\,\mathrm{cm}^{-3}$, before
reaching local thermodynamic equilibrium in its ro-vibrational levels
\citep{bromm02, klessen_glover23}. 
The Jeans mass at this point, $M_J \sim 10^3\,\Msun$, is orders of
magnitude larger than that of present-day star-forming cores, leading
to the early expectation that Pop III stars were predominantly massive
and formed in isolation within each minihalo
\citep{abel02, bromm02, yoshida08}.
These early simulations bridged cosmological collapse down to
protostellar densities ($n \sim 10^{21}\,\mathrm{cm}^{-3}$), resolving
the characteristic stages of primordial collapse: H$_2$ three-body
formation heating at $n \gtrsim 10^8\,\mathrm{cm}^{-3}$,
collision-induced emission cooling at $n \gtrsim 10^{14}\,\mathrm{cm}^{-3}$,
and a second collapse driven by H$_2$ dissociation at
$n \sim 10^{16}\,\mathrm{cm}^{-3}$ \citep{omukai00, yoshida08, wollenberg20}.

This picture of isolated, monolithic Pop III star formation was
altered as a result of simulations that continued to follow the system
after the initial protostar formed.
The infalling gas inevitably carries angular momentum, seeding a
self-gravitating accretion disk prone to gravitational instability and
fragmentation, so that a protostellar cluster arises instead of a single object \citep{stacy10, clark11, greif11, stacy12, wollenberg20}.

The protostellar cluster is shaped by competitive accretion
\citep{bonnell01}, fragmentation-induced starvation
\citep{peters10, prole22a}, protostellar mergers, and dynamical ejections,
with $31$--$43\%$ of the surviving sinks on escaping
trajectories \citep{Stacy_Bromm_2013, stacy16}.

Turbulence plays a fundamental role in shaping fragmentation and the
resulting mass distribution.
Cosmological infall into dark-matter
potential wells drives supersonic turbulence in Pop III
minihalos, with Mach numbers $\mathcal{M} \approx 1.8$--$5$ that increase
with halo mass \citep{ho25, ho25b}.
\citet{wollenberg20} showed that the degree of initial turbulence
strongly affects the number of fragments and the resulting mass
distribution, with higher initial turbulent velocities producing more
fragments and more stochastic outcomes.
\citet{prieto11} showed that turbulent fragmentation in a larger,
$3 \times 10^7\,\Msun$ halo at $z \sim 11$ produces $\sim 25$
gravitationally unstable clumps, proposing that infall-driven turbulence
may be the dominant fragmentation mechanism when halo masses exceed the
canonical $10^6\,\Msun$ minihalo scale.

Magnetic fields can further modify Pop III star formation, though their net effect remains debated: even trace seed fields are amplified during collapse to 3--100\% of equipartition \citep{higashi24} and can become dynamically important within $\sim 10^3$ yr of protostar formation \citep{sharda21, hirano_machida22, machida25}.
Strong fields can suppress disk fragmentation
\citep{hirano_machida22, stacy22}, but fragmentation persists, if reduced, in other runs \citep{prole22, tanaka24}, and early on magnetic fields limit protostellar masses more effectively than radiation \citep{sharda_menon25, prole25}.
The outcome depends on initial field strength, geometry,
resolution, run duration \citep{klessen_glover23} and even the
numerical MHD scheme \citep{ugalino26}.

The Pop III initial mass function (IMF) inferred from these simulations
is generally thought to be top-heavy, spanning $\sim 1$ to several hundred $\Msun$,
with single stars in some halos and multiple systems in others \citep{susa14}.
When radiative feedback is included, accretion onto a single protostar
can be halted at $\sim 43\,\Msun$ \citep{hosokawa11}, though
episodic accretion from a fragmenting disk can produce intermittent
feedback and final masses from $\sim 10$ to several hundred $\Msun$
\citep{hosokawa16}.
The large statistical survey of \citet{hirano_sakai23} across
$>100$ cosmological halos confirms that the Pop III IMF is not
universal: the outcome depends on halo mass, angular momentum, and
baryon streaming velocity.

While many Pop III simulations now use cosmological initial conditions
\citep{greif11, stacy16, latif_khochfar22, hirano_sakai23, meziani26},
simulations that simultaneously combine cosmological ICs,
full radiation-MHD, and resolved individual protostellar formation remain
few. \cite{meziani26} use the FORGE'd in FIRE framework \citep{FORGE_d_in_FIRE_Hopkin_2023} to study Pop III star formation in an UFD (Ultra Faint Dwarf) galaxy progenitor, the m09 halo \citep{wetzel23} ($\sim 2\times10^{9}\,\Msun$ at $z=0$), where the first Pop III star forms at $z \sim 14$ in a halo of $\sim 10^{6}\,\Msun$. A minihalo embedded in a strong overdensity destined to grow to Milky Way (MW) mass may experience qualitatively different infall rates, tidal torques, and angular momentum injection compared to one in a quieter
environment destined to remain a dwarf galaxy. Therefore, in this paper, we zoom-in to the formation of the first stars in a massive halo of $2\times10^{7}\,\Msun$ (above the atomic-cooling threshold at $z \approx 21$) that eventually grows to MW mass at z=0.

\begin{figure*}[!htbp]
	\centering
	\includegraphics[width=\textwidth]{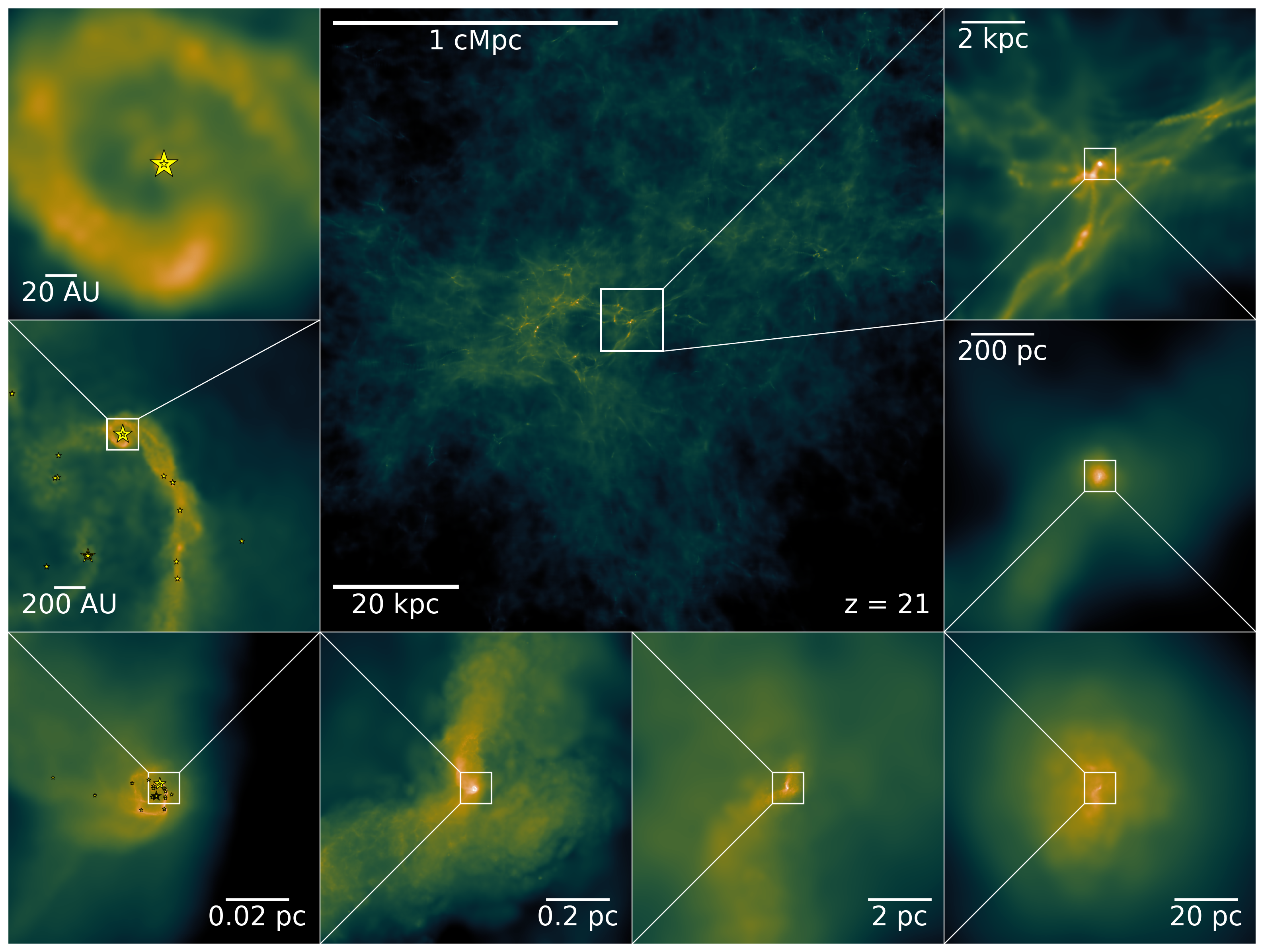}
	\caption{Multi-scale density projection of the simulation at
		$\Delta t \approx 15 \kyr$, illustrating the dynamic range
		from the cosmic web (box size of $ \approx 100 \kpc$) to the center of the
		protostellar disk (resolving AU scales).
		White boxes mark the region shown in the next panel.
		Yellow star symbols mark sink particles, with size of the marker scaling with the final mass of the star.  The progenitor halo has a dark matter mass of $\sim 2\times10^{7}\,\Msun$ at the first collapse, $z \sim 21$.}
	\label{fig:zoom}
\end{figure*}

We perform a cosmological radiation-magnetohydrodynamic (RMHD) zoom-in
simulation of Pop III star formation in the MW--progenitor halo 
\citep{FIRE_2, wetzel23}, adopting the cosmology and the $60\,h^{-1}$ comoving Mpc parent box size of the AGORA project
\citep{AGORA2014}. 
We use the FORGE'd in FIRE framework \citep{FORGE_d_in_FIRE_Hopkin_2023}, combining FIRE-3 \citep{FIRE_3_Updated_Hopkin} large-scale
physics with STARFORGE \citep{STARFORGE_Towa_Grudic_2021} protostellar
physics, and evolve the system for $\sim 15\kyr$ after the first
protostar forms, reaching a gas mass resolution of
$\Delta m \sim 10^{-3}\,\Msun$.
We study the thermodynamic collapse, disk formation and fragmentation,
magnetic field amplification, and protostellar mass assembly in this
environment, providing the first cosmological RMHD simulation that
resolves individual Pop III protostellar formation in a MW-progenitor halo. \autoref{fig:zoom} shows a multi-scale view of the simulation at the end of the run (at $\Delta t \approx 15\kyr$).

The rest of the paper is organized as follows.
\hyperref[sec:methods]{Section \ref*{sec:methods}} describes the simulation setup, physics
modules, refinement strategy, and analysis methods.
\hyperref[sec:results]{Section \ref*{sec:results}} presents the disk morphology, thermodynamic
and energetic evolution, disk stability, star formation history,
accretion rates, protostellar mass function, and magnetic field structure.
\hyperref[sec:disc_comparison]{Section \ref*{sec:disc_comparison}} compares our results to previous
Pop III simulations.
\hyperref[sec:caveats]{Section \ref*{sec:caveats}} presents caveats of our simulation.
We state our conclusions in \autoref{sec:conclusions}.

\section{Simulation and Methods}
\label{sec:methods}
\subsection{Simulation details}

For this paper we utilize the FORGE'd in FIRE framework of
\citet{FORGE_d_in_FIRE_Hopkin_2023}, extended to the primordial
star-formation regime similar to \citet{meziani26}.
We employ FIRE-3 \citep{FIRE_3_Updated_Hopkin} physics modules at
low resolution ($m_\mathrm{DM} \approx 3.5\times10^{4}\,\Msun, \; m_\mathrm{b} \approx 7\times10^{3}\,\Msun$) and STARFORGE \citep{STARFORGE_Towa_Grudic_2021}
physics to resolve individual protostellar formation. We briefly review these below \enlargethispage{-\baselineskip}and refer the reader to \citet{STARFORGE_Towa_Grudic_2021, FIRE_3_Updated_Hopkin, FORGE_d_in_FIRE_Hopkin_2023}. Numerical parameters not specified below are identical to those described in the above references.

\subsubsection{Code and gravity solver}
\label{sec:hydro}

The governing equations solved by the code are the ideal MHD equations
in conservative form \citep{Hopkins_Raives_2015}:
\begin{equation}
	\frac{\partial \mathbf{U}}{\partial t} + \nabla \cdot \mathbf{F} = \mathbf{S},
	\label{eq:mhd_conservation}
\end{equation}
where the state vector of conserved quantities and the flux tensor are
\begin{equation}
	\mathbf{U} =
	\begin{pmatrix} \rho \\ \rho\mathbf{v} \\ \rho e \\ \mathbf{B} \end{pmatrix},
	\qquad
	\mathbf{F} =
	\begin{pmatrix}
		\rho\mathbf{v}                                                                     \\
		\rho\mathbf{v}\otimes\mathbf{v} + P_{T}\,\mathcal{I} - \mathbf{B}\otimes\mathbf{B} \\
		(\rho e + P_{T})\mathbf{v} - (\mathbf{v}\cdot\mathbf{B})\mathbf{B}                 \\
		\mathbf{v}\otimes\mathbf{B} - \mathbf{B}\otimes\mathbf{v}
	\end{pmatrix}.
	\label{eq:mhd_flux}
\end{equation}
Here $\rho$ is the mass density, $\mathbf{v}$ the velocity,
$\mathbf{B}$ the magnetic field (in Gaussian units with $4\pi$ absorbed),
$e = u + |\mathbf{B}|^{2}/(2\rho) + |\mathbf{v}|^{2}/2$ the total specific energy,
$P_{T} = P + |\mathbf{B}|^{2}/2$ the total (thermal plus magnetic) pressure,
and $\mathcal{I}$ the identity tensor.
$\mathbf{S}$ includes source terms for self-gravity and the Powell 8-wave and Dedner divergence-cleaning
terms \citep{Hopkins_Raives_2015} that maintain $\nabla\cdot\mathbf{B}\approx 0$.

The FORGE'd in FIRE framework is based on GIZMO Meshless Finite Mass (MFM) code
\citep{A_new_class_of_Hopkin_2015}.
Gravity is solved with the TreePM algorithm \citep{2002JApA...23..185B, Springel_2005},
which combines a particle-mesh grid for large-scale forces with an oct-tree
for short-range forces \citep{Barnes_Hut_1986}, as in \citet{Springel_2010b}.
Gas particles use adaptive Lagrangian gravitational softening matching the
hydrodynamic kernel radius.
Sink particles use a fixed Plummer-equivalent softening
$\epsilon \approx 0.4$\,AU (gravitational kernel radius $\approx 1.2$\,AU).
Direct $N$-body summation is used for all sink--sink and sink--gas
gravitational interactions, with direct summation applied for
sink--sink separations below $1000\,\mathrm{AU}$ to accurately
resolve close binaries and dynamical ejections
\citep{STARFORGE_Towa_Grudic_2021}.

\subsubsection{Initial Conditions}
\label{sec:ics}

We use initial conditions (ICs) developed as part of the FIRE
collaboration \citep{FIRE_2}, specifically the m12f halo from the
Latte suite of MW-mass zoom-in simulations
\citep{wetzel23}.
The m12f halo has a virial mass $M_\mathrm{vir} \sim 1.5 \times 10^{12}\,\Msun$
at $z = 0$ and was selected as an isolated halo of this mass, without regard to its formation history \citep{wetzel23}. It is one of the four Latte halos (with m12i, m12m and m12b) that are most similar to the MW across a range of properties \citep{wetzel23}.
At the redshift of first collapse in our simulation ($z \sim 21$),
CDM part of the progenitor halo has a mass of $2 \times 10^{7}\,\Msun$.
This is above the atomic-cooling threshold at this redshift: the
virial temperature is $T_\mathrm{vir} \approx 2\times10^{4}$\,K, whereas $T_\mathrm{vir} = 10^4$\,K corresponds
to $\approx 10^{7}\,\Msun$ at $z = 21$, and the halo is $\sim 50$ times
more massive than the minimum mass for H$_2$ cooling, a few
$\times10^{5}\,\Msun$ \citep{yoshida03, kulkarni21}.  As in other atomic-cooling halos,
the dense gas still cools through H$_2$, whose formation there is largely
insensitive to external UV \citep{oh_haiman02}: later work finds that H$_2$ cooling in such halos is suppressed only when the Lyman--Werner (11.2--13.6\,eV) background exceeds a critical intensity. Expressed through $J_{21}$, the specific intensity in units of $10^{-21}\,\mathrm{erg\,s^{-1}\,cm^{-2}\,Hz^{-1}\,sr^{-1}}$, this critical value is $J_{21} \sim 30$--$300$ for soft spectra and $\sim 10^{3}$ for realistic spectra of metal-poor galaxies, and up to $10^{4}$--$10^{5}$ for metal-free stellar spectra \citep{shang10, sugimura14}. This is far above the external UV background in our run, which is negligible at $z \approx 21$ (\autoref{sec:thermochem}), and gas becomes fully molecular
through three-body H$_2$ formation here by $n \sim
	10^{11}$--$10^{12}\,\mathrm{cm^{-3}}$ (\autoref{fig:phases}; \citealt{abel02}).
The late, massive first collapse is most plausibly a consequence of the
parent-simulation resolution. With $m_\mathrm{DM} \approx 3.5\times10^{4}\,\Msun$,
a $10^{6}\,\Msun$ minihalo contains only $\sim 30$ DM particles, while $\gtrsim 500$ dark-matter particles are needed to resolve the gas content of a halo
\citep{kulkarni21}, i.e.\ $\gtrsim 2\times10^{7}\,\Msun$ here. Earlier
H$_2$-cooling minihalos in this region are therefore unresolved, and
star formation can only begin once the halo is this massive. The
hyper-Lagrangian refinement (\autoref{sec:refinement}) does not change this: it is switched
on only at the restart, at $z \approx 21$, after the parent run has already set the
epoch and the host of the first collapse; it splits only gas particles, so a
$10^{6}\,\Msun$ minihalo would still be resolved by $\sim 30$ DM particles; and
its Jeans criterion is triggered only once the gas is already dense and cold
($n \gtrsim 10$--$100\,\mathrm{cm^{-3}}$ at a few hundred K for the parent gas
particle mass), so it can follow a collapse but not start one in an unresolved
halo.

We adopt pre-generated ICs at $z = 99$ in a $60\,\mathrm{Mpc}/h$ comoving
cosmological box.
This large box contains a zoom-in region of comoving size $\sim 3\,\mathrm{Mpc}/h$,
within which the FIRE galaxy forms by $z = 0$.
Only the zoom-in region contains baryonic particles and
high-resolution dark matter (DM) particles; the remainder of
the volume is populated with low-resolution DM particles.
As in the FIRE-2 Latte simulations, we adopt cosmological parameters
from WMAP-7 \citep{komatsu11}:
$h = 0.702$,
$\Omega_m = 0.272$,
$\Omega_\Lambda = 0.728$,
$\Omega_b = 0.0455$,
$\sigma_8 = 0.807$, and
$n_s = 0.961$.

The full simulation snapshot contains $\sim 88$ million CDM particles
and FIRE-resolution gas particles whose smoothing lengths
($\sim$\,kpc) far exceed the protostellar disk scale.
We extract spatial cutouts centerd on the most-massive sink particle
with a physical radius of $1 \kpc$ (22 ckpc).
The cutout sphere retains only the gas, high-resolution DM, and sink particles; we then split gas particles as described below, but leave the DM particles unsplit.
All analysis presented in this manuscript is of this cutout simulation.

Over the $\sim 15$ kyr of post-collapse evolution studied here, the
sound-crossing time across the cutout radius
($r_\mathrm{cut} \approx 1$ kpc) is
$\sim r_\mathrm{cut}/c_s$ $\gtrsim 90$ Myr for $T \lesssim 10^4$ K ($c_s \lesssim 11\,\mathrm{km\,s^{-1}}$),
more than three orders of magnitude longer than our integration time.
The large-scale structure outside the cutout therefore has no causal
influence on the disk dynamics on these timescales, and the cutout
fully captures the relevant physics.

\begin{figure}[!htbp]
	\centering
	\includegraphics[width=\columnwidth]{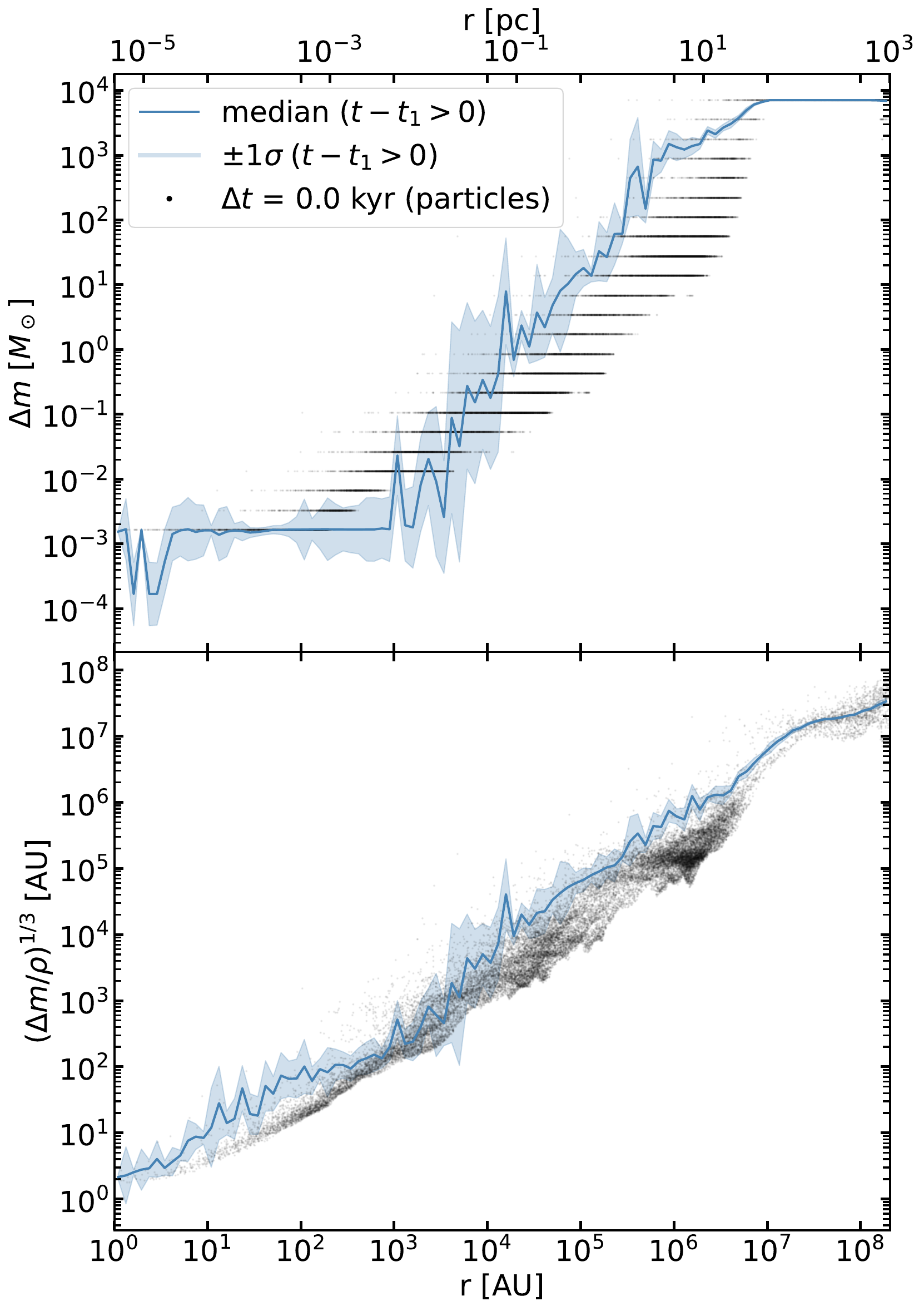}
	\caption{
		Top panel: Radial profile of gas particle mass (resolution), centered around the most massive sink.
		Scatter plot shows particle distribution at the time of first sink formation;
		solid line and shaded region correspond to the mean and $\pm 1\sigma$ for
		radial profiles across all snapshots after first sink formation.
		Horizontal lines correspond to $0.5 \: \Delta m$ of previous resolution region,
		since our particle splitting algorithm is subdividing parent particles in 2
		for each new level of split.
		Bottom panel: Same as top panel, but for radial profile of equivalent cell size
		$\Delta x = (\Delta m / \rho)^{1/3}$.
	}
	\label{fig:profile_resolution}
\end{figure}

\subsubsection{Refinement strategy}
\label{sec:refinement}

We first evolve the full cosmological simulation with FIRE-3 physics until
first $10^5 M_\odot$ stellar population (SSP) particle forms.
We then restart from the preceding snapshot with an additional
``hyper-Lagrangian'' Jeans refinement layer, applied globally; \citet{FORGE_d_in_FIRE_Hopkin_2023} use a similar ``hyper-refinement'' layer centred on the central black hole.

We refine gas cells (splitting into pairs of equal-mass daughter cells)
whenever their mass exceeds a fraction $f_J$ of the local Jeans mass:
\begin{equation}
	\Delta m > f_J M_J = f_J M_0 \left( \frac{T}{10^4 \text{ K}} \right)^{3/2} \left( \frac{n_\mathrm{H}}{\text{cm}^{-3}} \right)^{-1/2},
	\label{eq:jeans_refinement}
\end{equation}
where $M_0 = 9\times10^{6}\,\Msun$ is the Jeans mass of gas at $T = 10^{4}$\,K and
$n_\mathrm{H} = 1\,\mathrm{cm^{-3}}$ adopted in GIZMO, and $T$ is the gas temperature for a mean
molecular weight of $1.23$. Only gas well above the cosmic mean density is refined, and
cells are not split below $\Delta m = 10^{-3}\,\Msun$.
We use $f_J = 10^{-2}$, corresponding to
$100$ resolution elements per Jeans mass and satisfying the
\citet{truelove97} criterion, since we resolve $\lambda_J/\Delta x =
	(6/\pi f_J)^{1/3} \approx 5.8 > 4$ cells
per Jeans length.
This is sufficient: Lagrangian methods need only $\sim 2$ particle widths
per Jeans length to avoid artificial fragmentation, and particle splitting in
GIZMO's MFM does not seed artificial fragmentation (spurious bound clumps that would go on to form sink particles), its effect on the outcome lying
within the run-to-run scatter \citep[][and references therein]{khullar2026}.
It is, however, below the
$\gtrsim 30$ cells per Jeans length that grid codes need to capture
small-scale dynamo amplification \citep{Federrath2011}. No equivalent
criterion has been calibrated for MFM: only magnetorotational instability (MRI) quality factors
\citep{deng19} and particle-number convergence tests of gravitoturbulent
dynamos \citep{deng20} exist. We therefore do not claim that a small-scale
dynamo is resolved here (\autoref{sec:caveats}).

Only gas particles get split, DM particles stay at the same resolution.
This is a fine approximation because dark matter's potential is smooth at these
scales, and splitting of these particles should not change the results of our
simulation after the restart. It does matter for the epoch of the first
collapse, which is set before the restart by DM halos resolved with
$m_\mathrm{DM} \approx 3.5\times10^{4}\,\Msun$ (\autoref{sec:ics}).

At our highest refinement level, the gas mass resolution is
$\Delta m \approx 10^{-3}\,\Msun$ and the maximum spatial resolution is
$\Delta x \approx 1$\,AU in most snapshots, reaching $0.6$\,AU briefly, sufficient to resolve densities
up to $n \sim 10^{14}$--$10^{15}\,\mathrm{cm}^{-3}$. This spatial resolution
applies only at the highest densities: at $15\kyr$ the median disk
particle has $\Delta x \approx 110\AU$ (the depleted late-time disk gas has
low density, which is also why it is resolved by $\sim 55$ cells per Jeans
length: the median $\lambda_J/\Delta x$ of the disk particles at $15\kyr$, with $10$th--$90$th percentiles of $26$ and $160$), so the disk scale height is covered by only a few resolution
elements at late times.

\autoref{fig:profile_resolution} shows the radial profile of the
length scale of gas particles $\lambda = \left( m / \rho \right)^{1/3}$.
It exhibits the refinement profile, described in \autoref{sec:refinement}
at all times, except the earliest one.
Initial profile at $\Delta t = t - t_1 =-0.5$ (0.5 kyrs before the first sink formation)
is  limited by the refinement scale inside of 100 AU radius (the center was not yet
refined to the highest resolution). But by the time of first sink
formation, the refinement reaches highest resolution that we expect from our
scheme, (around $10^{-3} M_\odot$ or a few AU) so the inner star forming disk
region is fully refined.

\subsection{Physics we model}
\subsubsection{Magnetic fields}
\label{sec:mhd}

We integrate the ideal MHD equations as described in \citet{Hopkins_Raives_2015},
using the constrained-gradient method and divergence cleaning
\citep{Hopkins_2016_CG_MHD} for divergence control, which has been
validated for magnetorotational instability dynamics
\citep{FORGE_d_in_FIRE_Hopkin_2023}.

We include Spitzer--Braginskii thermal conduction and viscosity via
\citet{Hopkins_2017}. For metal-free gas at the temperatures
($T \sim 200$--$10^{4}$ K) and free-electron fractions relevant here,
the non-ideal MHD timescales (Ohmic, Hall and ambipolar
diffusion) exceed the local dynamical timescale by many orders of
magnitude \citep{meziani26}, so ideal MHD dominates throughout
the simulation. Spitzer conduction and Braginskii viscosity are also
negligible: even for fully ionized gas at $10^{3}$\,K and $n = 10^{10}\,\mathrm{cm^{-3}}$,
the conduction and viscous timescales across $100\AU$, $t \sim L^{2}/D$ with $D$ the Spitzer thermal or momentum diffusivity, are $\gtrsim 10^{15}$\,yr, and 
the gas here is almost neutral.

The initial magnetic field is a trace cosmological seed field of
$B_x = B_y = B_z = 10^{-10}$\,G (physical), imposed at the start of the
parent simulation ($z \simeq 99$), i.e.\ $|B_0| \approx 1.7\times10^{-10}$\,G, or
$\approx 1.7\times10^{-14}$\,G comoving, uniform in magnitude and direction,
amplified self-consistently by gravitational compression,
differential rotation, and (possibly) turbulent dynamo. The seed lies about two orders of magnitude below the present-day field strengths of $2\times10^{-12}$--$2\times10^{-10}$\,G (comoving) that \citet{koh21} consider, and below the CMB upper limits on primordial fields they quote, $\lesssim 10^{-10}$\,G for inflationary and $\lesssim 2\times10^{-11}$\,G for causal fields; it therefore does not test the possibility that a strong primordial field delays or prevents the collapse. The original orientation is $(1,1,1)$, i.e.\ $\mathbf{B}_0 \propto \hat{\mathbf{x}}+\hat{\mathbf{y}}+\hat{\mathbf{z}}$, in the cosmological box, but the protostellar disk is oriented randomly in the simulation.

\subsubsection{Radiation}
\label{sec:rad}

We solve radiation transport with the M1 moments method
\citep{Levermore_1984}, as implemented in {\small GIZMO} and described in
\citet{Hopkins_Grudic_2019, Hopkins2020, STARFORGE_Towa_Grudic_2021, FORGE_d_in_FIRE_Hopkin_2023}
across five frequency bands: H-ionizing ($>13.6$\,eV), Lyman--Werner (LW)/photo-electric ($8.0$--$13.6$\,eV), near-UV (NUV; $3.4$--$8.0$\,eV), optical--NIR ($0.41$--$3.4$\,eV), and an adaptive-wavelength FIR band ($<0.41$\,eV) that tracks the evolving dust emission temperature \citep{FIRE_3_Updated_Hopkin}. 
Gas, dust, and radiation temperatures are evolved separately with
appropriate coupling terms, allowing self-consistent treatment of
both optically thin and optically thick regimes.
We use a reduced speed of light $\tilde{c} = 10^{-3} c$,
i.e.\ $\approx 300\,\mathrm{km\,s^{-1}}$, much lower than the $\tilde{c} = 0.1\,c$ adopted for most runs in \citet{FORGE_d_in_FIRE_Hopkin_2023}. Tests in galaxy simulations find small effects for $\tilde{c} \gtrsim 500\,\mathrm{km\,s^{-1}}$, with slightly weaker radiative feedback at lower values down to $300\,\mathrm{km\,s^{-1}}$ \citep{Hopkins2020}, and $\tilde{c} \gtrsim 30\,\mathrm{km\,s^{-1}}$ suffices to follow expanding ionization fronts \citep{STARFORGE_Towa_Grudic_2021}. A radiation-hydrodynamic simulation of Pop III star formation that adopts the same $\tilde{c} = 10^{-3}\,c$ found no significant change in its results for $\tilde{c}$ between $10^{-1}\,c$ and $10^{-4}\,c$, with the fastest outflows at only $\sim 100\,\mathrm{km\,s^{-1}}$ \citep{Park_2022}. Our value is therefore at the lower edge of the galaxy-scale range and well above the ionization-front limit, but we have not tested the sensitivity of our results to $\tilde{c}$. Fast jets would be the feature most likely to be affected, but the jets of this run do not propagate as intended because of an implementation error (\autoref{sec:stars}). No ionization fronts are followed either: the Str\"omgren radii of the sinks are below the resolution, so their ionizing photons are absorbed within a cell (\autoref{sec:caveats}).

The Lyman-Werner (LW) band is combined with the photo-electric (PE)
heating band into a single radiation frequency bin. This follows
the five-band scheme of \citet{Hopkins2020} and \citet{FORGE_d_in_FIRE_Hopkin_2023},
in which a single far-UV band serves both purposes. Our gas is metal-free and
dust-free, so there is no photo-electric heating , and
the band acts only as the Lyman--Werner band.
H$_2$ photo-dissociation by LW photons is accounted for through the
$\Gamma_{\mathrm{H_2}}$ destruction term in \autoref{eq:nh2}.

\subsubsection{Thermochemistry}
\label{sec:thermochem}

We solve gas thermodynamics and chemistry as in \citep{FIRE_3_Updated_Hopkin}.
The following cooling functions are employed and are relevant to our setup:
$\Lambda_{\rm H}$ (atomic H cooling);
$\Lambda_{\rm HD}$ (HD cooling, \citealt{Galli_Palla_1998});
$\Lambda_{\rm H_2}$ (H$_2$ cooling);
$\Lambda_{\rm H_2}^{\rm thin},\, \Lambda_{\rm H_2}^{\rm thick}$
(H$_2$ interactions with $e^-,\,{\rm H}^+,\,{\rm H}_2,\,{\rm H},\,{\rm He}$
below and above the critical density at which the H$_2$ levels reach LTE, combined per molecule as $\Lambda_{\rm H_2}^{\rm thin}/(1 + \Lambda_{\rm H_2}^{\rm thin}/\Lambda_{\rm H_2}^{\rm thick})$ \citep{FIRE_3_Updated_Hopkin}). The network does not include H$_2$ line opacity; in GIZMO, optically thick cooling is only limited by a gray slab photosphere cap).
 The adiabatic index accounts for H, He, free electrons and H$_2$ \citep{boley07}.
We follow non-equilibrium H and He chemical networks.
Default FIRE-3 thermochemistry includes cooling due to metal lines, but we do
not evolve our simulation past the point of supernova-driven metal enrichment,
thus making all metal cooling contributions effectively zero (as expected in
the Pop III formation regime).

Since we simulate metal-free primordial gas, the dust-to-gas ratio is zero
and the dust term in the H$_2$ formation equation is absent.
H$_2$ formation proceeds exclusively via gas-phase channels:
the two-body H$^{-}$ and H$_2^+$ pathways at low densities
($n \lesssim 10^8\,\mathrm{cm}^{-3}$) and the three-body
reaction (3H $\to$ H$_2$ + H) at high densities.

Molecular fraction of H$_2$ evolves in every cell according to (eq. 6 from
\citealt{FIRE_3_Updated_Hopkin}), with the dust formation term set to zero:
\begin{equation}
	\begin{split}
		\frac{d \bar{n}_{\rm H_2}}{dt} ={} & \alpha_{\mathrm{GP}}C_2 \bar{n}_{\rm HI} \bar{n}_{\rm H} + \alpha_{3\mathrm{B}}C_3 \bar{n}_{\rm HI}^{2}\!\left( \bar{n}_{\rm HI}+\tfrac{ \bar{n}_{\rm H_2}}{8}\right) \\
		                                   & - \Gamma_{\mathrm{\rm H_2}} \bar{n}_{\rm H_2} - \xi_{\mathrm{\rm H_2}} \bar{n}_{\rm H_2} - \sum_{i}\beta_{\mathrm{\rm H_2},i}C_2 \bar{n}_{\rm H_2} \bar{n}_{i}
	\end{split}
	\label{eq:nh2}
\end{equation}
where the gas-phase, three-body, photodissociation, cosmic-ray and collisional terms and their coefficients are defined in \autoref{app:chem}.

The chemical energy of H$_2$ ($4.48$\,eV per molecule) is not
tracked in the gas energy equation: neither the heating released when H$_2$
forms nor the cooling when it is collisionally dissociated is included.
This matters below the densities we resolve. Three-body H$_2$ formation
heating is the dominant heat source in primordial gas at
$n \sim 10^{8}$--$10^{12}\,\mathrm{cm^{-3}}$ \citep{turk11}, and
in real gas dissociation cooling thermostats gas heated above $\sim 2000$\,K, which our network cannot do. The
FIRE-3 network is therefore a reduced, not a complete, primordial network
(see \autoref{sec:thermo} and \autoref{sec:caveats}). In addition, the
protostellar evolution tracks of \citet{Offner_2009} self-consistently evolve
the stellar radius without requiring the gas to track dissociation energy
explicitly.

The meta-galactic UV background is that of \citet{Faucher-Giguere_2020}; we expect it to be
negligible at $z \sim 21$ compared to local protostellar radiation resolved in the simulation.
H$_2$ self-shielding against Lyman-Werner photons is included using the
shielding function of \citet{Gnedin_Draine_2014} (their Eq. 3), applied with a
local Sobolev column density $N\mathrm{H} \sim \rho / |\nabla \rho|$, with
turbulent line broadening accounted for following a Burgers velocity spectrum.

The simulation does not include a cosmic-ray (CR) module: CRs are neither
transported nor injected by stellar feedback (which would in any case be negligible
over $\sim 15$\,kyr without supernovae). The thermochemistry instead uses GIZMO's
default for runs without such a module, a uniform and constant CR energy density
$u_\mathrm{CR} = 1\,\mathrm{eV\,cm^{-3}}$, the local interstellar value inferred from
Voyager 1 data \citep{cummings16}, which corresponds to an ionization rate
$\zeta_\mathrm{CR} = 1.6\times10^{-17}\,\mathrm{s^{-1}}$.
$200$ times the cosmic mean baryon density and is fully on above $2\times10^{5}$ times
the mean ($n \gtrsim 400\,\mathrm{cm^{-3}}$ at $z \approx 21$), which includes all the gas
analyzed here. Without attenuation, it enters the H$_2$ network (\autoref{eq:nh2}), the
free-electron fraction and the gas heating. We discuss its effect in
\autoref{sec:caveats}.

\subsubsection{Stellar formation and evolution}
\label{sec:stars}

We resolve the formation of individual protostars in the densest region
of our simulation using the STARFORGE \citep{STARFORGE_Towa_Grudic_2021}
formation and evolution modules.
A gas particle becomes a sink (protostar) particle when
it meets the full STARFORGE formation criteria: the cell must be
(i) locally self-gravitating (virial parameter $\alpha_\mathrm{vir} < 1$),
(ii) Jeans unstable,
(iii) in a converging flow ($\nabla \cdot \mathbf{v} < 0$),
(iv) experiencing compressive tides in all directions,
(v) a local maximum of the density field, and
(vi) not within the accretion radius of a pre-existing sink
\citep{STARFORGE_Towa_Grudic_2021}.
For brevity we refer to these protostellar particles as ``sinks''
throughout this paper. When a gas particle is converted into a sink, the
magnetic field of that gas is deleted.

Once formed, the sink particles accrete the bound gas whose circularized orbit falls
within the sink radius. Accreted gas is first added to the sink's internal reservoir, which  represents the unresolved envelope and disk inside the sink radius. Mass flows from the reservoir onto the protostar on a smoothed accretion timescale, and when jets are enabled a fraction $f_w = 0.3$ of it is diverted into a jet reservoir and returned to the gas as outflow \citep{STARFORGE_Towa_Grudic_2021}. In our run the reservoirs hold $\lesssim 0.3\%$ of the mass of most sinks at $15\kyr$ (at most $2.4\%$), so the sink masses we quote are effectively stellar masses.
Each sink evolves along protostellar tracks following \citet{Offner_2009},
explicitly tracking the stellar radius, luminosity, and effective temperature
as a function of the instantaneous mass and accretion rate.

 Jets are the least tested part of this feedback. Most Pop III simulations omit them, and the few that include them treat them as mass-removal prescriptions rather than resolved outflows \citep{meziani26}. Whether Pop III protostars can launch jets at all depends on whether a trace magnetic field is amplified enough for the mass-to-flux ratio to permit it; \citet{meziani26} find that the ratios around a Pop III protostar fall well within the range required for strong jet launching.  Protostellar jets are launched with a mass-loading proportional to the
accretion rate onto the protostar and a launch velocity of $0.3$ times the Keplerian speed at the protostellar surface \citep{STARFORGE_Towa_Grudic_2021}.
The jets module is active in
the present run, but the jet material is coupled to the gas closer to the protostars than intended. STARFORGE launches jets by spawning new gas cells near the sink \citep{STARFORGE_Towa_Grudic_2021}, and in the code version we use these jet cells are merged with the neighboring ambient gas cells too readily. Their mass and momentum are therefore spread into the surrounding gas near the protostar, and the jets propagate less far than intended. Because the internal structure of the jets is not resolved in any case, this amounts to a choice of where the jet feedback is deposited rather than an error in the injected mass or momentum. The same merging is more serious for stellar-wind bubbles, where the contact discontinuity between the shocked wind and the ambient gas must be followed, but STARFORGE winds come only from main-sequence stars \citep{STARFORGE_Towa_Grudic_2021}, and none of our protostars reaches the main sequence (\autoref{sec:massevol}). We defer
the analysis of jet feedback to future work.

\subsection{Postprocessing methods}
\subsubsection{Disk identification}
\label{sec:diskid}

\begin{figure*}[!htbp]
	\centering
	\includegraphics[width=\textwidth]{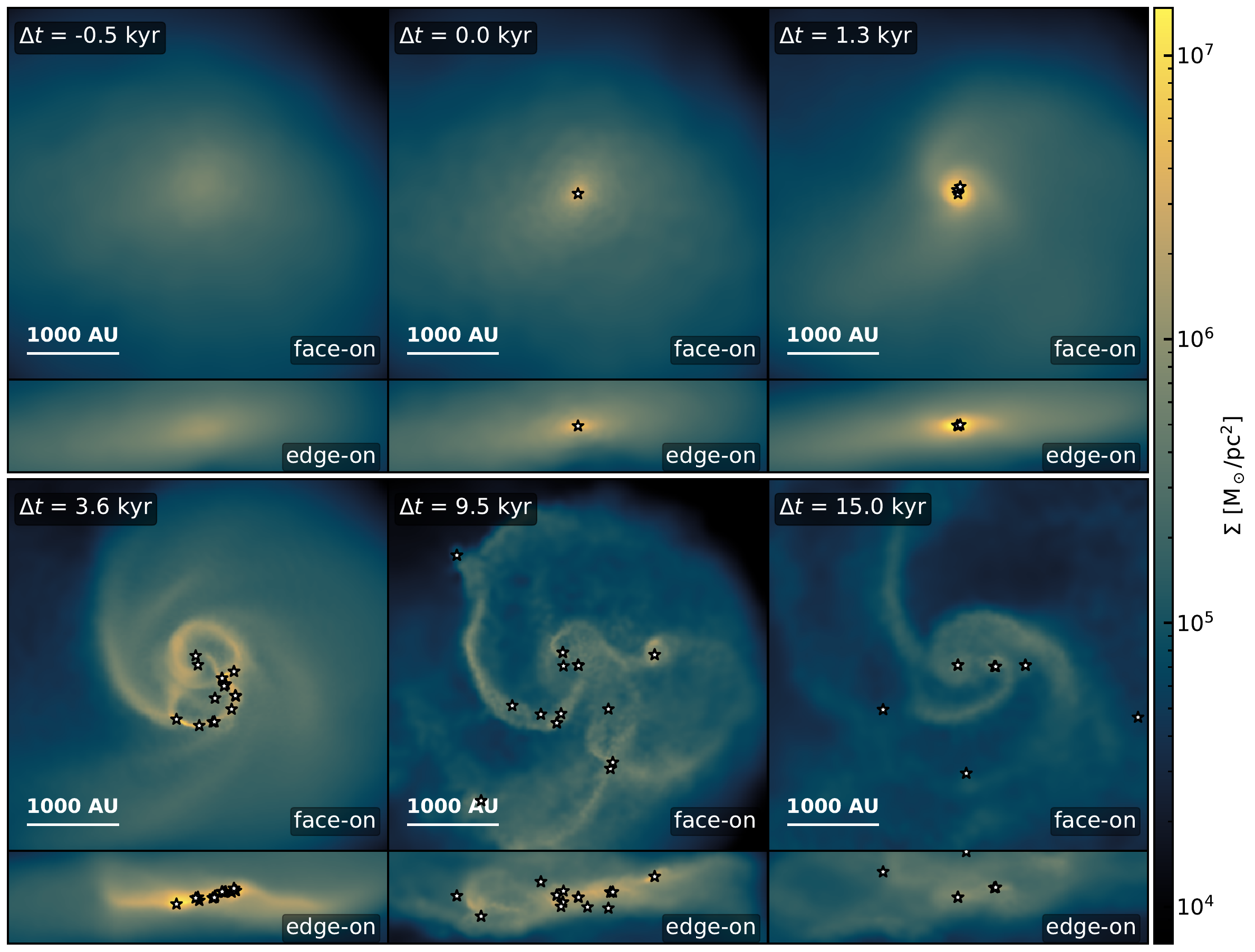}
	\caption{ Face-on and edge-on surface density projections at six epochs
		spanning the $\sim 15\kyr$ after the formation of the first sink.
		Each panel shows a $4100 \times 4100\AU$ region centerd on the
		most massive sink. Star symbols mark sink particles within the
		image box.
		The sinks are scattered throughout the disk rather than
		concentrated in a single cluster, tracing the spiral arms. }
	\label{fig:combined_density}
\end{figure*}

To compute velocity dispersion, local Toomre $Q$, and other disk
properties, we identify the gas particles belonging to the disk at
each snapshot. We center on the densest gas particle before the first sink
forms and on the most massive sink afterwards, and rotate into the frame of the
total angular momentum within $r_\mathrm{search} = 2000\AU$. Particles are
selected by cuts on cylindrical radius, aspect ratio ($|z|/r_\mathrm{cyl} < 0.3$),
density ($\rho > 10^{-15}\gcm$) and rotation ($v_\phi / v_K > 0.3$), and the
selection is extended to its bounding cylinder. The disk radius $\rdisk$ is the
90th-percentile cylindrical radius of the disk members. The full
algorithm and tests of its robustness are given in \autoref{app:diskid}.

\autoref{fig:combined_density} shows the gas surface density $\Sigma$ of the disk and its surroundings at the six epochs used throughout the paper ($\Delta t = -0.5$, $0$, $1.3$, $3.6$, $9.5$ and $15\kyr$ relative to the formation of the first sink). Each epoch has a face-on panel, seen along the angular momentum axis $\Lhat$ of the procedure above, and a narrower edge-on panel below it that shows the vertical structure. All panels are centerd on the most massive sink, span $4100\AU$ on a side, and share one logarithmic color scale ($10^{4}$--$10^{7}\,\mathrm{M_\odot\,pc^{-2}}$); the bar in each face-on panel marks $1000\AU$, and stars mark the sinks inside the image box.
Before the first sink forms ($\Delta t = -0.5\kyr$) the gas is a smooth, centrally concentrated core with no sinks. From $\Delta t \approx 3.6\kyr$ onward the disk develops spiral arms along which the sinks that formed by fragmentation are scattered, rather than gathering in one central cluster (\autoref{sec:stability}). By $15\kyr$ the disk is depleted, and only a few sinks and a faint spiral remain in the field.

\section{Results}
\label{sec:results}

In this section we present results from our cosmological RMHD simulation of Pop III star formation. We organize the discussion as follows: \autoref{sec:thermo} discusses the thermodynamics and chemistry of the collapse; \autoref{sec:prestellar} describes the structure and kinematics of the pre-stellar envelope and shows that its collapse proceeds outside-in; \autoref{sec:disk} characterizes the disk structure and energetics; \autoref{sec:stability} covers its gravitational stability; \autoref{sec:massevol} covers the star formation history, accretion rates, and protostellar mass function; and \autoref{sec:bfield} describes the magnetic field evolution.

\begin{figure*}[!htbp]
	\centering
	\includegraphics[width=\textwidth]{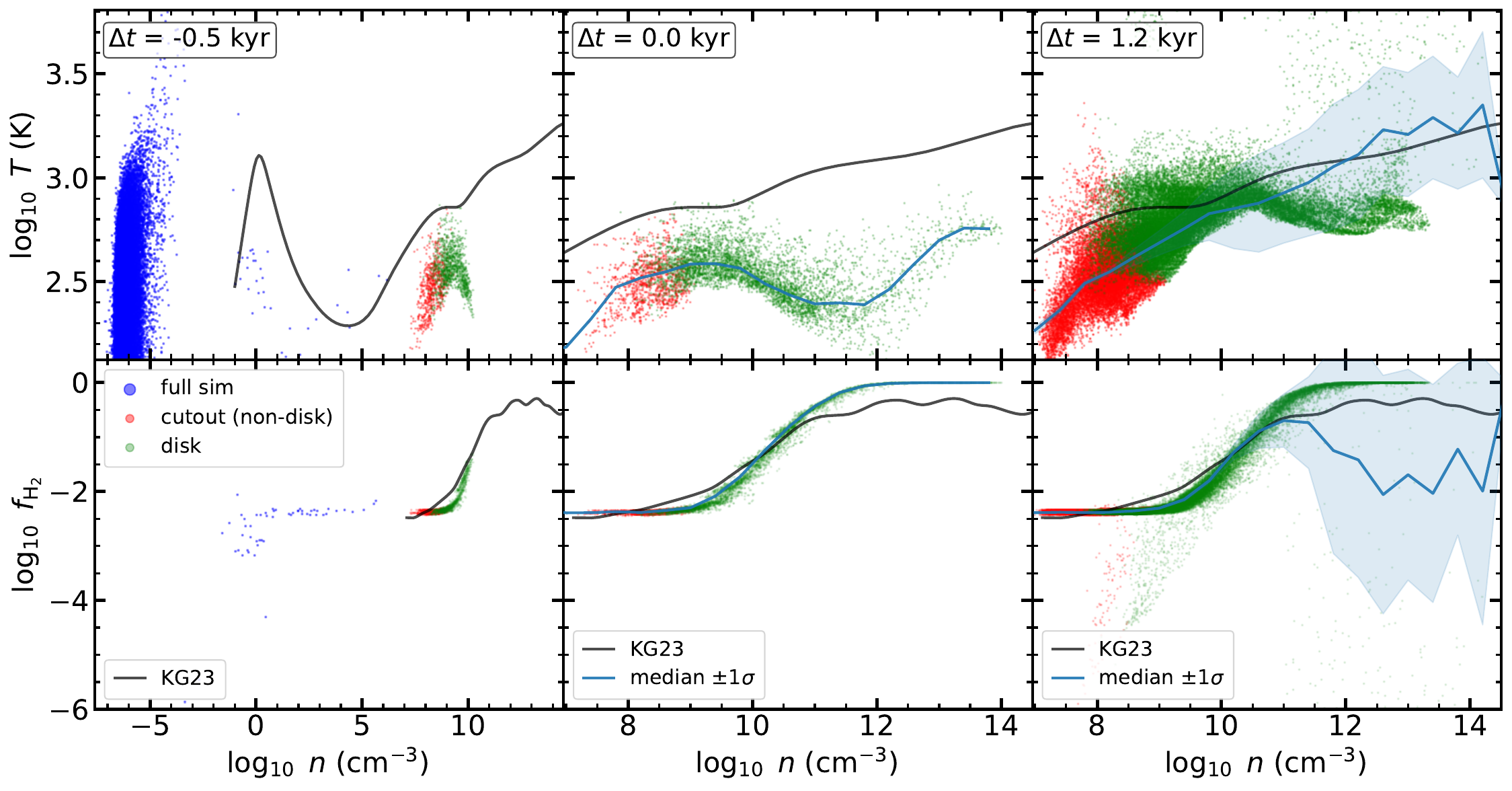}
	\caption{ Phase diagram (temperature and H$_2$ fraction vs.\ number density)
		at three representative epochs: pre-stellar ($\Delta t \approx -0.5\kyr$,
		left), first sink formation ($\Delta t = 0$, middle), and
		post-first-sink ($\Delta t = 1.2\kyr$, right).
		Green points correspond to disk-identified particles, red points correspond
		to points in our cutout simulation that have not been identified as disk,
		and blue points correspond to points in the full-box simulation right before
		the cutout was applied (only included in the leftmost plots). $f_{\rm H_2}=0$
		for most of particles in the low-density region (blue particles), which
		therefore do not appear in the bottom-left panel.
		The curves in the temperature panels show the theoretical $T$--$n$
		tracks for primordial ($Z=0$) and metal-enriched ($Z=10^{-4}\,Z_\odot$) gas
		from \citet{klessen_glover23} (KG23); the shaded bands indicate the median $\pm$
		standard deviation of the fits over the time intervals
		$-0.5\,\mathrm{kyr} \le \Delta t \le 0$ (middle) and
		$0 \le \Delta t \le 1.2\,\mathrm{kyr}$ (right), respectively.
		\label{fig:phases}%
		The spread of the fit is narrow at early times and broadens at
		later times. KG23 $Z=0$ curve follows similar curve from \citep{omukai00} closely.}
\end{figure*}

\subsection{Thermodynamics}
\label{sec:thermo}

The thermal and chemical state of the gas
reflects the competition between compressional heating and radiative cooling.
The temperature--density phase trajectory is therefore diagnostic of which physical process dominates at each stage of collapse, while the H$_2$ fraction $f_{\mathrm{H}_2}$ shows how much of the main coolant is available. We present phase space plots in \autoref{fig:phases} for three representative epochs, pre-stellar,
first sink formation, and post-first sink formation. Disk particles are identified by the algorithm described in \ref{sec:diskid}; disk particles are green, particles in the cutout but not in the disk are red, while particles outside the cutout region are shown in blue.

Before the first star forms ($\Delta t = -0.5\kyr$; \autoref{fig:phases}, left
panels), the gas occupies a narrow track spanning $n \sim 10^7$--$10^{10}\,\mathrm{cm}^{-3}$ and $T \sim 500$ K. 
The low temperature reflects the fact that at densities $n\lesssim 10^4{\rm cm}^3$, the molecular radiative cooling $\Lambda\sim n^2$, while the adiabatic heating $\Gamma\sim n^{3/2}$, and cooling dominates. Above $n \sim 10^{4}\,\mathrm{cm^{-3}}$ the H$_2$ levels are in LTE, so that $\Gamma\sim n$, with the result that heating dominates.
Our network follows the same scaling, because at high density it uses the LTE cooling rate per molecule (section \ref{sec:thermochem}).
 
The H$_2$ fraction (lower panel) rises modestly from $f_{\mathrm{H}_2}
	\sim 10^{-2.5}$ at $n \sim 10^7\,\mathrm{cm}^{-3}$ to $f_{\mathrm{H}_2} \sim
	10^{-1.3}$ at $n \sim 10^{10}\,\mathrm{cm}^{-3}$, driven primarily by two-body
formation channels (H$^-$ and H$_2^+$ pathways). The increase in H$_2$
fraction enhances the cooling of the gas, since H$_2$ is a much more efficient
coolant than atomic H. 

The low-density (blue) particles in the left column ($n \leq 10^{-5}$ cm$^{-3}$; see the caption) lie outside our region of interest; the dynamical time scale is long in that region, so the thermodynamic state of the gas does not change significantly over the time interval shown in the figure.

By $\Delta t = 0$ (\autoref{fig:phases}, middle panels), the gas has
collapsed to much higher densities
($n \sim 10^{13}\,\mathrm{cm}^{-3}$) and the thermal track has
extended significantly.
At intermediate densities ($n \sim 10^8$--$10^{10}\,\mathrm{cm}^{-3}$),
compressional heating outpaces H$_2$ line cooling, which is saturated at the LTE value, and the temperature rises
steadily through $T \sim 650$ K.
The H$_2$ fraction increases steeply in this regime as three-body
formation reactions (3H $\to$ H$_2$ + H) become efficient at
$n \gtrsim 10^8\,\mathrm{cm}^{-3}$, converting the gas to fully
molecular ($f_{\mathrm{H}_2} \to 1$) by
$n \sim 10^{11}$--$10^{12}\,\mathrm{cm}^{-3}$.
Note that unlike in the pre-stellar regime, we see an increase in temperature associated with increase in H$_2$ fraction. With the levels in LTE, increasing H$_2$ adds little cooling, so it cannot offset the compressional heating of the continuing infall. In real primordial gas the H$_2$ lines also become optically thick at these densities \citep{ripamonti_fragmentation_2004}, but our network does not model line opacity (\autoref{sec:thermochem}).

At $\Delta t = 1.2\kyr$ (\autoref{fig:phases}, right panels), the temperature range in the lower density $n\lesssim 10^{10}{\rm cm}^{-3}$ part of the phase diagram has broadened substantially. At higher densities, the gas has separated into 2 parts: a branch with $T < 2000$ K and a hot branch at $T \sim 5000$ K, which we interpret as the result of shock-heated and adiabatically compressed gas.
 This hot branch is capped at $T \approx 8000$\,K, where Lyman-$\alpha$ cooling of atomic hydrogen becomes efficient \citep{oh_haiman02, wise_abel07}.
We note that while our thermochemical model does not couple
H$_2$ dissociation to thermodynamics, the $\beta_{{\rm H}_2, i}$
terms in \autoref{eq:nh2} can still drive partial dissociation of H$_2$.

At higher densities attained at later times, $H_2$ is formed primarily by three-body interactions. This results in constantly increasing fraction of molecular hydrogen in the disk with time.

\subsection{Pre-stellar envelope and collapse}
\label{sec:prestellar}

Here we describe the mass distribution and infall rate of the pre-stellar envelope (\autoref{sec:envelope_mass}) and its kinematics (\autoref{sec:kinematics}), and then compare both with self-similar collapse solutions to show that the simulated collapse is driven from the outside in (\autoref{sec:outside_in}).
This contrasts with the inside-out expansion wave picture of \citet{Shu_1977}, which applies only when a core is marginally critical.

\begin{figure}[!htbp]
	\centering
	\includegraphics[width=\columnwidth]{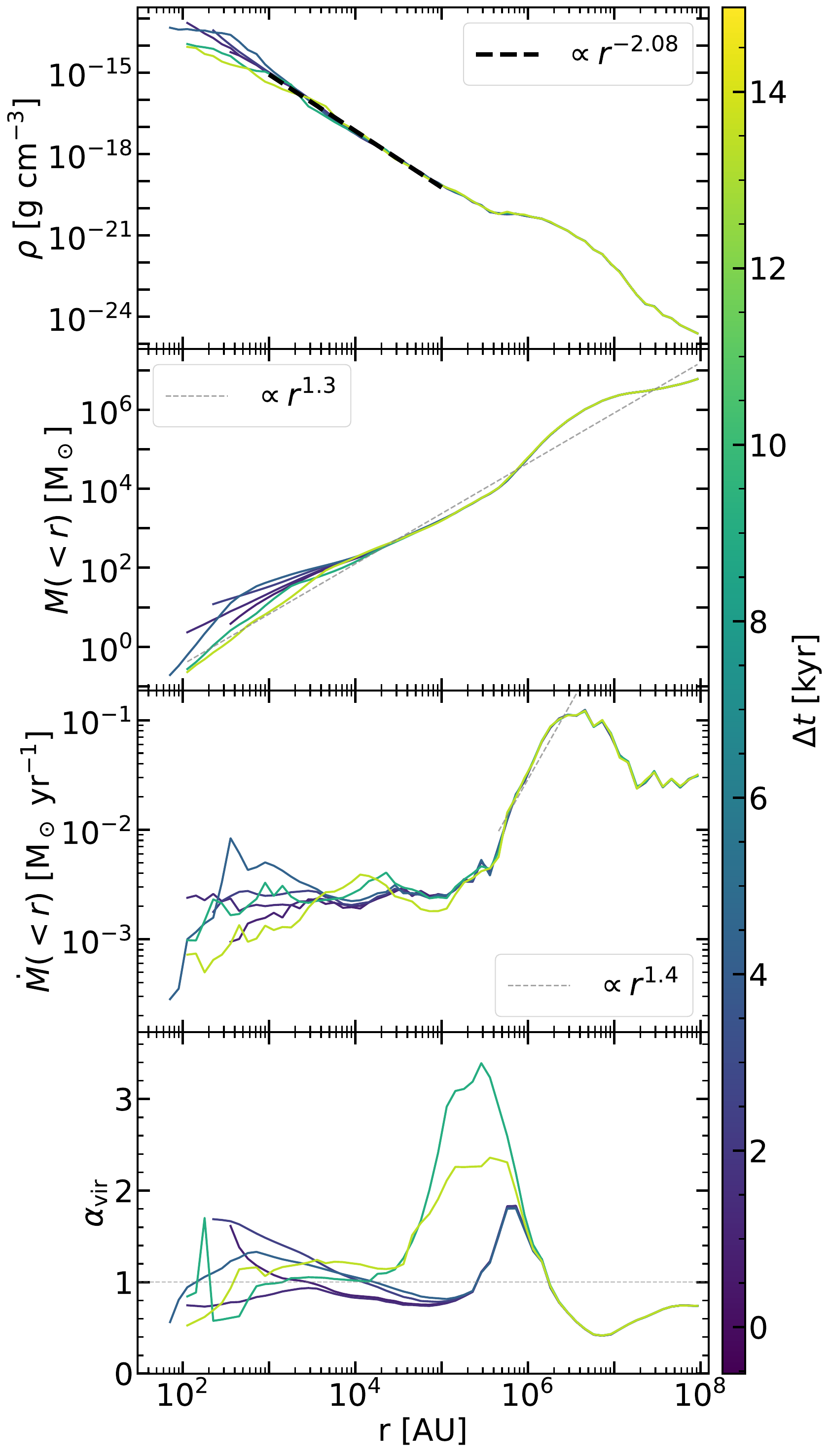}
	\caption{\label{fig:shell_mass_accretion}
		Top panel: Radial density profiles through simulation time.
		The x-axis begins at $\sim 1\,\mathrm{AU}$, corresponding to the
		resolution limit of the simulation. The dashed line is a single power law fitted to all epochs together between $10^{3}$ and $10^{5}\AU$, where the profiles coincide: $\rho \propto r^{-2.08}$.
		Top-middle panel: Enclosed gas mass $M(<r)$ in a sphere vs radius $r$ of the sphere. The fit is roughly $M \propto r^{1.3}$.
		Bottom-middle panel: the infall rate into a sphere of radius $r$,
		$\dot{M}_{\mathrm{infall}}(<r) = -\sum_{\mathrm{shell}} m_i v_{r,i}/r_i$,
		counting only infalling gas ($v_r < 0$) on the surface of the sphere of
		radius $r$. At later times
		the inner region gains mass slower than at the earlier times
		(mostly close to the time of the first sink forming). There is a characteristic
		``flattening'' at smaller radii inside of the disk before the first star has formed.
		Outer shells gain mass faster at all times -- a characteristic of ``outside-in''
		collapse.
        Bottom panel: evolution of radial profile of kinematic virial parameter
with time, calculated with the energies defined in \autoref{sec:energyevol} for the gas and sinks inside
a sphere of radius $r$ (dark matter and stars included in the potential). Each
curve starts at the resolution limit of its epoch. Inside $10^{5}\AU$,
$\alpha_\mathrm{vir} \approx 1$. }
	
\end{figure}

\begin{figure*}[!htbp]
	\centering
    \includegraphics[width=1\textwidth]{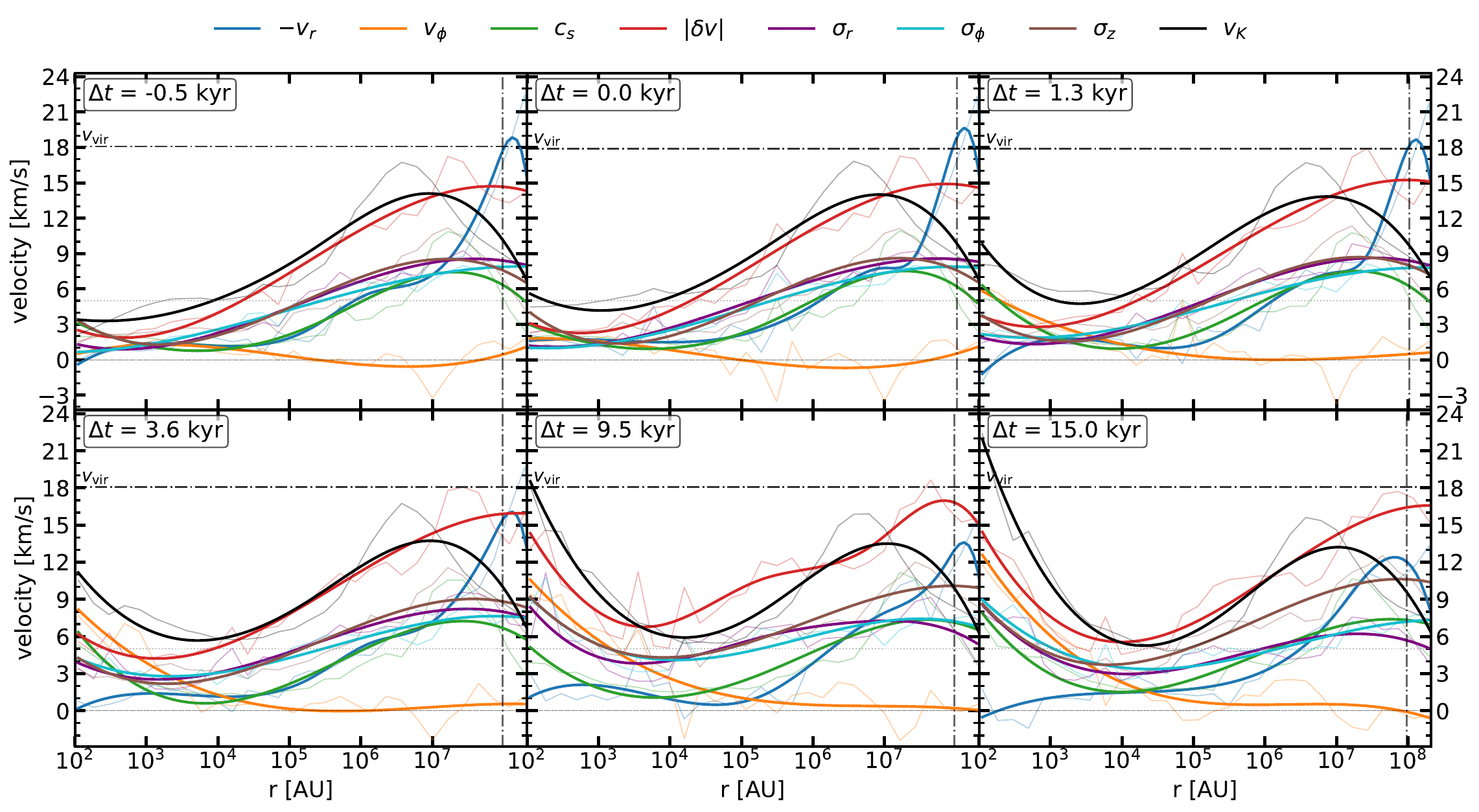}
	\caption{\label{fig:kinematics_wide}  Radial profiles of inward radial velocity $-v_r$, azimuthal velocity
		$v_{\phi}$, sound speed $c_s$, magnitude of turbulent velocity
		$\left| \delta v \right|$, radial/azimuthal/vertical turbulent velocity dispersion
		$\sigma_r/\sigma_\phi/\sigma_z$,
		Keplerian speed $v_K$ at each radius, computed from the enclosed mass.
		Bright lines show functions with spline smoothing, dim lines show actual data.
		$\sigma_r$ grows with time, as well as $\left| \delta v \right|$. Radial
		infall $(v_r)$ speeds up until the bulk star formation events  ($1.3-3.6\kyr$),
		and slows down afterwards. Disk rotation $(v_{\phi})$ is constantly speeding
		up, leveling off outside of the disk ($r \gtrsim 2500$ AU).
		$\sigma_r, \sigma_{\phi}, \sigma_z$ track each other very closely, but
		$\sigma_z$ decouples from the other two by the $\Delta t \gtrsim 9$ kyr time
		outside of the disk. Turbulence is supersonic at all times, tracking Keplerian
		speed $v_K$ closely for most times outside of the disk, but falling below it
		inside of the disk. Dot-dashed lines correspond to virial radius $r_\mathrm{vir}$ and virial velocity $v_\mathrm{vir}$ of the halo in the x and y axes respectively.
		The turnover seen in the innermost bins of these wide profiles
		is a numerical artifact of plotting data in the cutout region -- it does not exist in \autoref{fig:vr_fullsim}, which uses the full simulation data. This is not a physical feature
		and thus is excluded from the interpretation.}
\end{figure*}

\subsubsection{Mass distribution and infall rate}
\label{sec:envelope_mass}

The density profiles of all epochs coincide outside $\sim 10^{3}\AU$, so only
the inner region changes with time (\autoref{fig:shell_mass_accretion}, top
panel). Between $10^{3}$ and $10^{5}\AU$ the envelope follows $\rho \propto r^{-2.08}$ (one fit to all epochs). Inside $\sim 10^{3}\AU$ the profile changes with time: it is shallower before the first sink forms, when the center is not yet fully refined, and it becomes irregular once the spiral arms and the sinks dominate the azimuthal average. 

The enclosed gas mass (\autoref{fig:shell_mass_accretion}, top-middle panel)
follows $M(<r) \propto r^{1.3}$ overall, a mean density slope of $\bar\rho \propto r^{-1.7}$ over the whole range, 
and the curves of all epochs again differ only
at small radii. The infall rate through each  sphere (\autoref{fig:shell_mass_accretion},
bottom-middle panel) is nearly independent of radius, at 
$\approx 3\times10^{-3}\,\Msun\,\mathrm{yr^{-1}}$ between $\sim 10^{2}$ 
and $\sim 10^{5}\AU$, at all epochs, and rises as $\dot M \propto r^{1.4}$ 
further out, to 
$\sim 10^{-1}\,\Msun\,\mathrm{yr^{-1}}$ at $\sim 5\times10^{5}\AU$. A flux
that does not change with radius means that the gas does not pile up on its way
in: the inner envelope is a quasi-steady flow fed from larger scales. 

The virial parameter $\alpha_\mathrm{vir}(r)$
(\autoref{fig:shell_mass_accretion}, bottom panel) is the ratio
$2(E_\mathrm{kin}+E_\mathrm{therm})/|E_\mathrm{pot}|$ of the energies defined
for the disk in \autoref{sec:energyevol} (\autoref{eq:Energies}), but
calculated for all the gas and sinks inside a sphere of radius $r$ instead of
the disk particles only (\autoref{fig:energy_evol}).  The ratio is
$\alpha_\mathrm{vir} \approx 0.85$ before the first sink forms and 
$\approx 1.1$ afterwards between $10^{3}$ and $10^{5}\AU$, so the gas is close
to virial balance in the collapsing region outside of the disk. 

\subsubsection{Gas kinematics}
\label{sec:kinematics}

\begin{figure}[!htbp]
	\centering
	\includegraphics[width=\columnwidth]{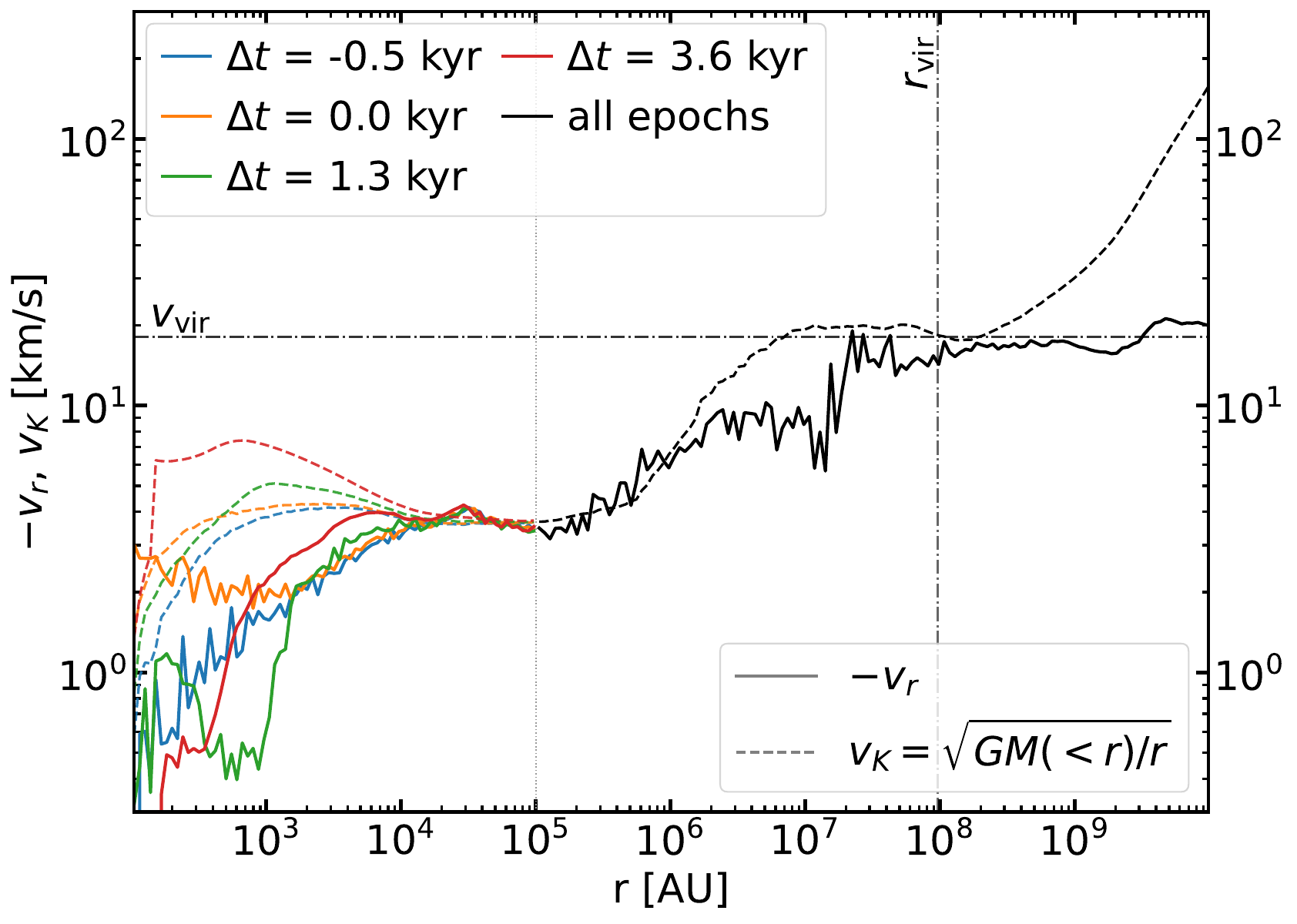}
	\caption{ Radial inflow profile over the full simulation volume, out to
		$\sim 10^{9}\,\AU$ (far beyond the virial radius), at four epochs
		($\Delta t = -0.5$, $0$, $1.3$, and $3.6\kyr$): the
		mean inflow velocity $-v_r$ (solid) and the Keplerian speed
		$v_K = \sqrt{G M(<r)/r}$ from the total enclosed mass, including dark matter
		(dashed), both in the color of each epoch, and in black beyond $10^{5}\AU$, where the epochs are combined. Beyond $\sim 10^{4}\AU$ the curves of
		the four epochs coincide, because the enclosed mass and the inflow there barely
		change over these $4\kyr$, with the virial velocity
		$v_\mathrm{vir}$ and the virial radius $r_\mathrm{vir}$ marked.
		\textit{Bottom}: enclosed gas mass $M_\mathrm{gas}(<r)$. Unlike the
		wide kinematics profiles of \autoref{fig:kinematics_wide}, this
		plot reaches the scale of the halo. The few bins inside $200\AU$ with $-v_r \le 0$ are not shown
		on the logarithmic axis. }
	\label{fig:vr_fullsim}
\end{figure}

\autoref{fig:kinematics_wide} shows radial profiles of the gas velocity
components, the turbulent velocity magnitude and dispersion, and the Keplerian
speed $v_K = \sqrt{G M(<r)/r}$ of the enclosed mass. We plot the radial velocity, $-v_{r}$ (negated
so that positive values indicate radially inward flow), to simplify comparison.
All quantities are mass-weighted averages in logarithmically spaced
cylindrical bins about the disk axis. Here $v_r$ and $v_\phi$ are the mean
radial and azimuthal velocities of a bin and $\sigma_r = \sqrt{\langle
v_r^2\rangle - \langle v_r\rangle^2}$ is the radial velocity dispersion. The
turbulent velocity $|\delta v| = \langle[(v_r-\langle v_r\rangle)^2 +
(v_\phi-\langle v_\phi\rangle)^2 + v_z^2]^{1/2}\rangle$ is the mean speed after
subtracting the bin's mean radial and azimuthal flow, and $c_s$ is the sound speed.

 The general trend is that $v_{\phi},\, \left| \delta v \right|, \sigma_r$ increase with time everywhere. $v_r$ is negative and thus gas is mostly infalling for the whole simulation, with $|v_r|$ increasing until stabilizing at around 1.3-3.6 kyr (bulk star formation events), and dropping after stars are fully formed. $c_s$ rises until 1.3 kyr, after which it stabilizes. It is clear that the turbulence starts off as sonic or sub-sonic, but becomes greatly supersonic once the sink forms. The turbulent velocity dispersion increases inward, tracking the increase in gravitational potential. Because the collapse compresses the gas faster than the turbulence can decay, the turbulence is continuously re-driven by the contraction itself rather than by stellar feedback \citep{robertson_goldreich12}; feedback only begins to contribute once the first protostar forms.

\autoref{fig:vr_fullsim} shows the same inflow over the full simulation
volume: beyond the disk, $-v_r$ rises with radius overall. Here $r_\mathrm{vir}$ is the radius inside which the mean total density (gas, dark matter, stars and sinks) equals $\Delta_c\,\rho_\mathrm{crit}(z)$, with the spherical-collapse overdensity $\Delta_c = 18\pi^2 + 82x - 39x^2$ and $x = \Omega_\mathrm{m}(z) - 1$ \citep[the fit for a flat universe of][]{bryan98}, $M_\mathrm{vir}$ is the mass inside it, and $v_\mathrm{vir} = \sqrt{GM_\mathrm{vir}/r_\mathrm{vir}}$. Between $10^{4}$ and
$10^{6}\AU$, where the gas dominates the enclosed mass, it stays close to the
local Keplerian speed ($-v_r \approx v_K$, i.e.\ $\approx 0.7$ of the local free-fall speed $\sqrt{2}\,v_K$), so this envelope
collapses at close to free fall. Further out $-v_r$ drops below $v_K$, to
$\approx 0.45\,v_K$ near $10^{7}\AU$, and reaches $\approx 0.8\,v_\mathrm{vir}$
at the virial radius, about half the free-fall speed of the whole halo. Beyond the virial radius $v_K$ rises steeply as the enclosed mass takes in the surrounding structure, while $-v_r$ stays at $\approx 18\,\mathrm{km\,s^{-1}}$. This
extends the outside-in signature seen in the cutout to the full range of scales
feeding the core. Inside $\sim 10^{4}\AU$, $-v_r$ falls below $v_K$ as rotation takes over, and within the disk ($r < 2500$ AU) the flow becomes predominantly subsonic. The deceleration is gradual rather than abrupt
(\autoref{sec:morphology}).

\subsubsection{Outside-in collapse}
\label{sec:outside_in}

The infall-rate profile (\autoref{fig:shell_mass_accretion}, bottom-middle panel), flat inside and rising outward, is the opposite of an inside-out
collapse, where $\dot M$ approaches a constant at small radii but falls as
$1/r$ at large radii \citep{Shu_1977, murray16}, and so is a signature of outside-in collapse. It is what the model of self-gravitating turbulent
collapse of \citet{murray15} predicts, with $\dot M$ independent of radius
inside the sphere of influence of the central mass (their Eq. 45), and what
\citet{murray16} find in their simulations, where the accretion rate is either nearly flat or increasing with radius. Our plateau extends over a
much wider range of radii and our density slope ($-2.08$) differs from their
$-3/2$ inside the sphere of influence, so we regard the comparison to their
solution as qualitative.

 We find that the outer layers gain mass faster
(\autoref{fig:shell_mass_accretion}, bottom-middle panel).
This contrasts with the inside-out model of
\citet{Shu_1977} and is qualitatively similar to the runaway,
Larson--Penston-type collapse \citep{Larson_1969, Penston_1969}. Our envelope slope, $\rho \propto r^{-2.08}$ between $10^{3}$ and $10^{5}\AU$, is close to the $\rho \propto r^{-2}$ of the isothermal Larson--Penston solution, although the correspondence remains approximate because our gas is not isothermal. Supersonic
infall alone does not distinguish the two pictures, since the interior of the
Shu expansion wave also falls in supersonically. The envelope allows to
differentiate between the two cases.  The envelope mass flux, 
$\dot M \approx 10^{-2}\, \Msun\,\mathrm{yr^{-1}} \approx 48\,c_s^3/G$,
is well above the Shu rate and comparable to the Larson--Penston values. 

The comparison above concerns the collapse
dynamics only: the core is not a Bonnor--Ebert sphere
\citep{Ebert_1955, Bonnor_1956} since 
(i) the core is never in hydrostatic equilibrium: it is assembled
continuously by supersonic inflow from the halo
(\autoref{fig:kinematics_wide}, \autoref{fig:vr_fullsim}) and its virial
parameter is $\alpha_\mathrm{vir} \approx 0.9$ between $10^{3}$ and $10^{5}\AU$
at $\Delta t = -0.5\kyr$ (\autoref{fig:shell_mass_accretion}, bottom panel).
The gas is therefore near virial balance, but because it is still infalling
there is no static equilibrium state that later becomes unstable.
(ii) The gas is not isothermal: across the pre-stellar core the
temperature changes between $\sim 200$ and $\sim 800$\,K as H$_2$ cooling
saturates and compressional heating takes over (\autoref{sec:thermo}).
(iii) There is no confining external pressure: the core merges smoothly
into the infalling halo envelope instead of ending at a pressure boundary.
Rotation, turbulence and magnetic fields also provide support that the
Bonnor--Ebert model does not include.
The Larson--Penston solution remains a useful comparison because it does not
require an equilibrium starting state: \citet{Larson_1969} found that the
inner density and velocity profiles approach the same self-similar form for
various initial conditions, provided the collapse time exceeds the
sound-crossing time.

\subsection{Disk evolution}
\label{sec:disk}

Once the accretion flow acquires sufficient angular momentum to form a centrifugal barrier, a rotationally supported disk grows around the central protostar. The disk radius $\rdisk$ (\autoref{sec:diskid}) stays at $\approx 1.8\times10^{3}\AU$ throughout the run, close to $r_\mathrm{search}$. The disk gas mass grows from $\approx 28\,\Msun$ just before the first sink forms to $\approx 65\,\Msun$ at $\Delta t \approx 3.5\kyr$, and then falls to $\approx 35\,\Msun$ at $9.5\kyr$ and $\approx 16\,\Msun$ at $15\kyr$ as it is converted into stars. Its mass-weighted thickness, $\langle z^2\rangle^{1/2}/\langle R\rangle$, is $0.23$. The disk is thus comparable in size to the $\sim 0.01$\,pc ($\approx 2000\AU$) disk of \citet{meziani26}, but unlike that Toomre-stable disk,  here the disk fragments (\autoref{sec:stability}).
The subsequent evolution of the system is governed by the competition between gravitational instability (which fragments the outer disk into secondary protostars) and the supply of mass and angular momentum by infall from the envelope; the magnetic energy stays well below the thermal and kinetic energies (\autoref{sec:bfield}).

\subsubsection{Energetics}
\label{sec:energyevol}

In this section, we consider the interplay of the gravitational, kinetic,
thermal and magnetic energies for the particles identified in the disk.
We compute the following energy components:
potential $E_{\text{pot}}$,
rotational $E_{\text{rot}}$,
turbulent $E_{\text{turb}}$,
thermal $E_{\rm therm}$,
and magnetic $E_{\text{mag}}$, each summed over the gas particles identified as disk members (\autoref{sec:diskid}); the explicit expressions are given in \autoref{eq:Energies} below. 

\begin{subequations}
	\label{eq:Energies}
	\begin{align}
		E_\mathrm{pot}   & = -G \sum_{\substack{i < j \\ i,j \in \mathrm{disk}}}
		\frac{m_i \, m_j}{\sqrt{|\mathbf{r}_i - \mathbf{r}_j|^2 + \epsilon^2}}
		\nonumber \\
		& \quad -\; G \sum_{s \in \mathrm{sinks}} \sum_{i \in \mathrm{disk}}
		\frac{m_s \, m_i}{\sqrt{|\mathbf{r}_s - \mathbf{r}_i|^2 + \epsilon^2}}, \\
		E_\mathrm{rot}   & = \tfrac{1}{2}\sum_{i \in \mathrm{disk}} m_i v_{\phi,i}^2, \\
		E_\mathrm{turb}  & = \tfrac{1}{2}\sum_{i \in \mathrm{disk}} m_i |\mathbf{v}_i
		- v_{\phi,i}\hat{\boldsymbol{\phi}} - v_{r,i}\hat{\boldsymbol{r}}|^2, \\
		E_\mathrm{therm} & = \sum_{i \in \mathrm{disk}} m_i u_i, \\
		E_\mathrm{mag}   & = \sum_{i \in \mathrm{disk}} \frac{|\mathbf{B}_i|^2}{8\pi}
		\frac{m_i}{\rho_i}
	\end{align}
\end{subequations}
where $m_i$ is the mass of particle $i$, $\textbf{v}_i$ its total velocity vector,
$v_{\phi, i}, v_{r, i}$ are bulk velocity profile fits for the azimuthal and radial
components in each annulus (both functions of $r$ only),
$\textbf{B}_i$ is the magnetic field vector,
$\rho_i$ is the density, and $u_i$ is the internal energy.

\begin{figure}[!htbp]
	\centering
	\includegraphics[width=\columnwidth]{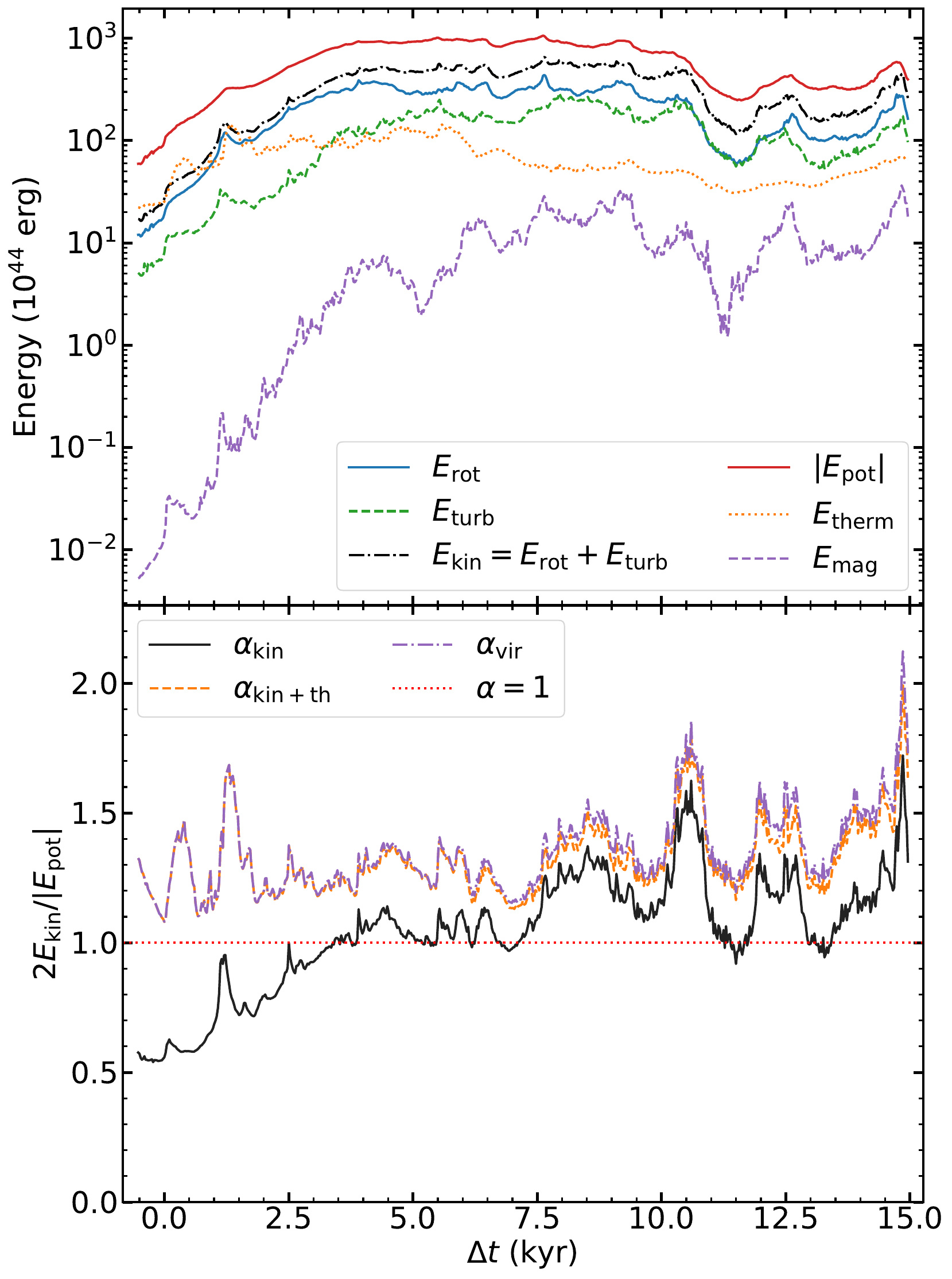}
	\caption{Top panel: Time evolution of the main energy components computed for
		disk-identified particles
		(the disk radius $r_{\mathrm{disk}}$ is taken as the
		90th-percentile cylindrical radius of disk-identified
		particles, see \autoref{sec:disk}):
		rotational kinetic energy $E_\mathrm{rot}$ (yellow),
		turbulent kinetic energy $E_\mathrm{turb}$ (cyan), total kinetic energy
		$E_\mathrm{kin} = E_\mathrm{rot} + E_\mathrm{turb}$ (green), gravitational
		potential energy $|E_\mathrm{pot}|$ (red), thermal energy $E_\mathrm{therm}$
		(orange), and magnetic energy $E_\mathrm{mag}$ (magenta).
		Bottom panel: pure kinematic virial ratio
		$\alpha_{\text{kin}} = 2 E_\mathrm{kin}/|E_\mathrm{pot}|$
		(black); virial ratio with added thermal component (orange)
		$\alpha_{\text{kin+th}} = 2 \left(  E_\mathrm{kin} + E_\mathrm{therm} \right)/|E_\mathrm{pot}|$
		, and with added
		thermal and magnetic components (purple),
		$\alpha_{\text{vir}} = 2 \left(  E_\mathrm{kin} + E_\mathrm{therm} + E_\mathrm{mag} \right)/|E_\mathrm{pot}|$
		. The dashed red line marks virial
		equilibrium ($= 1$). The disk starts sub-virial and reaches equilibrium by
		$\sim 2\kyr$.
		Including the thermal energy matters mainly at early times, when
		$E_\mathrm{therm}$ is comparable to the kinetic terms; at late times the
		disk is supported mainly by rotation and turbulence.
		Magnetic energy provides negligible
		additional support to the virial ratio. }
	\label{fig:energy_evol}
\end{figure}

\autoref{fig:energy_evol} shows the time evolution of the main energy
components computed for disk-identified particles. Throughout the simulation,
$|E_\mathrm{pot}|$ dominates over all other components, confirming that the
disk is gravitationally bound. $E_\mathrm{rot}$ and $E_\mathrm{turb}$
are the main support terms and track each other to within a factor of $\sim 2$,
with $E_\mathrm{rot}$ slightly exceeding $E_\mathrm{turb}$. $E_\mathrm{therm}$
starts comparable to the kinetic energies at early times but becomes
subdominant at late times as the gravitational potential well deepens faster than the gas heats.

The magnetic energy,
$E_\mathrm{mag}$, grows
over the $\sim 15\kyr$ simulation, from $\sim 10^{42}$ erg at
$\Delta t \approx -0.5\kyr$ to $\sim 10^{45.3}$ erg at $15\kyr$. This growth
is consistent with a combination of flux-freezing compression during the collapse,
field-line winding
by differential rotation, and potentially both the small-scale turbulent dynamo and the large-scale mean-field dynamo
\citep{sharda21}, but we cannot separate these channels (\autoref{sec:bfield}). Despite this rapid amplification, $E_\mathrm{mag}$ remains
below $E_\mathrm{therm}$ throughout the simulation, though the gap narrows at
late times -- consistent with the declining mass-to-flux ratios discussed in
\autoref{sec:bfield}.

The bottom panel shows the virial ratio $2 E_\mathrm{kin} / |E_\mathrm{pot}|$.
At early times ($\Delta t < 0$), the virial ratio is $\sim 0.65$,
indicating a sub-virial, contracting state, i.e. the system has
insufficient kinetic energy to balance gravity.
The virial ratio rises as the disk forms and spins up, reaching $\sim 1.0$ by
$\Delta t \approx 2\kyr$. From $2$ kyr onward, the virial ratio oscillates
around unity, indicating that the disk has achieved and maintains approximate
virial equilibrium through a balance of rotational support, turbulent pressure,
and gravitational collapse. 
The virial ratio never approaches the unbound threshold of $2$, except towards the end of our run, confirming that the system remains gravitationally bound throughout.

\subsubsection{Disk morphology and evolution}
\label{sec:morphology}

As the disk accretes mass and angular momentum, its morphology undergoes a characteristic sequence: an initially smooth, centrally concentrated structure gives way to spiral arms as the Toomre parameter drops below 1 ( the disk becomes unstable), and those arms subsequently fragment into secondary sinks. Face-on and edge-on surface density projections at six representative epochs
spanning the $\sim 15\kyr$ after the formation of the first sink ($t_1$) are shown in
\autoref{fig:combined_density} (\autoref{sec:diskid}).

The disk does not begin as a rotationally supported
structure. At the pre-stellar epoch $v_\phi$ lies well below the local
Keplerian speed at every radius in \autoref{fig:kinematics_wide}, so the
centrifugal support of the collapsing core is weak and the infall is close
to radial. Rotational support is acquired only gradually: $v_\phi$ rises
monotonically with time while $-v_r$ falls after the bulk star formation
events, and only then does $v_\phi / v_K$ approach the
$f_\mathrm{kep} = 0.3$ threshold used for disk membership
(\autoref{sec:diskid}) over an extended range of radii. The disk we
identify is therefore assembled from an initially slowly rotating,
predominantly infalling core. The disk starts thick, with aspect ratio $h/r$  
at $\approx 0.9$ during the pre-stellar epoch ($\Delta t = -0.5\kyr$), drops to
a minimum of $\sim 0.3$ around the first bulk star formation event 
($\Delta t \approx 3.5\kyr$) as the disk settles and gains rotational
support, and then rises again at later times, to $\approx 0.8$ at $15\kyr$.
This late-time increase is most plausibly a consequence of the disk being
disrupted: as gas is converted into stars it is removed from the disk, leaving
material that is less rotationally supported and vertically more
extended.

\subsection{Disk stability and fragmentation}
\label{sec:stability}

 In this section, we analyze the stability properties of the disk. Whether a self-gravitating disk fragments into bound clumps, or instead disperses the accumulated mass through spiral-arm shocks and turbulent transport depends on the ratio of stabilizing pressures (thermal, turbulent) to the destabilizing self-gravity of the gas. The Toomre parameter, defined in \autoref{eq:toomreq} \citep{Toomre_1964}, quantifies this balance locally: $Q < 1$ marks regions where perturbations on the disk scale grow faster than orbital shear can wind them up.
We track the spatial and temporal evolution of $Q$ to identify when and where fragmentation occurs, and how the velocity dispersion stirred up by the forming sinks subsequently self-regulates the disk back toward marginal stability.

\subsubsection{Toomre-Q maps and radial profiles}
To characterize non-thermal motions in the disk, we decompose
particle velocities into radial and azimuthal components
$(v_r, v_\phi)$ in the disk-plane coordinate system.
We subdivide the disk into concentric cylindrical annuli and compute
the mass-weighted mean velocity in each annulus, yielding radial
profiles $\bar{v}_r(r)$ and $\bar{v}_\phi(r)$. The residual (streaming-subtracted) velocity for each particle is
$\vec{\delta v} = (v_r - \bar{v}_r,\; v_\phi - \bar{v}_\phi,\; v_z)$.
The velocity dispersion $\sigma_v$ is computed as the root-mean-square
of $|\vec{\delta v}|$ within each annulus.

We calculate the Toomre $Q$ parameter as:
\begin{equation}
	\label{eq:toomreq}
	Q = \frac{\zeta \kappa}{\pi G \Sigma}, \; \zeta = \sqrt{\sigma_r^2 + c_s^2}
\end{equation}
where $c_s$ is the sound speed, $\sigma_r$ is the radial velocity dispersion, $\kappa$ the epicyclic frequency, computed numerically from the angular velocity profile as $\kappa^2 = (2\Omega/r)\,\mathrm{d}(r^2\Omega)/\mathrm{d}r$ (so $\kappa = \Omega$ for Keplerian rotation) ,
and $\Sigma$ the surface density. 
In hot disks $\zeta \approx c_s$, while in N-body systems (like a galactic disk) $\zeta \approx \sigma_r$. 
$\zeta = \sqrt{\sigma_r^2 + c_s^2}$ thus includes both thermal
and turbulent contributions, since supersonic turbulence provides
additional effective pressure support. We measure $\Sigma$, $\sigma_r$, $\Omega$ and $\kappa$ in 20 radial annuli and construct a pixel-by-pixel map in the same way; the estimators are given in \autoref{app:toomre}.

\begin{figure*}[!htbp]
	\centering
	\includegraphics[width=\textwidth]{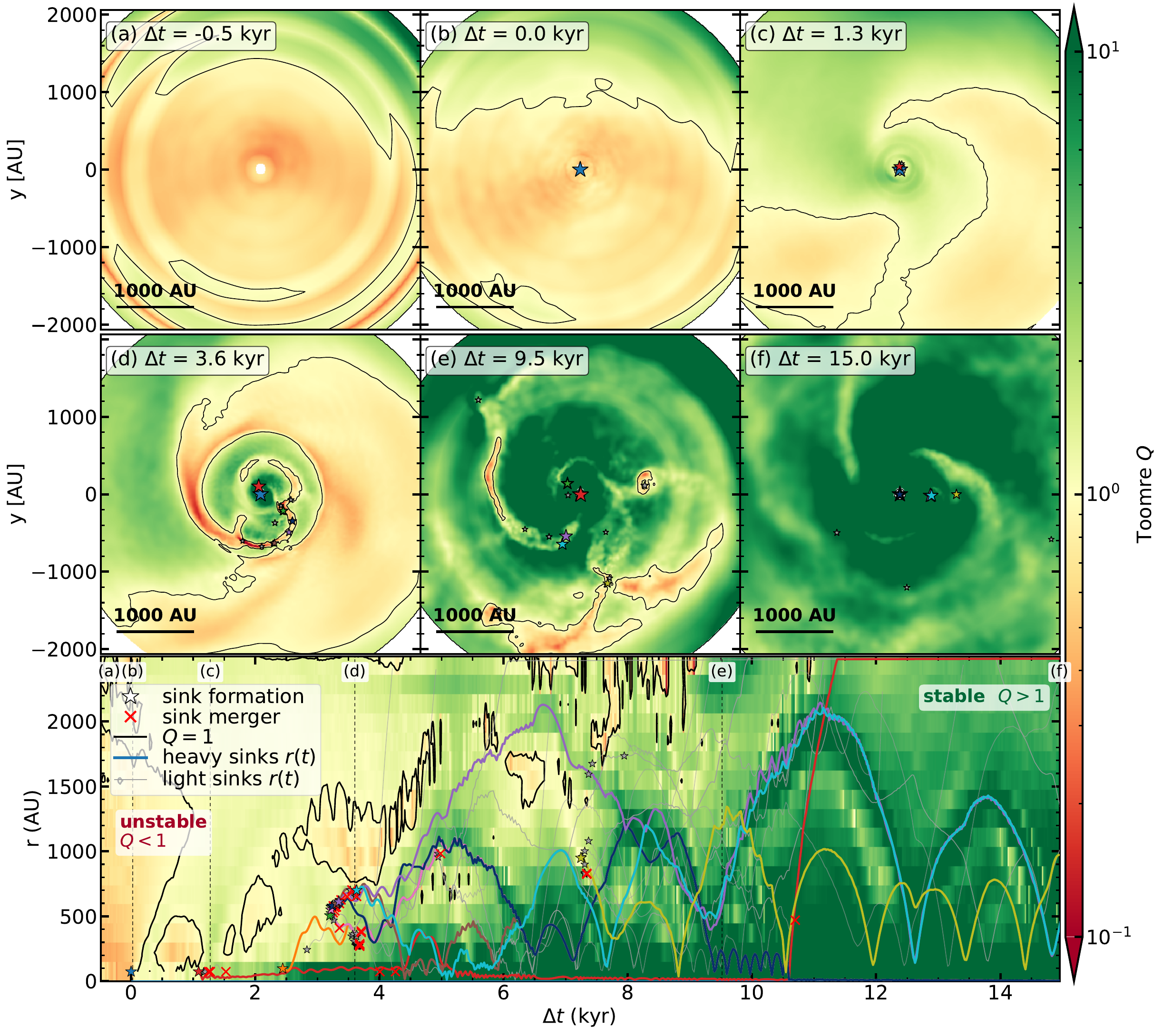}
	\caption{Top: Face-on maps of the local Toomre $Q$ parameter at six epochs.
		The black contour marks $Q = 1$.
		Bottom: Evolution of radial profile of Toomre $Q$ (heatmap), and sink formations,
		mergers, and radial positions' evolution (lines).
		At early times the gas is globally unstable ($Q < 1$); at late times
		the inner disk is strongly stabilized while the outer spiral arms
		remain near or below $Q = 1$.
		Radial profile of Toomre $Q$ at six epochs. The horizontal
		dashed line marks $Q = 1$. At early times the entire region has
		$Q < 1$; at late times a radial gradient develops, with
		$Q \sim 60$ at $r < 100\AU$, falling to $Q \sim 1$ at $r \sim 2700\AU$.
		Heavy sinks are the 10 most massive sinks at $\Delta t = 15\kyr$ (sinks that have merged away by then do not count);
		all other sinks are light and drawn in gray. In the maps, the size of the sink markers scales with sink mass.
		In the bottom panel, each heavy sink's trajectory $r(t)$ and formation marker have the same color as that
		sink in \autoref{fig:imf} and \autoref{fig:mass_evol_individual}. Most massive sink always stays at $r=0$ by definition of the center of our disk.
	}
	\label{fig:toomre_Q_merged}
\end{figure*}

\autoref{fig:toomre_Q_merged} (top) shows face-on maps of the local $Q$.
At the pre-stellar epoch ($\Delta t = -0.5\kyr$), the gas is
broadly unstable with $Q \lesssim 1$ throughout most of the region.
By $\Delta t = 0$, the central region near the first sink begins to
stabilize, while the surroundings remain marginally unstable.
At $\Delta t \approx 1.3\kyr$, a clear boundary emerges between
a stabilized inner region and an unstable outer envelope.
By the time spiral arms have developed ($\Delta t \approx 3.6\kyr$),
a ring of $Q < 1$ material traces the spiral arms, while the inner disk
is stabilized ($Q > 1$) by the strong velocity dispersion generated
near the central sinks.

\begin{figure}[!htbp]
	\centering
	\includegraphics[width=\columnwidth]{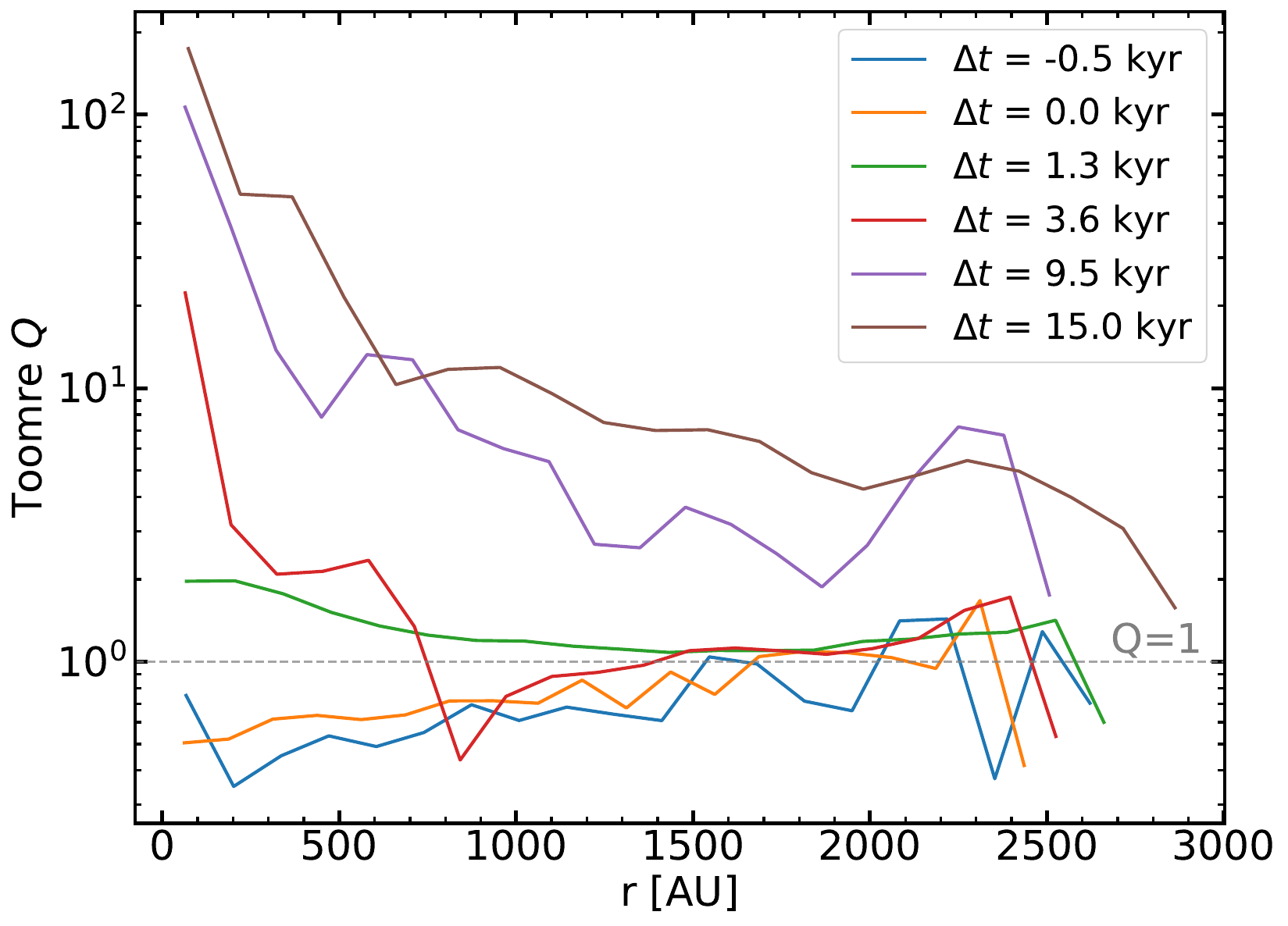}
	\caption{Toomre Q radial profiles time evolution.
		The disk starts with $Q < 1$ inside and $Q \approx 1$ near the edge,
		but $Q$ rises with disk evolution, since gas is removed from the disk because of
		star formation events. Center of the disk is extremely stable ($Q \sim 60$)
		by the 15 kyr mark, preventing any further star formation in the middle. }
	\label{fig:profile_toomre_Q}
\end{figure}

The radial Toomre $Q$ profile (\autoref{fig:profile_toomre_Q}) quantifies
its evolution.
At the pre-stellar epoch ($\Delta t = -0.5\kyr$), $Q$
is below unity inside $\sim 1500\AU$ ($Q \approx 0.55$) and close to unity
further out, reflecting a
globally unstable collapsing core.
As the first sink forms and accretes, the inner $Q$ rises:
by $\Delta t = 1.3\kyr$ (green), $Q \approx 1$ throughout, consistent
with a marginally stable state.
By $\Delta t = 9.5\kyr$ (brown),
$Q$ is mostly stable in the whole disk, with the central region ($r \lesssim 150\AU$, where $Q \gtrsim 60$) being the most stable. At later times a radial gradient develops, with the inner disk more stable ($Q \gg 1$); the outer spiral arms seen at earlier epochs remain near or below $Q \sim 1$. By the latest epochs the disk ($r \lesssim 2000\AU$) is Toomre-stable throughout, and the $Q$ profile remains near or above unity out to $r \sim 2700\AU$.

\subsubsection{What sets Q: thermal versus turbulent support}

The shape of these profiles follows directly from the disk kinematics
of \autoref{fig:kinematics_wide}. Using $\kappa \approx \Omega =
	\bar{v}_\phi / r$ and $\Sigma \approx M_\mathrm{disk}/(\pi r^2)$ for the disk
mass enclosed at $r$, and eliminating $G$ through
$v_K^2 = G M_\mathrm{enc}/r$, \autoref{eq:toomreq}
can be rearranged into \autoref{eq:toomre_scaling}:
\begin{equation}
	Q \;\approx\; \frac{\zeta}{v_K}\,
	\frac{\bar{v}_\phi}{v_K}\,
	\frac{M_\mathrm{enc}}{M_\mathrm{disk}},
	\label{eq:toomre_scaling}
\end{equation}
where $\zeta = \sqrt{\sigma_r^2 + c_s^2}$ as above and
$M_\mathrm{enc} = M_* + M_\mathrm{disk}$. The balance between these two
terms in $\zeta$ changes with time. Near marginal stability ($0.3 < Q < 3$),
thermal pressure dominates ($\Delta t \lesssim 4\kyr$).
The radial velocity dispersion only becomes
comparable to $c_s$ at $\Delta t \approx 5 \kyr$ and dominates afterwards.

The $Q<1$ at early times is easy to explain from \autoref{eq:toomre_scaling}:
at that point the first two terms are $<1$ (turbulence is sonic/subsonic, and 
rotational support is weak), and $M_* \approx 0$, making the last term 1.
The product is well below  unity at every radius, which is why the core starts
out globally unstable.

After the first sinks form, all three factors from \autoref{eq:toomre_scaling}
move upward. $\sigma_r$ and $\left| \delta v \right|$ (mostly tracing $\zeta$)
grow monotonically. The inner $v_K$ rises slightly as $\sqrt{M_\mathrm{enc}}$
is growing from the infalling gas; $\bar{v}_\phi$ rises steadily as the disk
spins up; and the stellar mass grows quadratically while the disk gas is
depleted by star formation, so $M_\mathrm{enc}/M_\mathrm{disk}$ climbs by more
than an order of magnitude.

The extreme central values ($Q \gtrsim 10^2$ inside
$\sim 150\AU$) are dominated by the last of these, since there the
enclosed mass is almost entirely stellar and very little gas is left. The
outer disk stays near $Q \sim 1$ because it remains gas-dominated and its
turbulent velocities stay a roughly fixed fraction of the local orbital
speed. The stabilization of the disk is therefore kinematic and structural
rather than thermal: turbulence rises relative to the orbital speed, the
disk spins up, and gas is converted into stars.

\subsubsection{Fragmentation, heating fronts, and self-regulation}

\autoref{fig:toomre_Q_merged} (bottom) shows the space--time evolution of the
azimuthally-averaged Toomre $Q$ radial profile.
The disk begins in a predominantly unstable state ($Q < 1$, red)
at all radii.
The first sink forms at the center, and within $\sim 0.2\kyr$ the
$Q = 1$ contour (black line) begins to propagate outward from
the center, tracing a wave of stabilization driven by heating of the gas, which we speculate is caused by the first sink.
A second stabilization wave starts at $\Delta t \approx 1.1\kyr$, together
with the formation of a second sink near the center.
The leading edge of the first wave,  $2.5$ times the sound speed of the gas
ahead of it. Both fronts are driven by a rise in the sound speed.
We speculate that this gas is heated by the first sink. . 

The second wave coincides with the formation of the second sink, but the second
sink doesn't start producing radiation until the "stability wave" has already
began traveling through the disk. We speculate that the second wave is also
driven mainly by the first sink. Because $Q$ is then already
$\lesssim 1$ throughout the inner $1000\AU$, the $Q = 1$ contour of this wave
sweeps the whole disk within $\lesssim 0.1\kyr$, while the maximum of $Q$ moves outward
at $\approx 4.5\,\mathrm{km\,s^{-1}}$. Only sinks forming at the center produce
such fronts, because later sinks form in gas that has already been heated. In
the second wave the gas stays at $\gtrsim 2000$\,K for a further $\sim
0.2 \kyr$, while at $150\AU$ the molecular fraction falls from $\approx
0.5$ to $0$ at $ \Delta t \sim 1.2\kyr$. This is consistent with
collisional dissociation by atomic hydrogen in the network, which at these
temperatures destroys H$_2$ in much less than $50$\,yr but, unlike in real gas,
does not absorb energy (\autoref{sec:thermochem}).

The early stabilization waves are therefore thermally driven. After
$\sim 3\kyr$ the inner disk stays stable because accretion onto sinks depletes
its gas, while $Q$ falls back to $\lesssim 1$ farther out as gas accumulates.
At $\approx 3\kyr$ a spiral arm fragments into $\sim 40$ closely spaced sinks
within $\approx 1\kyr$; many merge soon after \citep[see][for the merging
criteria]{STARFORGE_Towa_Grudic_2021}, leaving $\sim 19$ stars. Locking gas into
stars lowers the disk mass and stabilizes it. 

The outer disk reaches $Q>1$ by
$\sim 8\kyr$, coinciding with the second and last bulk star formation event,
which converts most of the available disk gas into stars. In the bottom panel of
\autoref{fig:toomre_Q_merged}, each successive event occurs farther from the
center because the inner gas has already been depleted, so stability propagates
inside-out.

\subsection{Star Formation and the protostellar mass function}
\label{sec:massevol}

In this section, we analyze the star formation outcome of the system by
quantifying the total stellar mass growth, the instantaneous star formation efficiency within the disk, and the shape of the proto-stellar mass function (PMF).
Because the jet feedback in this run is deposited close to the protostars and does not propagate as far as intended (\autoref{sec:stars}), the stellar masses, accretion rates and star formation efficiencies below  represent upper limits (\autoref{sec:caveats}) .

\subsubsection{Total stellar mass growth and accretion rates}

\begin{figure}[!htbp]
	\centering
	\includegraphics[width=0.8\columnwidth]{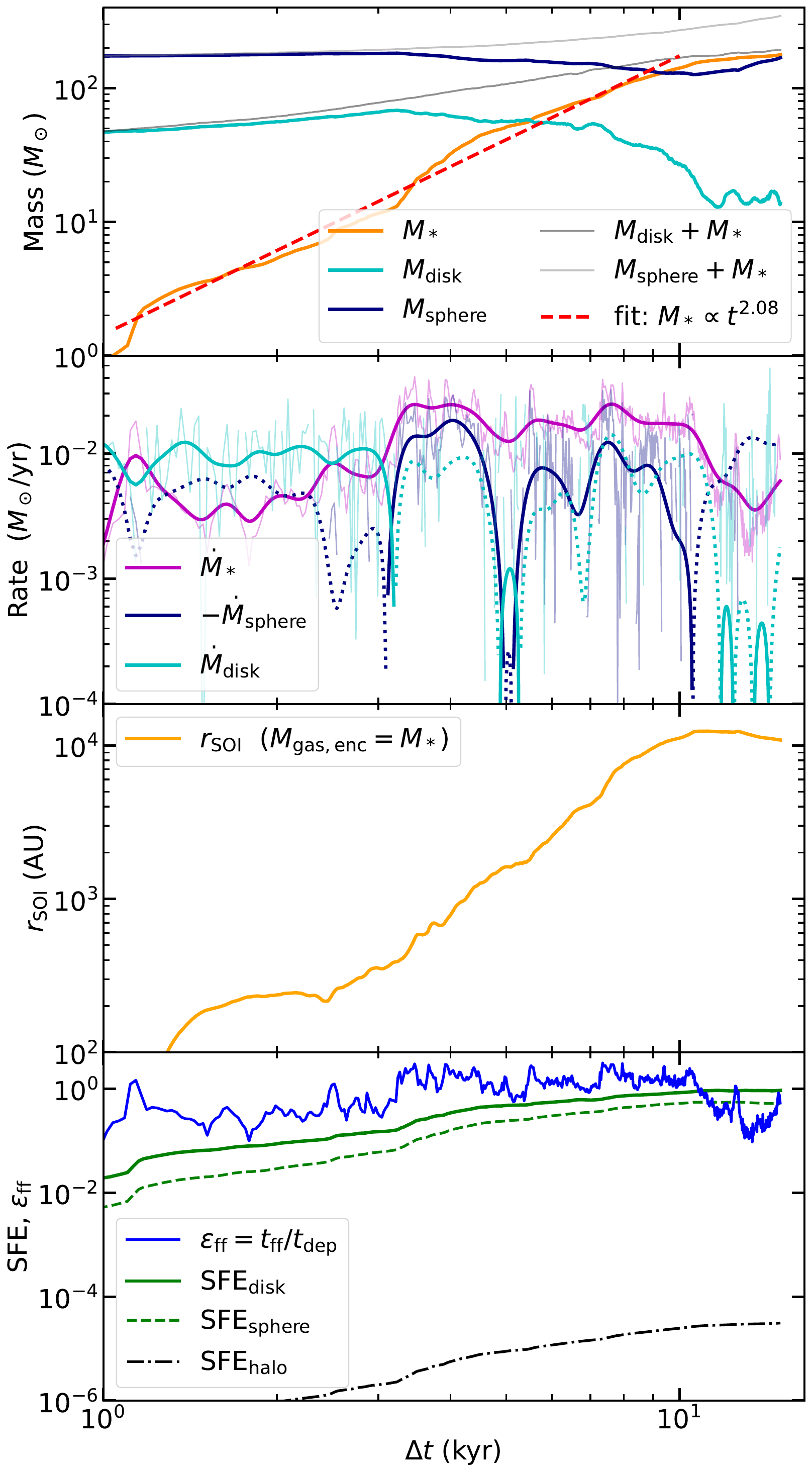}
	\caption{\label{fig:mass_evol}Top panel: time evolution of the stellar mass (all sinks; orange),
		disk gas mass (cyan), gas within a $0.1$ pc sphere,
		$M_\mathrm{sphere}$ (navy), and total mass ($M_\mathrm{disk} + M_*$, dark gray;
		$M_\mathrm{sphere} + M_*$, light gray).
		The best-fit power law $M_* \propto t^{2.08}$ is shown as a
		dashed red line.
        Top-middle panel: three net rates of change between consecutive
snapshots: the growth rate of the total sink mass, $\dot M_* =
\mathrm{d}M_*/\mathrm{d}t$ (magenta); the rate at which the sphere loses gas,
$-\dot M_\mathrm{sphere} = -\mathrm{d}M_\mathrm{sphere}/\mathrm{d}t$ (navy);
and the rate of change of the disk gas mass, $\dot M_\mathrm{disk} =
\mathrm{d}M_\mathrm{disk}/\mathrm{d}t$ (cyan). Dim lines show the rates at each
snapshot, bright lines the same rates smoothed. Where the smoothed rate is negative, its absolute value is shown dotted. These are
changes of the mass inside a region, not fluxes through a surface, so 
$\dot M_\mathrm{disk}$ is not measured at the disk edge; it is the infall minus
the accretion onto sinks. The infall rate through a sphere of radius $r$ is
shown in \autoref{fig:shell_mass_accretion}.
		Bottom-middle panel: sphere of influence radius $r_\mathrm{SOI}$
		($M_\mathrm{gas,enc} = M_*$; orange).
		Bottom panel: star formation efficiency within the disk,
		$\mathrm{SFE}_\mathrm{disk} = M_*/(M_\mathrm{disk}+M_*)$ (green solid), and within the $0.1$ pc sphere,
		$\mathrm{SFE}_\mathrm{sphere} = M_*/(M_\mathrm{sphere}+M_*)$ (green dashed), and within the whole halo, $\mathrm{SFE}_\mathrm{halo} = M_*/(M_\mathrm{gas,halo}+M_*)$ (black dash-dotted), where $M_\mathrm{gas,halo} = 5.7\times10^{6}\,\Msun$ is the gas inside the virial radius ($0.46$ kpc, physical), held constant because it changes by less than $0.01\%$ over the run, together
		with the star formation efficiency per free-fall time $\epsilon_{\rm ff}$ (blue), $\epsilon_\mathrm{ff} = t_\mathrm{ff}/t_\mathrm{dep}$, where $t_\mathrm{dep} = M_\mathrm{sphere}/\dot M_*$ and $t_\mathrm{ff}$ is the free-fall time at the mean density of gas and sinks inside the sphere.   }
	
\end{figure}

The top panel of \autoref{fig:mass_evol} shows the evolution of gas, stellar
(sink), and total mass within a $0.1$ pc sphere centered on the most massive star in the disk. The total stellar mass (orange, solid line) reaches $\sim 200\,\Msun$ by $15\kyr$, growing as $M_* \propto \Delta t^{2.08}$ (dashed red line). This is consistent with the $M_* \propto t^2$ scaling predicted by the theory of collapse in self-gravitating turbulent fluids \citep{murray15} and confirmed in high-resolution AMR simulations by \citet{lee15, murray16}. The density inside the stellar sphere of influence evolves to a time-independent attractor $\rho(r,t) \to \rho(r) \propto r^{-3/2}$, which, combined with infall at a fixed fraction of the Keplerian speed, yields a mass accretion rate $\dot{M} \propto t$ and thus $M_* \propto t^2$.

The disk gas mass (cyan) grows as infall from the envelope outweighs gas consumed by sink accretion (values in \autoref{sec:disk}). Both the infall rate and the total stellar accretion rate are of order $10^{-2}$--$10^{-3}\,\Msun\,\mathrm{yr}^{-1}$ over this period, with infall initially the larger of the two
(\autoref{fig:shell_mass_accretion}; top-middle panel of \autoref{fig:mass_evol} ). It then declines as the star formation efficiency rises and the envelope supply is depleted.
The gas mass within the $0.1$ pc sphere (navy) rises from $\approx 160$ to $\approx 180\,\Msun$ over the first $\sim 3\kyr$, declines to $\approx 130\,\Msun$ by $\sim 10\kyr$ as accretion drains it, and then rises again to $\approx 160\,\Msun$ by $15\kyr$ as infalling gas enters the sphere. The reason for this is a large
amount of infalling gas suddenly entering the sphere, resulting in a
positive $\dot{M} _{gas,sphere}$ (i.e. more gas is entering the system
than is being converted to stellar mass). This sudden rise in the gas mass
also ``ignites" the stellar accretion again after its slow-down, driven by the
depletion of initial accessible reservoirs in the original disk,
and one can see this ``bounce-back" in the top and bottom panels of \autoref{fig:mass_evol}: $M_*$ grows slowly, with $M_\mathrm{apert}$ increasing faster after $\Delta t \approx 10$ kyr, while $M_\mathrm{disk}$ stays roughly constant.

\autoref{fig:mass_evol} (top-middle panel) shows the corresponding mass rates.
The stellar accretion rate (magenta) fluctuates around
$\sim 5 \times 10^{-3}\,\Msun\,\mathrm{yr}^{-1}$ at early times,
rising to $\sim 10^{-2}\,\Msun\,\mathrm{yr}^{-1}$ at late times.
The median accretion rate is
$\dot{M}_* \approx 0.011\,\Msun\,\mathrm{yr}^{-1}$, and the disk-wide
gas infall rate is $\approx 0.016\,\Msun\,\mathrm{yr}^{-1}$.
These rates are $\sim 16\,c_s^3/G$ for $600$\,K gas, an order
of magnitude above the Shu rate $0.975\,c_s^3/G \sim 10^{-3}\,\Msun\,\mathrm{yr^{-1}}$
\citep{Shu_1977}. They are comparable to the Larson--Penston rates ($29\,c_s^3/G$
in the envelope at core formation, $47\,c_s^3/G$ after it; \citealt{mckee_ostriker07})%
and to the $\sim 10^{-2}\,\Msun\,\mathrm{yr^{-1}}$ found by \citet{stacy10},
as expected from the envelope kinematics (\autoref{sec:results}).

\autoref{fig:mass_evol} (third panel) shows the sphere of influence radius
$r_\mathrm{SOI}$, defined as the radius where the enclosed
gas mass equals the total stellar mass
($M_\mathrm{gas,enc} = M_*$).
This grows from $\sim 80\AU$ at $1\kyr$ to $\sim 10^4\AU$ by
$10\kyr$. It drops slightly with the new gas entering the sphere. The growth of
$r_\mathrm{SOI}$ quantifies the outward expansion of the region from which the
central cluster can draw its gas: as $M_*$ grows superlinearly, the scale on
which the cluster potential dominates the gravitational field expands at $\sim
10^3\AU\,\mathrm{kyr}^{-1}$, roughly as $r_\mathrm{SOI} \propto t^{2}$. This is
faster than the $r_* \propto t^{4/3}$ predicted for a self-similar envelope
\citep[][Eq. 42]{murray15}, which would give only $\approx 1700\AU$ at
$10\kyr$. The difference is consistent with the finite disk reservoir being
consumed rather than continuously replenished, since the gas reigniting star
formation after 10 kyrs only had time to enter the 0.1 pc sphere and did not
yet fall into the disk.

\subsubsection{Star formation efficiency}

We now examine the star formation efficiency, $\mathrm{SFE}$, and the efficiency per free-fall time, $\epsilon_{\rm ff}$. Both are instantaneous quantities, evaluated at each snapshot for a region $X$ (the disk, the $0.1$ pc sphere or the whole halo), and shown in the bottom panel of \autoref{fig:mass_evol}:
\begin{align}
    \mathrm{SFE}_{*,X}(t) &= \frac{M_{*,X}(t)}{M_{\mathrm{gas},X}(t) + M_{*,X}(t)} \\
    \epsilon_{\mathrm{ff},X}(t) &= \frac{\dot{M}_{*,X}(t)\, t_{\mathrm{ff},X}(t)}{M_{\mathrm{gas},X}(t)}
\end{align}
where $X \in \{\mathrm{disk}, \mathrm{sphere}, \mathrm{halo}\}$, $M_{\mathrm{gas},X}$ and $M_{*,X}$ are the gas and stellar (sink) masses in $X$, $\dot{M}_{*,X}$ is the rate at which the stellar mass in $X$ grows, and $t_{\mathrm{ff},X} = \sqrt{3\pi/(32 G \bar\rho_X)}$ is the free-fall time at the mean density $\bar\rho_X$ of gas and sinks in $X$.

The SFE remains below unity throughout our run. Within the disk, $\rm SFE$
rises steeply and saturates at $\sim 80\%$ by $10\kyr$. Within the $0.1$ pc
sphere, however, $\rm SFE$ only reaches a maximum of $\sim 50\%$ by $10\kyr$.  Over the whole halo the efficiency is only $\approx 3\times10^{-5}$ at $15\kyr$, where $M_\mathrm{gas,halo} = 5.7\times10^{6}\,\Msun$ is the gas inside the virial radius ($0.46$ kpc, physical), which stays constant to better than $0.01\%$ over the run.
This is not the case for $\epsilon_{\rm ff}$, which exceeds far above unity,
and exhibits a sensitive dependence on the star formation rate and ongoing gas
accretion. While $\epsilon_{\rm ff}$ is only $\sim 1\%$ in Galactic clouds and
galaxies from the local universe to high redshift \citep{krumholz12}, it can exceed unity
locally when a star forming region is being fed by multiple filaments or
streamers \citep[see e.g.][]{matzner_jumper15}. This is indeed the case here. Between $4$ and $10\kyr$ the median $\epsilon_{\rm ff} \approx 1.4$ (and $\lesssim 1$ before and after); \autoref{app:eff} shows that this is what the stellar accretion rate of $\approx 1.7\times10^{-2}\,\Msun\,\mathrm{yr^{-1}}$ implies for the free-fall time of a Larson--Penston-like envelope.

\subsubsection{Protostellar mass function}

\begin{figure}
	\centering
	\includegraphics[width=\columnwidth]{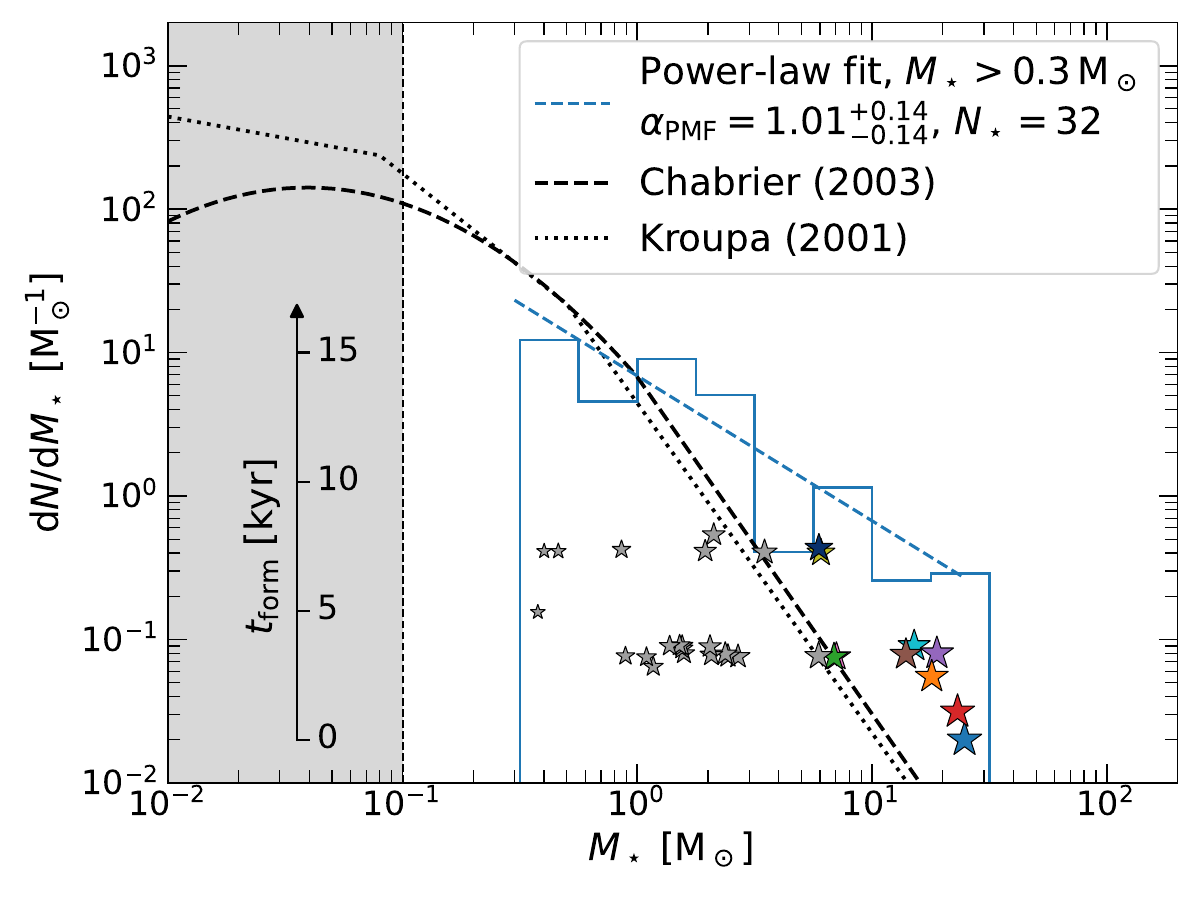}
	\caption{\label{fig:imf}Protostellar mass function of the 32 sinks present at	$\Delta t = 15\kyr$ (blue histogram), compared to the \citet{chabrier03}
		system IMF (dashed) and the \citet{kroupa01} IMF (dotted), both scaled to
		the number of sinks above $0.1\,\Msun$.
		The blue dashed line is the power-law fit $\mathrm{d}N/\mathrm{d}M \propto M^{-\alpha_\mathrm{PMF}}$ above $0.3\,\Msun$, with $\alpha_\mathrm{PMF} = 1.01^{+0.14}_{-0.14}$. Star symbols mark individual sinks: the $x$ position is the sink mass and the height is its formation time along the arrow ($\Delta t = 0$ at the bottom, $15\kyr$ at the top), not a value of $\mathrm{d}N/\mathrm{d}M_\star$. Star size grows with mass. Heavy sinks, the 10 most massive sinks at $\Delta t = 15\kyr$, are colored and light sinks are gray; each heavy sink has the same color as in \autoref{fig:toomre_Q_merged} and \autoref{fig:mass_evol_individual} (bottom). The gray shaded region lies below the completeness limit ($\sim 100\,\Delta m \approx 0.1\,\Msun$).
	}
\end{figure}

The PMF uses the current masses of the $32$ sinks present at $\Delta t =
15\kyr$, not their maximum masses. Most low-mass sinks merged early, so only one sink lies below
$0.1\,\Msun$ and the PMF at $15\kyr$ is not skewed by them.

\autoref{fig:imf} shows the protostellar mass function (PMF) of the
sinks at $\Delta t = 15\kyr$, compared to the
present-day \citet{kroupa01} and \citet{chabrier03} initial mass
functions (IMFs).
The high-mass slope of the PMF is well fitted by a power law
$\mathrm{d}N/\mathrm{d}M \propto M^{-\alpha_\mathrm{PMF}}$ with
$\alpha_\mathrm{PMF} = 1.01^{+0.14}_{-0.14}$  
much shallower than the Salpeter slope ($\alpha = 2.35$) and the
Kroupa high-mass slope ($\alpha = 2.3$). A slope of $\approx 1$ means
equal numbers of sinks per logarithmic mass interval, i.e.\ a flat
$\mathrm{d}N/\mathrm{d}\log M$. We compare this slope with other Pop III simulations in \autoref{sec:disc_pmf}.

The low-mass end of the distribution must be read with care.
Because the PMF is a $15\kyr$ snapshot, many of the lowest-mass sinks
have only recently formed or have been ejected from the disk, and have
not yet had time to accrete; they populate the gray region below the
completeness limit rather than a genuine low-mass tail. The only sink below that limit, of $1.7\times10^{-3}\,\Msun$, formed at $\Delta t \approx 3\kyr$ and was ejected before it could accrete. In \autoref{fig:imf} the star symbols form two rows, the sinks of the two fragmentation bursts at $\Delta t \approx 3.5$ and $\approx 7.5\kyr$, and ten of the 13 sinks below $2\,\Msun$ have moved beyond $2500\AU$ from the most massive sink, after which they barely accrete. None of the sinks have reached the main sequence by the end of our simulation.

In the protostellar model \citep{Offner_2009, STARFORGE_Towa_Grudic_2021} all 18 sinks above $2\,\Msun$ are in the shell deuterium-burning phase at $15\kyr$, the ten sinks of $1.5\,\Msun$ burn deuterium in their cores at fixed core temperature, three of $0.45\,\Msun$ have not yet ignited deuterium, and the ejected $1.7\times10^{-3}\,\Msun$ sink is still a pre-collapse first core. None has reached the main sequence.

\subsubsection{Individual accretion histories and runaway}

\begin{figure}[!htbp]
	\centering
	\includegraphics[width=\columnwidth]{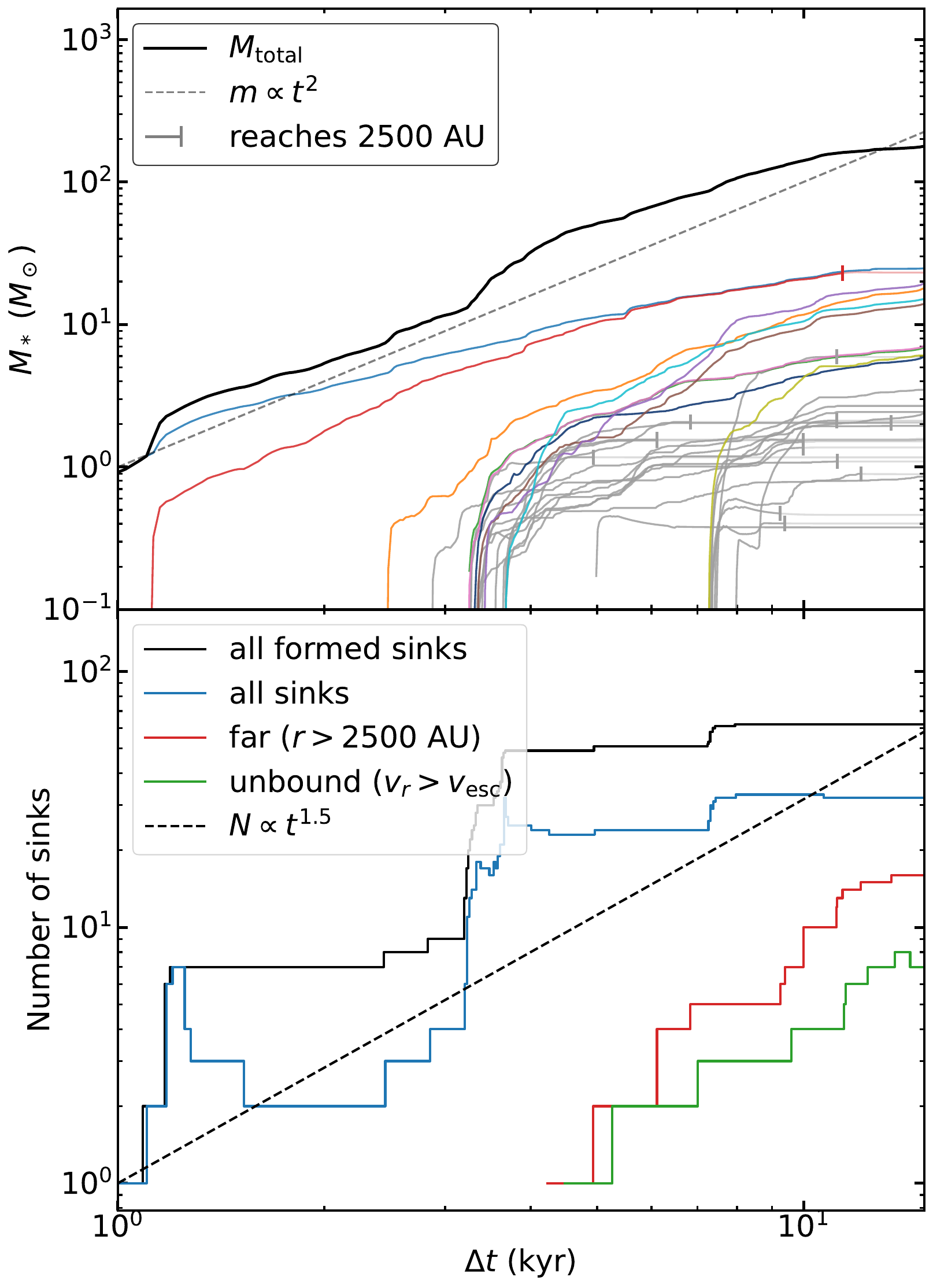}
	\caption{Evolution of masses of individual sinks. Dashed line represents
		ideal $M_{*} \propto t^2$ law, which is being followed for total mass of all
		the sinks (black line). At earlier times, before the first bulk star formation
		event at 2-3 kyr, all of stellar mass is concentrated in the 2 central sinks
		(top blue lines). These massive sinks reach mass of around $20 M_\odot$ by
		the 15 kyr mark, with their accretion slowing down compared to less massive
		sinks, which still accrete rapidly.
		Small vertical ticks mark the time at which each sink first moves more than
		$2500\AU$ from the most massive sink (``far'' sinks).
		The mass history for these far sinks is drawn fainter because their mass doesn't change by more than 3\% afterwards.
		The lower panel shows the time evolution
		of the number of all formed sinks, including those that merged later (black, cumulative), of the sinks present at that time (blue), of far sinks (red, cumulative), and of
		gravitationally unbound sinks (green): sinks beyond $2500\AU$ from the center of
		mass of the central cluster with $v_\mathrm{rad} > v_\mathrm{esc} =
			\sqrt{2GM_\mathrm{enc}/r}$, where $M_\mathrm{enc}$ includes the gas and the other
		sinks. A sink is counted as unbound (or bound again) only after the
		criterion has held for 10 consecutive snapshots ($\approx 0.25\kyr$), with a $N \propto t^{1.5}$ reference
		\citep{li_klein_mckee18}. Heavy sinks, the 10 most massive sinks at $\Delta t = 15\kyr$, are colored with the same color as in
		\autoref{fig:toomre_Q_merged} (bottom) and \autoref{fig:imf}; all other (light) sinks are gray. }
	\label{fig:mass_evol_individual}
\end{figure}

\autoref{fig:mass_evol_individual} shows integrated accretion history of
individual sinks.  A lot of low-mass ($< 3 M_\odot$) sinks stop accreting
relatively early because they get ejected out of the disk, thus limiting
the amount of gas, available to them. The accretion histories in
\autoref{fig:mass_evol_individual} (with the times at which sinks move beyond $2500\AU$ marked as ticks at the end of the lines) shows that
accretion is active for most of the time for most of the sinks until they leave the disk.

Of the 62 sinks formed during the run, 30 ($48\%$) were lost to
mergers at low mass: STARFORGE merges two sinks when they are bound with a semimajor axis below the sink radius and the lighter one has less than $10\,\Delta m \approx 0.01\,\Msun$, to remove spurious sinks that form next to existing ones \citep{STARFORGE_Towa_Grudic_2021}.

In our simulation, 14 of the $32$ surviving sinks are far ($r > 2500$ AU) by $15\kyr$. Applying the escape criterion of \citet{Stacy_Bromm_2013},
$v_\mathrm{rad} > v_\mathrm{esc}$ (lower panel of \autoref{fig:mass_evol_individual}),
only 7 of the 32 survivors are unbound at $15\kyr$ ($\approx 20\%$). All of them have
$M \lesssim 2\,\Msun$, as expected for dynamical ejection, which preferentially
removes the lowest-mass members of a few-body system.

The unbound count peaks at 8 at $\Delta t \approx 13.6\kyr$, and these sinks
are on wide orbits within the envelope rather than escaping. Far sinks
typically retain close to their mass at ejection. 

\autoref{fig:mass_evol_individual} also shows that the 2 heaviest sinks are the first 2 sinks that formed in the simulation, both in the center of the disk. At the same figure it can be seen that the accretion in the center of the disk
(heaviest 2 sinks) mostly halts, which is a consequence of depletion of
gas reservoir. This gas depletion is shown by a drop in stellar accretion rate
in \autoref{fig:mass_evol}. The most massive sink reaches
$25\,\Msun$ by $15\kyr$, but its accretion has slowed to
$\dot M \approx 2.5\times 10^{-4}\,\Msun\,\mathrm{yr}^{-1}$ over the final
kiloyear as the local gas reservoir is depleted. The ``bounce-back'', described in detail in \autoref{sec:massevol}, seems to mostly target the non-central sinks, since we still observe their growth after $\Delta t \approx 11\kyr$.

The second-most massive sink
($23\,\Msun$) is at $\sim 4500\AU$ and its accretion has stalled
entirely by $15\kyr$. It is a far sink, but it is
gravitationally bound to the central group of five sinks
($\approx 60\,\Msun$ within $200\AU$ of the primary). Between $12$ and $15\kyr$
its two-body orbit has $E \approx -0.5\,GM/r$, a semi-major axis of
$\approx 4000\AU$, an eccentricity of $\approx 0.6$ and a
period of $\approx 30\kyr$. It is a wide, eccentric companion near
apocenter, and it returns to $\approx 1200\AU$ at $\approx 18\kyr$ in the
extended run. Its accretion stalls because its orbit lies largely outside
the gas disk. 

\begin{figure}[!htbp]
	\centering
	\includegraphics[width=\columnwidth]{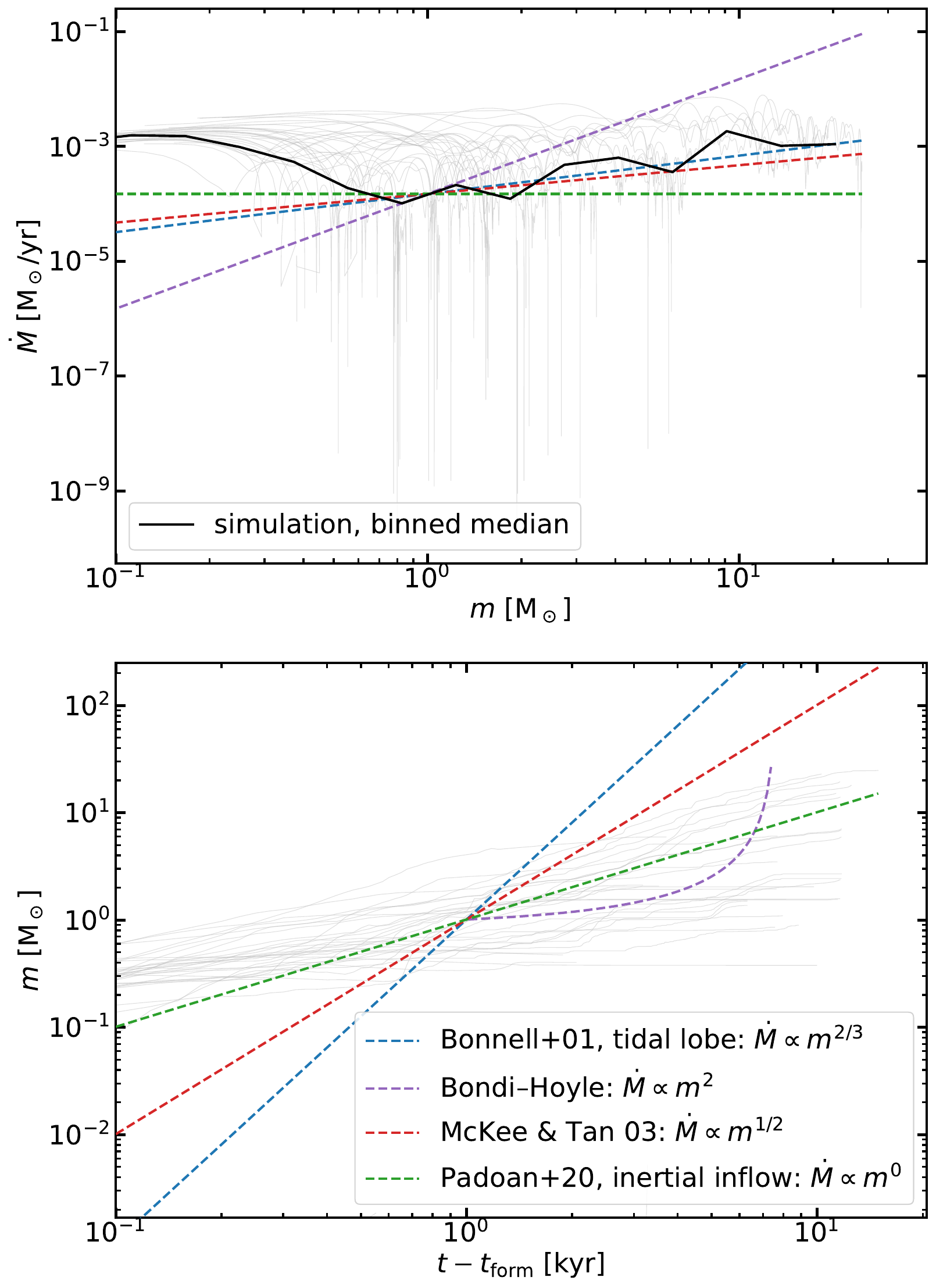}
	\caption{Top: accretion rate $\dot M$ against current mass $m$ for all sinks
		(gray) with the binned median (black), compared with tidal-lobe
		($\dot M \propto m^{2/3}$) and Bondi--Hoyle ($\dot M \propto m^{2}$)
		competitive accretion \citep{bonnell01}, turbulent-core accretion
		($\dot M \propto m^{1/2}$; \citealt{mckee03}) and inertial inflow
		($\dot M \propto m^{0}$; \citealt{padoan20}), normalized to the median (line colors as in the legend of the bottom panel). Bottom:
		mass of each sink against time since its formation, with the integrated form
		of each scaling anchored to the median track at $1\kyr$. Individual sinks grow
		at most linearly in time, with no sign of the runaway growth of Bondi--Hoyle
		accretion.
	}
	\label{fig:mdot_vs_mass}
\end{figure}

Competitive accretion in a cluster environment would preferentially feed the
most massive objects \citep{bonnell01}, flattening the mass function slope. To
distinguish competitive accretion from simple disk fragmentation, we show in
\autoref{fig:mdot_vs_mass} (top) the accretion rate $\dot M$ against the current mass
$m$ for all sinks, with the binned median, compared with the scalings of
tidal-lobe ($\dot M \propto m^{2/3}$), Bondi--Hoyle ($\dot M \propto m^{2}$)
competitive accretion \citep{bonnell01}, accretion from a turbulent core
($\dot M \propto m^{1/2}$; \citealt{mckee03}) and inertial inflow ($\dot M$
independent of $m$; \citealt{padoan20}). Above $\sim 1\,\Msun$ the median rises
roughly as $\dot M \propto m^{0.8\text{--}0.9}$, between the tidal-lobe and
Bondi--Hoyle cases. Discarding the
$\approx 7$ per cent of rates below $10^{-5}\,\Msun\,\mathrm{yr^{-1}}$, from sinks that have
almost stopped accreting, flattens the slope to $\approx 0.65$, between the turbulent-core ($1/2$) and tidal-lobe ($2/3$) values; the
data cannot distinguish between these two.
Below $1\,\Msun$ it falls with mass,
because newly formed sinks accrete fastest just after formation.

This ensemble
relation mixes sinks of different ages, however: the most massive sinks are the
oldest, central ones with the best access to gas. The bottom panel therefore
tests the scalings on individual sinks, comparing their mass growth since
formation with the integrated form of each model. Individual sinks grow at most
linearly in time, close to the constant-rate (inertial-inflow) track and far 
slower than the runaway growth expected from Bondi--Hoyle accretion. We find no
evidence of runaway competitive accretion: each sink grows at a roughly constant
or declining rate set by its local gas supply, and growth is limited by
competition for a shared, finite reservoir.

\subsection{Magnetic fields}
\label{sec:bfield}

\begin{figure*}[!htbp]
	\centering
	\includegraphics[width=\textwidth]{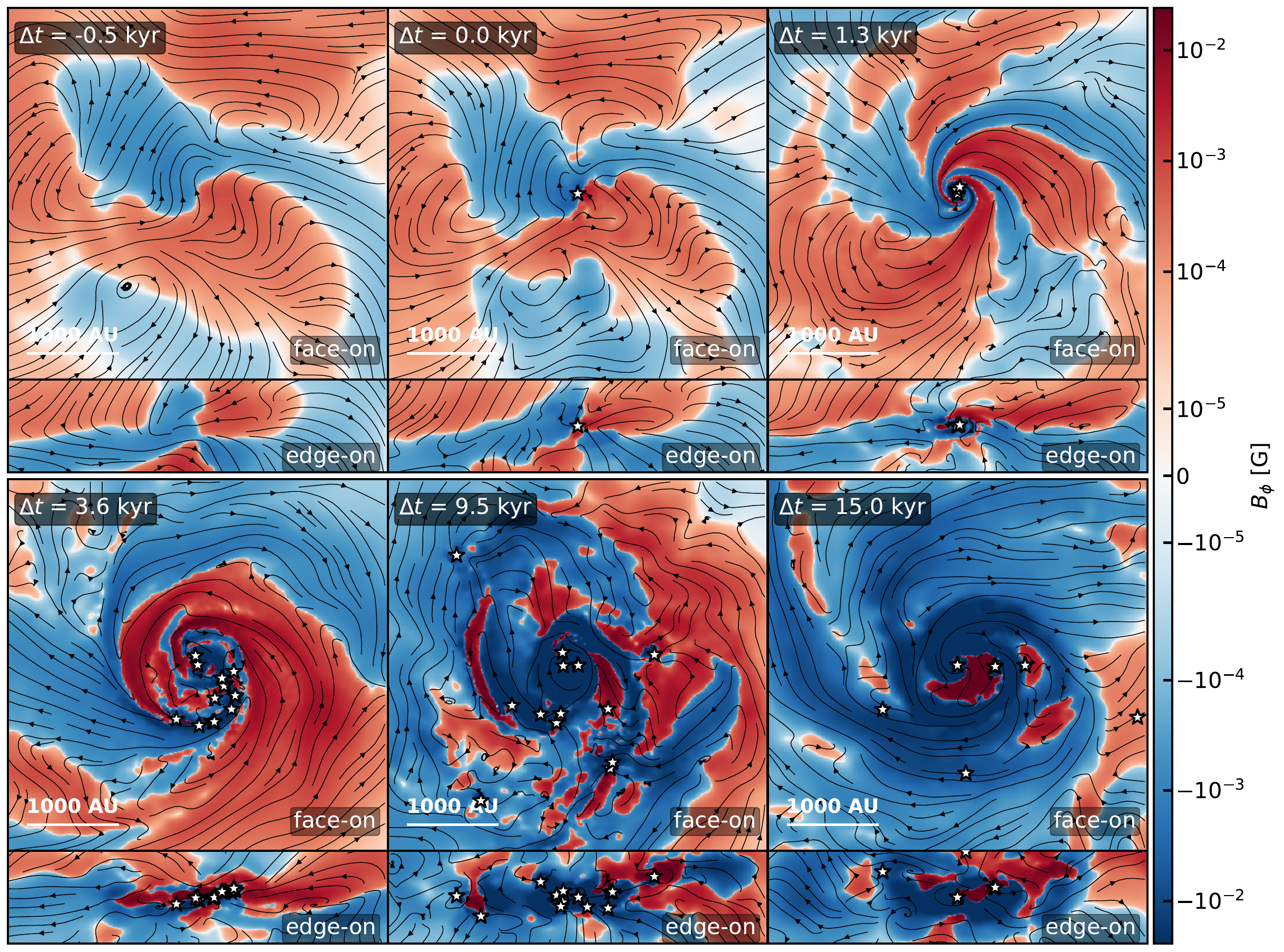}
	\caption{Thin face-on ($|z|<\Delta$) and edge-on ($|y|<\Delta$)
		slices through the most massive sink, with $\Delta$ equal to 5\% of
		the panel width ($\approx 200\AU$ for the $4100\AU$ panels) and the field mass-weighted across the slice
		thickness, of the toroidal
		magnetic field component $B_{\phi}$ at six epochs, with in-plane
		magnetic field lines overlaid (face-on panels: $B_x$, $B_y$;
		edge-on panels: $B_x$, $B_z$). The $m=1$ sign pattern
		of $B_\phi$ is present before the first sink forms and is inherited
		from the nearly uniform seed field; differential rotation winds it
		into the two-armed spiral visible by
		$\Delta t \approx 1.3\kyr$.
	}
	\label{fig:combined_Btor}
\end{figure*}

\begin{figure*}[!htbp]
	\centering
	\includegraphics[width=\textwidth]{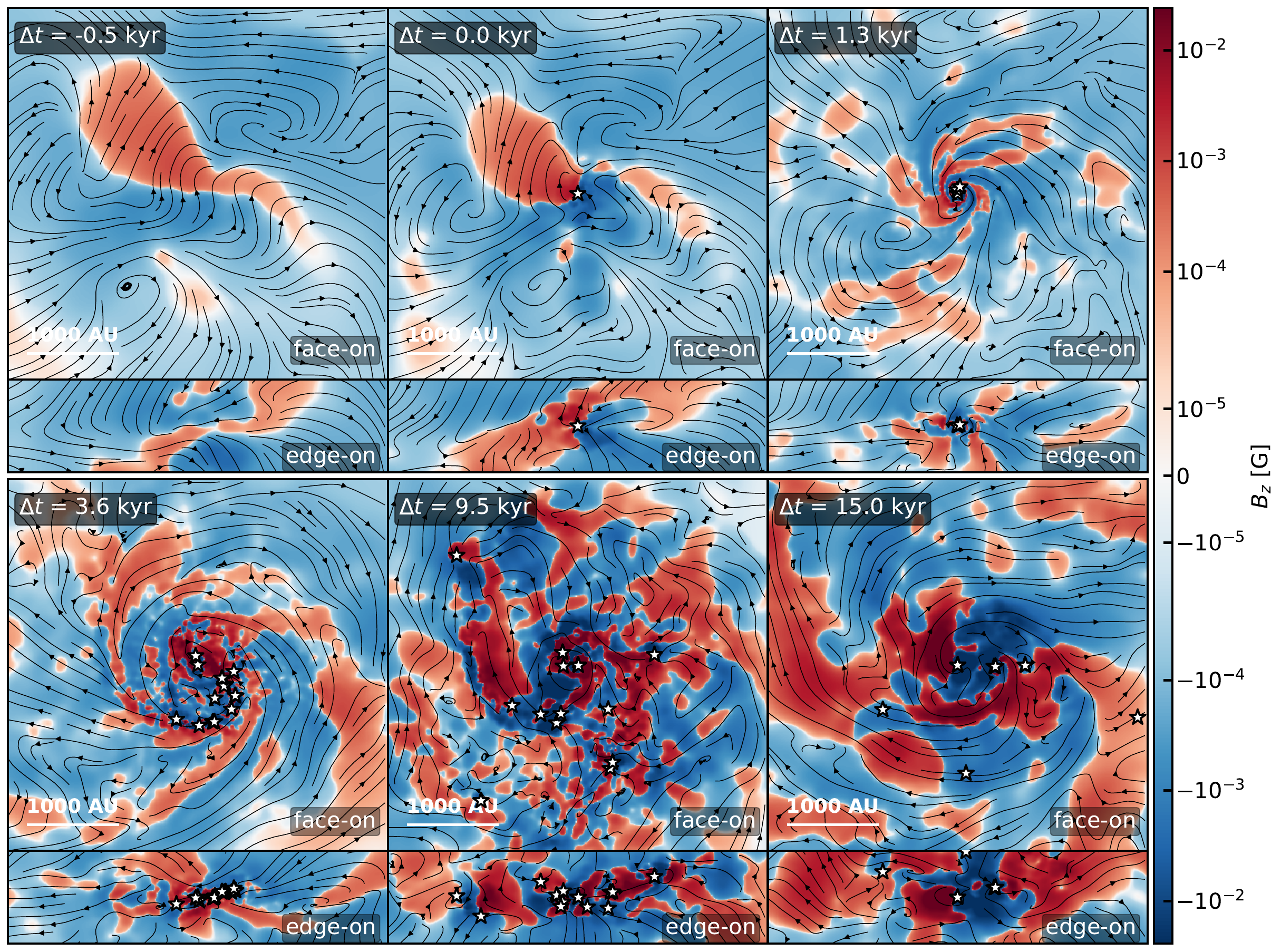}
	\caption{Thin face-on and edge-on slices, as in
		\autoref{fig:combined_Btor}, of the vertical
		magnetic field component $B_{z}$ at six epochs.
		By the late epochs shown here, most of the stars in the
		simulation have already formed, so the field structure reflects
		the post-formation disk rather than the initial collapse.
	}
	\label{fig:combined_Bz}
\end{figure*}

  In this section, we analyze the evolution of the magnetic field as the disk collapses and fragments.
The magnetic fields are amplified during the collapse, and within the disk once it forms.
We characterize the resulting field morphology and its evolution through the vertical ($B_z$) and toroidal ($B_\phi$) components.

\subsubsection{Field morphology and topology}

Figures \ref{fig:combined_Btor}--\ref{fig:combined_Bz}
show face-on and edge-on thin slices of the
toroidal and vertical components of the magnetic field at six epochs.
In the pre-stellar epoch, $B_z$ shows large-scale coherent structure
with a preferred sign, aligned with the imposed initial field orientation.
As the disk forms and develops spiral structure, $B_z$ is amplified
in the disk midplane and develops sign reversals that trace the
spiral arms, with field strengths reaching
$|B_z| \sim 10^{-2}$ G at late times.

The toroidal component $B_\phi$ (\autoref{fig:combined_Btor})
has a two-tone appearance before the first sink forms
($\Delta t = -0.5\kyr$, top left panel): roughly half of the disk plane has $B_\phi>0$, shown as red, and the other $B_\phi<0$, shown as blue, with the two colors inhabiting clearly distinct regions. Differential rotation winds the two regions into a two-armed
spiral pattern of alternating color by $\Delta t \approx 1.3\kyr$, with the field gaining in strength close to the center of the spiral.

By $\Delta t \approx 3.6\kyr$, the toroidal field component reaches strengths of $|B_\phi| \gtrsim 10^{-2}$ G over a region of size 1000 AU. 
The field winding due to the differential rotation of the disk has produced a field that is primarily toroidal. This is the expected geometry: winding an
in-plane seed produces a toroidal field that is symmetric about the
midplane \citep[Sec. 11.5.2]{brandenburg_subramanian05}. By late times ($15\kyr$), the toroidal
field in the inner disk is predominantly negative (in our chosen reference frame,
where the disk rotates with $\Omega>0$).

The alternating sign of $B_\phi$ between the two arms is similar to the well known
in-plane bisymmetric spiral (BSS) configuration seen in
galactic magnetic fields; the field reverses direction from one arm
to the next across a neutral line in the interarm region
\citep{sofue_fujimoto83, beck15, oswald25}. In our simulation, the reversal is inherited
from our choice of a uniform initial magnetic field, i.e., a field with an azimuthal $m=1$ pattern.
As long as differential rotation and collapse remain approximately axisymmetric, they amplify the field while preserving the initial $m=1$ pattern
\citep{roberts_yuan70, brandenburg_ntormousi23}. The approximate axisymmetry of the flow holds until
$\Delta t \sim 3\kyr$. 

At later times the spiral arms and the lopsided potential of
the off-center sink cluster make the flow non-axisymmetric;  the
induction term transfers field from the $m=1$ pattern into other azimuthal
modes, including an axisymmetric ($m=0$) net toroidal component. 
We see only weak evidence of growth in the radial field, discussed below, suggesting that this early-time pattern results from linear amplification of the seed field rather than arising from a dynamo \citep[][Secs. 4.2, 5.1]{sofue86}.

\citet{kim_ostriker01} showed that magnetic fields modify the
growth of non-axisymmetric structure in self-gravitating disks when
the fields are dynamically important ($\beta \sim 1$). Here
$E_\mathrm{mag} < E_\mathrm{therm}$ throughout
(\autoref{sec:energyevol}), so on energetic grounds the field is not expected to shape the
spiral arms strongly, although only a run without magnetic fields can confirm this.
The field does not remain wound up: the ratio of toroidal to radial field stays at $\approx 2$ rather than growing with the shear, the net toroidal flux has a definite sign set by the more massive spiral arm, and the $|B|$--$n$ relation steepens from $\alpha \approx 2/3$ to $\approx 1$. We cannot say whether this is anisotropic compression, winding or a small-scale dynamo; the evidence and the amplification channels are discussed in \autoref{app:bamp}.

\subsubsection{Mass-to-flux ratio}
\begin{figure}[!htbp]
	\centering
	\includegraphics[width=\columnwidth]{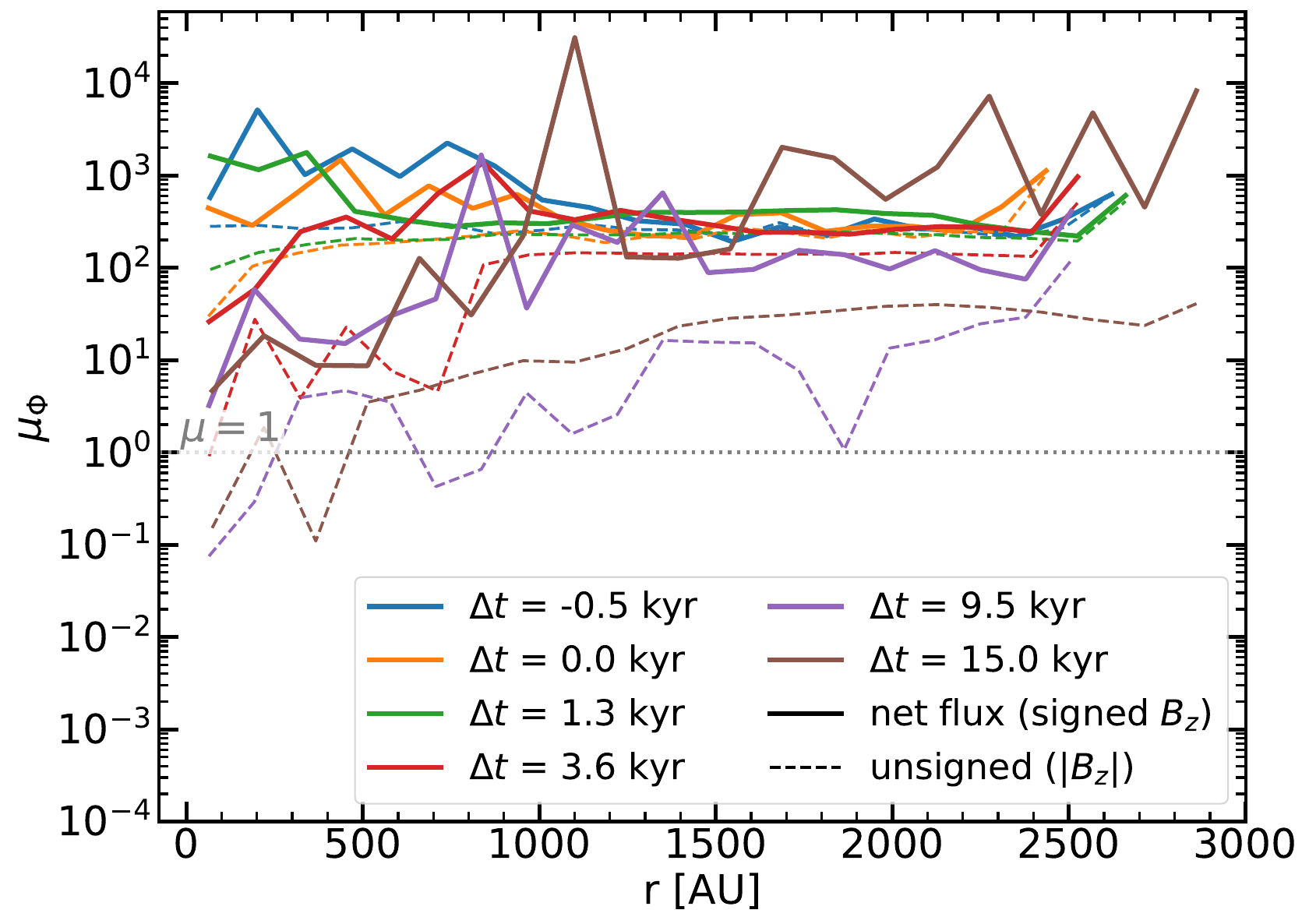}
	\caption{Dimensionless mass-to-flux ratio $\mu_\Phi$ as a function
		of radius at six epochs. Solid lines use the net vertical flux
		(volume-weighted signed $B_z$ in each annulus), which sets the magnetic
		support; dashed lines use the unsigned flux (mass-weighted $|B_z|$), which
		also counts field reversals. The dotted line marks the critical value
		$\mu_\Phi = 1$. The net-flux ratio falls from $\sim 10^{2}$--$10^{3}$ at early
		times to $\gtrsim 3$ at late times, whereas
		the unsigned ratio drops below unity at some inner radii by $9.5\kyr$ as the
		field becomes tangled.
	}
	\label{fig:profile_mf}
\end{figure}

We define mass-to-flux ratio per annulus $b$ as:
\begin{equation}
	\mu_{\Phi,b} = \frac{2\pi\sqrt{G}\,\Sigma_b}{|B_{z,b}|}
\end{equation}
where, in \autoref{fig:profile_mf}, $|B_{z,b}|$ is the mass-weighted
mean of $|B_z|$ in the annulus. This measures the unsigned flux.%

We also compute $\mu_{\Phi,\mathrm{net}}$ from the
volume-weighted signed mean of $B_z$ in each annulus, and its cumulative
version $\mu_{\Phi,\mathrm{net}}(<R) = 2\pi\sqrt{G}\,M_\mathrm{gas}(<R)/|\Phi(<R)|$.
The mass-to-flux ratio $\mu_\Phi$ (\autoref{fig:profile_mf})
evolves significantly over time.

At early times ($\Delta t \leq 1.3\kyr$), $\mu_\Phi \sim 200$
at all radii, indicating an extremely magnetically supercritical
state in which gravity overwhelms magnetic support by more than
two orders of magnitude.
The mass-to-flux ratio then declines steadily as the unsigned
vertical field $|B_z|$ grows relative to $\Sigma$:
by $\Delta t = 3.6\kyr$ (red), $\mu_\Phi$ has dropped to
$\sim 10$--$100$.
By $\Delta t = 6.3\kyr$ (purple), $\mu_\Phi$ exhibits large
radial scatter, with values ranging from $\sim 1$ to $\sim 100$
and dipping below the critical value ($\mu_\Phi < 1$) at some
inner radii. At the latest epoch ($\Delta t = 9.5\kyr$, brown), significant
portions of the inner disk ($r \sim 350\AU$) have
$\mu_\Phi < 1$ in this unsigned measure.

The net-flux ratio tells a different story. At $\Delta t \lesssim
	1.3\kyr$, $\mu_{\Phi,\mathrm{net}}(<R) \approx 400$ within
$2500\AU$, consistent with the unsigned values because the field is still
coherent. By $3.6\kyr$ the unsigned ratio has dropped to $\sim 0.2$--$90$ in
individual annuli, while $\mu_{\Phi,\mathrm{net}} \approx 50$--$600$ in the
same annuli. At $9.5\kyr$ and $15\kyr$, $\mu_{\Phi,\mathrm{net}}(<R) \approx 28$ and $\approx 16$, respectively. In individual annuli
(solid lines in \autoref{fig:profile_mf}) the net ratio never falls below
$\approx 3$, reached in the innermost annuli at $9.5\kyr$, where little gas is
left. The unsigned flux inside $2500\AU$
exceeds the net flux by a factor of $\sim 2$ at early times and $\sim 15$
at late times.

\begin{figure}[!htbp]
	\centering
	\includegraphics[width=\columnwidth]{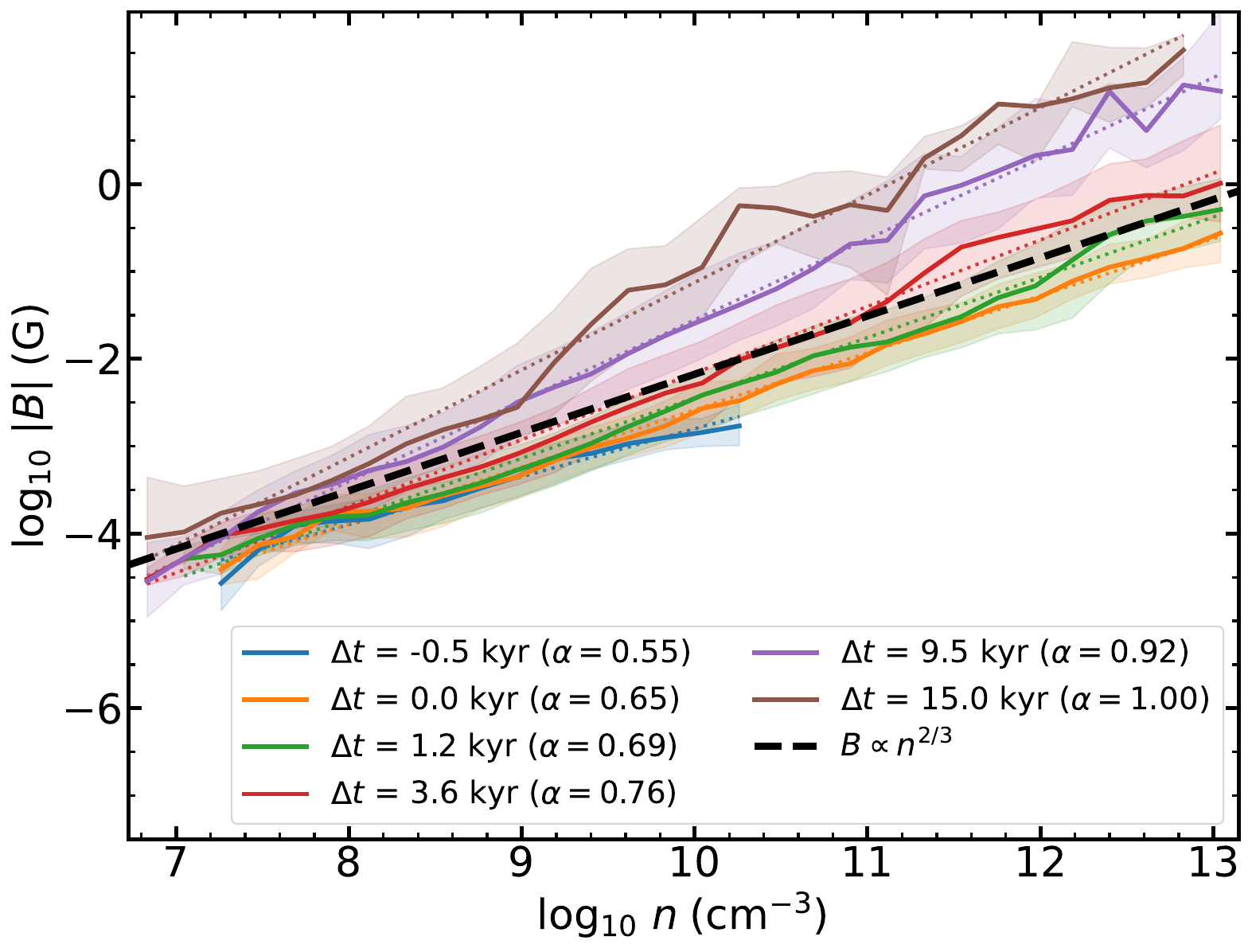}
	\caption{\label{fig:phase_Bfield} Evolution of phase plot of magnitude of magnetic field
		$\left| B \right|$ vs number density $n$.
		Solid lines correspond to mass-weighted median profiles at different times, with the shaded
		regions spanning the 16th--84th percentiles.
		$\alpha$ corresponds to the exponent in power-law fit
		$ \left|  B \right| \propto n^\alpha$.
		$\left| B \right|$ spans 3 orders of magnitude at the point of first sink
		formation, and up to 4 orders of magnitude at later times.
		The best-fit power-law slope $\alpha = \mathrm{d}\log|B|/\mathrm{d}\log n$
		rises from $\alpha = 0.55$ at the pre-stellar epoch to
		$\alpha = 0.65$ (first sink), $0.69$ (early disk), $0.76$ (GI onset),
		$0.92$ (late time) and $1.00$ (extended), i.e.\ from $\sim 2/3$
		(flux-freezing compression) toward $\sim 1$ at late times; each fitted slope is given in the legend and drawn as a dotted line.}
\end{figure}

The net ratio $\mu_\Phi$ (solid lines in \autoref{fig:profile_mf}) falls by one to two orders of magnitude as time increases.
In ideal MHD, neither compression nor winding can do this: compression
conserves the mass-to-flux ratio of a flux tube, and winding converts $B_r$
into $B_\phi$ without changing $B_z$. The decline instead comes from the
flux budget of the disk region. Between $-0.5$ and $15\kyr$ the net flux inside
$2500\AU$ grows by a factor of $\sim 6.5$ as the inflow drags field
inward, while the gas mass there falls by a factor of $\sim 7$ as
$\sim 200\,\Msun$ is accreted onto sinks. Sink accretion removes the accreted
gas together with its field, but the accreted gas is the densest, and since
$B \propto n^{\alpha}$ with $\alpha < 1$ (\autoref{fig:phase_Bfield}) the flux per
unit mass, $\propto B/n$, is smallest there. Accretion therefore removes mass
much more efficiently than flux. (In the sink implementations of
\citealt{hirano_machida22} and \citealt{stacy22} accretion removes the mass but
not the flux, which lowers $\mu_\Phi$ even more strongly.)

The much larger growth of the unsigned flux requires some other growth mechanism, for example, vertical motions that tilt the
strong toroidal field into $B_z$ of both signs, as at spiral-arm shocks, and
possibly a turbulent dynamo.

\section{Comparison to previous studies}
\label{sec:disc_comparison}

The collapse and subsequent disk fragmentation picture in this work is
consistent with the broad expectations of Pop III star formation theory. In
this section, we compare our results  with other works and highlight
differences.

\subsection{Role of the large-scale environment}

The properties of the first collapsing core are set by the large-scale
accretion history of the host minihalo, which in turn depends on where that
halo sits within the cosmic web. A halo embedded in a stronger overdensity
accretes faster and with more angular momentum than one in a quieter region of
the cosmic web, biasing the collapse toward a more massive, rotationally
supported structure. Our halo has a mass of $2\times10^7\Msun$ when the first star forms, which is above the atomic-cooling threshold. The halo is a progenitor of a MW-mass
system (with $M_{halo}\approx 10^{12}\Msun$ at $z=0$) and sits in a large-scale overdensity, unlike the field minihalos studied in most prior work.%

Our halo is a typical Pop III host in this kind of
environment: in the MEGATRON simulations \citep{storck26} of a MW progenitor, the very first
Pop III stars form in minihalos, but most form above the atomic-cooling
threshold, in halos of $3\times10^{6}$--$3\times10^{7}\,\Msun$, once a Lyman--Werner
background is established \citep{storck26}. 

The most directly comparable simulation, \citet{meziani26}, targets
the dwarf galaxy progenitor halo,
and finds a single Pop III protostar growing to
$\sim 27\,\Msun$ over $31\kyr$ -- in stark contrast to our
$\sim 60$-fragment cluster.
The two runs differ in more than environment, however: that star forms at
$z \sim 14$ in a minihalo of $\sim 10^{6}\,\Msun$ ($T_\mathrm{vir} \sim 10^{3}$\,K), 
with $\approx 55$ times finer DM resolution, but similar baryonic resolution, and in a disk that
stays Toomre-stable ($Q \gg 1$) with jets active from the start. We speculate 
that the difference in the outcome is partially dictated by the environment in 
which the gravitational collapse is happening.

The overdense filamentary infall geometry of the MW-progenitor
naturally seeds higher angular momentum, more complex disk structure,
and stronger turbulence than the quieter m09 environment.
\citet{ho25b} showed that infall into dark-matter potential wells
drives supersonic turbulence ($\mathcal{M} \approx 1.8$--$4.2$) in 15
minihalos, with $\mathcal{M}$ increasing with halo mass. 

In our simulation, at $r\sim 0.1$--$1$ pc, the velocity dispersion about
the mean infall gives $\mathcal{M} \approx 3.5$ at all times, similar to the value $\mathcal{M}\approx4$ seen in the slightly less massive ($M_h=8\times10^6\Msun$) halo in \citet{ho25b}.
Within the disk in our simulation, $\mathcal{M}\equiv|\delta v|/c_s\approx 0.75$ 
before $\Delta t \approx 1.3\kyr$ and $\mathcal{M} \approx 3$ after the first fragmentation burst (\autoref{fig:kinematics_wide}).

\subsection{Fragmentation and sink counts}

Comparing sink counts across simulations is difficult due to differences in the sink formation and accretion algorithm used, the spread in integration times, or the stochastic nature of large simulations.
Early simulations with
cosmological initial conditions found 4--5 protostars within $\sim 100$ yr in
one halo \citep{clark11} and of order ten per halo after $1\kyr$ in five
halos \citep{greif11}. Single-halo integrations over $\sim 5\kyr$ found 5--8
sinks \citep{stacy12}, whereas \citet{stacy16} formed 102 sinks within $2\kyr$,
and \citet{jaura22} formed 40--150 sinks over $20\kyr$.
The radiation-hydrodynamic cosmological zoom-ins of
\citet{latif_khochfar22} across five minihalos produced up to 23
stars per halo (55 total), with masses $0.1$--$37\,\Msun$ and a
slope $\mathrm{d}N/\mathrm{d}M \propto M^{-1.18}$.
The systematic study of \citet{wollenberg20} with 45 controlled
simulations (varying rotation and turbulence) found 15 to $\sim 120$ sinks
(mean 30--59 per setup) within $\sim 1\kyr$, without sink mergers.  Thus our $\sim 60$ sink formation events over the first $\sim 8\kyr$ fall within the range of other high-resolution runs, confirming that fragmentation is highly stochastic and sensitive
to initial conditions.

Several factors contribute to our fragment count.
First, our $15\kyr$ integration is substantially longer than most
published runs at similar resolution in cosmological environment, allowing many cycles of instability, collapse, and
re-accretion.
Second, cosmological initial conditions from the MW progenitor
provide asymmetric, turbulent infall that continually replenishes
the outer disk and seeds new unstable structures.

Of the 62 sinks formed, 30 ($48\%$) were lost to mergers, and 14 of the 32
survivors ($44\%$) have moved more than $2500\AU$ from the primary, of which
7 ($\approx 20\%$ of the survivors) are unbound by the
$v_\mathrm{rad} > v_\mathrm{esc}$ criterion. The merger fraction is in line with
\citet{Stacy_Bromm_2013} ($59\%$) and \citet{stacy16} ($64\%$); the unbound fraction ($\approx 20\%$ of survivors) is about half of their escaping fractions ($43\%$ and $31\%$ of survivors), while the fraction beyond $2500\AU$ ($44\%$) is comparable. Other works report merger
fractions of $\sim 10$--$50\%$ depending on the merging criterion
\citep{wollenberg20} and ejection fractions of up to $\sim 70\%$
\citep{latif_khochfar22}.

\subsection{Stellar mass growth and accretion rates}

The total stellar mass assembled by a given time is the time-integrated sum of the accretion rates of all active sinks, so it reflects both the efficiency of angular momentum transport (which determines how quickly gas drains inward through the disk) and the number of competing accretion regions (which determines how the common reservoir is partitioned among the protostars).
A large sink count spreading the accretion over many objects can produce a similar total mass to a single rapidly accreting protostar while leaving a completely different PMF. We compare our integrated mass growth and accretion rates to prior work to disentangle these effects.

The total stellar mass of $\sim 200\,\Msun$ at $15\kyr$ is broadly
consistent with canonical Pop III accretion rates.
\citet{stacy10} found $\sim 10^{-2}\,\Msun\,\mathrm{yr}^{-1}$ onto the star--disk
system over $5\kyr$, similar to our time-averaged rate, and a $43\,\Msun$ main
sink at $5\kyr$ growing as $M \propto t^{0.55}$. Single-star growth in other
runs is similarly sub-linear: \citet{sharda_menon25} find $M \propto t^{0.47}$--$t^{0.57}$. Our $M_* \propto t^{2}$ scaling is consistent with that found by \citet{murray16} and \cite{li_klein_mckee18}, and is a property of the whole
cluster, not of individual sinks, which indeed grow sub-linearly
(\autoref{fig:mass_evol_individual}).

The POPSICLE simulations of \citet{sharda_menon25} produce a single
most-massive protostar of $57\,\Msun$ within $5\kyr$ in their
RMHD run; our total mass of $\sim 200\,\Msun$ is distributed across
$\sim 30$ surviving sinks at $15\kyr$, implying a mean sink mass of $\sim 6\,\Msun$, consistent with
fragmentation-induced starvation redistributing mass from the most
massive object to the cluster population \citep{peters10, prole22a, sharda_menon25}.
\citet{sugimura20} and \citet{sugimura23} produced massive Pop III
binaries with individual masses $>30\,\Msun$ at separations
$>2000\,\AU$. In their runs the first protostar's disk fragments, most
fragments merge in few-body encounters, and only  2-4  stars
survive. They accrete from a shared circumbinary disk, which drives the mass
ratio towards unity and widens the orbit, because the accreted gas has high
angular momentum. Radiative feedback ends accretion after $\sim 10^{5}$ yr, at
final mass ratios $q \approx 0.3$--$0.8$ and separations of
$2\times10^{3}$--$2\times10^{4}\AU$ \citep{sugimura23}. 

Our two most massive
sinks ($24.8$ and $23.1\,\Msun$, $q = 0.93$) also form a bound pair.  After $\approx 12 \kyr$, the secondary orbits the central group on a wide, eccentric orbit
($a \approx 4000\AU$, $e \approx 0.6$), with a separation similar to that of the
\citet{sugimura20} binary at $\approx 2\times10^{4}$ yr, when it has widened from
$\sim 10^{3}\AU$ to a few $\times10^{3}\AU$.
Unlike their pair, our
secondary has stopped accreting because its orbit carries it outside the gas
disk, and the separation was set by few-body dynamics rather than by
circumbinary accretion. Without radiative feedback, and with a different sink
radius and merger criterion, our pair is therefore at most a qualitative
analogue of their early binary phase.

\subsection{Star formation efficiency}
\label{sec:sfe}

The star formation efficiency we find (${\rm SFE} \approx 90\%$ within the disk
and $\approx 50\%$ within a $0.1$\,pc sphere at $15\kyr$) is high compared to
present-day values ($\sim 1$--$10\%$), but these are typically measured for entire
giant molecular clouds rather than individual protostellar disks, so this
comparison is only qualitative. Other Pop III simulations measure the efficiency across the halo.
Over the whole halo our efficiency is $\approx 3\times10^{-5}$ at $15\kyr$
(\autoref{fig:mass_evol}), below the $\epsilon_\mathrm{III} \sim 10^{-2}$--$10^{-4}$ of
\citet{hirano_sakai23}, who divide the core mass by the gas mass of the virial halo.

\subsection{Protostellar mass function and runaway fraction}
\label{sec:disc_pmf}

A top-heavy, roughly log-flat PMF is commonly found in numerical studies of Pop III star formation , although most of them, like ours, follow only the first $\sim 10^{4}$--$10^{5}$\,yr.  Several authors attribute it to fragmentation together with mergers and ejections \citep[e.g.][]{stacy16, latif_khochfar22}. Two effects favor a top-heavy PMF: the primordial gas temperature ($\sim 10^3$ K) sets a large Jeans mass at the fragmentation scale, seeding comparatively massive protostars \citep{bromm02}, and low-mass objects are preferentially ejected or starved, which flattens the slope at the low-mass end \citep{wollenberg20}.

Our protostellar mass function (PMF) at $15\kyr$ has a high-mass
slope $\alpha_\mathrm{PMF} = 1.01^{+0.14}_{-0.14}$, shallower than the Salpeter
slope ($\alpha = 2.35$) but broadly consistent with other
Pop III simulations
that find top-heavy, log-flat distributions.
\citet{Stacy_Bromm_2013} find $\alpha = 1.17$ and \citet{latif_khochfar22} find $\alpha \approx 1.18$ across five
halos with RHD; \citet{stacy16} and \citet{greif11} both
report
roughly log-flat mass functions extending from sub-solar to
$\gtrsim 10\,\Msun$ ($\alpha = 1.13$ in \citealt{stacy16}). These slopes drift
with time and resolution \citep{stacy16, prole22a}.

We emphasize that our PMF is a snapshot at $15\kyr$, not a final IMF:
all sinks are still actively accreting, no radiative feedback has
terminated accretion, and the population will be further shaped by
continued mergers, ejections, and eventual UV feedback
\citep{hosokawa11, hosokawa16, stacy16}.

The two most massive sinks, a bound wide pair,
have each reached $\sim 25\,\Msun$ by $15\kyr$, about half of the mass at which
accretion was halted by feedback in the cloud of \citet{hosokawa11}. At late times, their accretion rates flatten due to local gas depletion in the inner disk 
The outer sinks, by contrast, continue to grow by accreting from the
surrounding envelope, so the cluster mass budget shifts progressively
outward.

The Pop III IMF is not necessarily universal:
its slope and peak can be modified by the large-scale environment,
the initial turbulence and angular momentum, and the strength and
geometry of the magnetic field, so different halos may produce
 different mass distributions \citep{wollenberg20, hirano_machida22, hirano_sakai23, sharda_menon25}.
With the same physics setup, albeit with a smaller halo mass, \cite{meziani26} form just a single $\sim 27\,\Msun$ protostar. This indicates that the large-scale environment is a key driver of the Pop III PMF.
\citet{meziani26b} simulate an ultra-faint dwarf progenitor from $z \sim 16$ to
$z \sim 7$ (about $500$ Myr) and form 56 stars across five minihalos. Although the slope of the IMF for all the stars taken together is $\alpha \simeq 1.9$, their most massive, first collapsing halo has a slope $\alpha \approx 1$.

\subsection{Magnetic field evolution}

The amplification of primordial seed fields during the collapse and disk phase is one of the most uncertain aspects of Pop III star formation: the initial field strength is poorly constrained, and the relevant dynamo mechanisms (turbulent small-scale, large-scale differential rotation) operate on overlapping timescales.
We compare the evolution of our mass-to-flux ratio and magnetic energy to  MHD simulations that  resolve the disk phase, to assess whether the rapid amplification we observe is generic or a consequence of our particular choice of seed field and numerical method.

Our net-flux mass-to-flux ratio falls from $\mu_{\Phi,\mathrm{net}} \sim 10^{2}$--$10^{3}$ at early times to $\sim 15$ by $15\kyr$ (\autoref{sec:bfield}), and the magnetic energy grows by $\sim 3.5$ orders of magnitude. Despite this amplification, disk fragmentation in our simulation
remains vigorous until $\sim 8\kyr$, in contrast to the strong suppression
found in some MHD simulations.
\citet{machida25} systematically varied initial field strength from
$B_0 = 10^{-20}$ to $10^{-4}$ G and found that even the weakest
field suppresses fragmentation and drives protostellar mergers within
the first $\sim 1000$ yr, once it has been amplified (weak-field models
fragment first and merge afterwards).
\citet{stacy22} found only two secondary sinks in cosmological MHD
runs.
 A key difference  is that our simulation starts from a cosmologically plausible highly
supercritical configuration ($\mu_\Phi \sim 300$).
We have not measured the magnetic torques directly.%
How much the field changes the fragmentation and the disk evolution can only be measured against an otherwise identical run without magnetic fields, which we leave to future work.

 \citet{sadanari23} found that
ambipolar diffusion is dynamically subdominant in primordial gas
(diffusion heating always smaller than compressional heating), broadly
consistent with our use of ideal MHD \citep{meziani26}.
\citet{tanaka24} similarly found in primordial disk simulations that
turbulent magnetic fields reduce but do not eliminate fragmentation,
consistent with the vigorous fragmentation of our magnetized disk.

\section{Caveats}
\label{sec:caveats}

Several limitations of this calculation should be kept in mind when
reading the results above.

\textit{Cosmic-ray background.} The cosmic-ray ionization and
dissociation rate entering the H$_2$ network is left at its default,
MW-like value rather than being scaled to a primordial environment.
Cosmic rays destroy H$_2$, but the free electrons they supply also increase
H$_2$ and HD formation through the H$^-$ channel. The net effect of the MW-like
rate is to make the $\approx 200$\,K gas that contracts slowly before runaway
collapse more molecular and colder, and the lowest-density gas hotter, than it
would be at the much lower rate expected at $z \sim 21$ \citep{nakauchi14,
hummel16}. Across a wide range of rates, the thermal path of the collapsing gas
converges to that without cosmic rays once runaway collapse begins
\citep{hummel16}. The cosmic-ray rate is therefore unlikely to cause the offset
from the primordial $T$--$n$ track at $n \gtrsim 10^{7}\,\mathrm{cm^{-3}}$
(\autoref{sec:thermo}), which is more likely set by the reduced chemistry
network (see below).

\textit{Chemistry network.} The primordial chemistry is followed with
the network described in \autoref{sec:thermochem} rather than with a dedicated
package such as GRACKLE, and H$_2$ dissociation is not coupled
self-consistently to the radiation transport. The network also omits the
chemical heating and cooling associated with H$_2$ formation and dissociation,
and ties HD to H$_2$ by a fixed ratio.  The three-body rate is a single prescription, whereas published fits are uncertain by nearly an order of magnitude and can change the disk accretion and fragmentation \citep{turk11b}. Our gas is 0.4 dex colder than the
primordial track, which lowers the Jeans mass by a factor of $\approx 4$.
Raising the temperature to the primordial track (a factor of 1.8--3.2) would
increase $Q$ by $\approx 43\%$ during the first fragmentation burst, when
thermal pressure supplies most of the support (\autoref{sec:stability}), and by
$\approx 20\%$ after $\sim 5\kyr$, when turbulence dominates. The disk would
still fragment, but this bias would make our sink count an upper
estimate, and the initial fragment masses may be low by a factor of a few.
 The total stellar mass is less affected,
because it is set by the envelope inflow at densities below those where
three-body H$_2$ formation heating matters. The molecular fractions and
temperatures we quote should therefore be treated as indicative rather than
converged.

\textit{Magnetic field limitations.} We cannot establish whether a small-scale
turbulent dynamo is resolved in this calculation.  The $B$--$n$
relation and the field normalization (\autoref{sec:bfield}) cannot separate
anisotropic compression, winding and dynamo amplification, and a resolved
dynamo could add to the field strengths we report. In addition to
that, the current STARFORGE implementation removes the magnetic field of a gas
particle when a sink particle is spawned, which is not a realistic
prescription. 

\textit{Parent-simulation resolution.} The epoch ($z \approx 21$) and host
($2\times10^{7}\,\Msun$, an atomic-cooling halo) of the first collapse are set by the
parent FIRE-resolution run, before the hyper-refinement is switched on. With
$m_\mathrm{DM} \approx 3.5\times10^{4}\,\Msun$, H$_2$-cooling minihalos of
$10^{5}$--$10^{6}\,\Msun$ are resolved by only $\sim 16$ DM particles, and
the gas-only Jeans refinement cannot compensate for this. In MW-progenitor regions the very first Pop III stars form earlier, in minihalos, but most form in atomic-cooling halos like ours once a Lyman--Werner background is established (\autoref{sec:disc_comparison}; \citealt{storck26}). 
Our results
therefore describe the first collapse in an atomic-cooling halo, and the
progenitor minihalos, their Lyman-Werner radiation and any pre-enrichment are
missing.

\textit{Protostellar radii and accretion luminosity.} The protostellar
model \citep{Offner_2009} gives our central sinks radii of only
$5.5\,\mathrm{R_\odot}$ while they grow from $0.2$ to $2.4\,\Msun$ during the first
$1.5\kyr$. Pop III protostars accreting at $10^{-3}$--$10^{-2}\,\Msun\,
	\mathrm{yr^{-1}}$ are instead swollen to $\approx 50\,\mathrm{R_\odot}$ at
$0.2$--$2\,\Msun$, $R_* \approx 48\,\mathrm{R_\odot}\,(M_*/\Msun)^{0.27}
	(\dot M/4.4\times10^{-3}\,\Msun\,\mathrm{yr^{-1}})^{0.41}$ \citep[][Eq. 21b]{stahler86}, as in the protostellar tracks of \citet{omukai_palla03}.
Since the accretion luminosity scales as $GM\dot M/R_*$,
it, and with it the early heating fronts of \autoref{sec:stability}, is likely
overestimated by a factor of $\sim 6$.

\textit{Integration time.} We follow only $\sim 15\kyr$ after the
first protostar forms. All the sinks are still accreting when the
calculation ends, no object has reached the main sequence (all sinks above $2\,\Msun$ are in the shell deuterium-burning phase of the protostellar model; \autoref{sec:massevol}), and the ionizing
feedback that would eventually terminate accretion has not had time to act.
The mass function we report is protostellar, and the final stellar masses
could differ substantially. Fragmentation in the disk has stopped by $\approx 8\kyr$, while the central sinks keep accreting, so in the near term the high-mass end should grow further. Over longer times the mass function is expected to steepen, as in the population run of \citet{meziani26b} (\autoref{sec:disc_pmf}).

\textit{Protostellar jets.} The jets are launched with the intended mass loading but couple to the gas close to the protostars, so they propagate less far than intended and entrain and clear little of the infalling envelope (\autoref{sec:stars}). Jets matter for the stellar masses: in STARFORGE
simulations of present-day clouds they reduce the stellar masses by about an order of
magnitude, both by removing accreted gas and by disrupting the accretion flow onto the
protostars \citep{guszejnov22}. In the FORGE'd in FIRE Pop III protostar of
\citet{meziani26}, jets regulate the accretion and the final mass of the star. With
jets that couple only close to the star, the part of the accreted gas they remove is still
taken out, but the disruption of the larger-scale accretion flow is underestimated, so our protostellar masses, accretion rates and star formation efficiency are
therefore overestimates, and the PMF may be shifted to  lower masses. 
\section{Conclusions}
\label{sec:conclusions}

We have presented a radiation-magnetohydrodynamic cosmological zoom-in
simulation of Pop III star formation, following the collapse of
primordial gas in the FIRE MW-progenitor halo from cosmological initial
conditions through the formation and evolution of a protostellar
accretion disk over $\sim 15\kyr$.
Our main conclusions are as follows:

\begin{enumerate}
	\item \textbf{Vigorous disk fragmentation.}\label{concl:frag}
	      The accretion disk becomes gravitationally unstable within
	      $\sim 3\kyr$ of the first protostar's formation, developing
	      spiral arms that fragment into $\sim 60$ sink formation events
	      over the first $\sim 8\kyr$, producing a total stellar mass
	      $> 150\,\Msun$.

    \item \textbf{Sub-virial start to virial equilibrium.}\label{concl:virial}
	      The disk starts in a sub-virial state
	      ($2 E_\mathrm{kin}/|E_\mathrm{pot}| \approx 0.6$) and reaches
	      virial equilibrium within $\sim 2\kyr$, thereafter self-regulating
	      around $2 E_\mathrm{kin}/|E_\mathrm{pot}| \approx 1$, robustly across disk identification criteria.
          
	\item \textbf{Magnetic field amplification and tangling.}\label{concl:mag}
	      Magnetic energy grows by $\sim 3.5$ orders of magnitude.%
          The net-flux mass-to-flux ratio
	      falls from $\sim 10^2$--$10^3$ to $\sim 15$, because the
	      inflow brings in flux while the dense gas accreted onto sinks carries
	      little flux per unit mass, while the unsigned
	      vertical flux grows 15 times more than the net flux. The
	      field becomes tangled into mixed polarity, with $B_z$ reversing
	      along the spiral arms.

	\item \textbf{Accelerating accretion and high SFE.}\label{concl:accr}
	      The total stellar mass grows as $M_* \appropto t^{2}$ at
	      accretion rates of $\sim 10^{-3}$--$10^{-2}\,\Msun\,\mathrm{yr}^{-1}$,
	      an order of magnitude above the Shu rate and close to the
	      Larson--Penston rate of an outside-in envelope, which falls in at a
	      nearly constant $\approx 3.5\,\mathrm{km\,s^{-1}}$.
	      The star formation efficiency reaches $\approx 90\%$ within the disk
	      (after $\approx 11\kyr$; $\approx 80\%$ at $9.5\kyr$)
	      and $\sim 50\%$ within a $0.1$ pc sphere.

	\item \textbf{Top-heavy protostellar mass function.}\label{concl:pmf}
	      The protostellar mass function has a high-mass slope
	      $\alpha_\mathrm{PMF} = 1.01^{+0.14}_{-0.14}$, i.e.\ roughly flat in
	      $\mathrm{d}N/\mathrm{d}\log M$, much shallower than the Salpeter/Kroupa
	      value ($\alpha = 2.35/2.3$) and similar to other Pop III
	      simulations ($\alpha \approx 1.1$--$1.2$).
	      Individual sinks grow at most linearly in time, with no sign of runaway
	      (Bondi--Hoyle) competitive accretion; the most massive
	      sinks are the oldest, central ones. The second most massive sink is on a wide, bound,
	      eccentric orbit.

	\item \textbf{The cosmological envelope sets the mass supply.}\label{concl:env}
	      The cosmological environment sets the mass supply, its duration and its
	      angular momentum: a Larson--Penston-like inflow of
	      $\approx 10^{-2}\,\Msun\,\mathrm{yr^{-1}}$ through 3000\,AU for most of the run
	      (falling to $\approx 7\times 10^{-3}\,\Msun\,\mathrm{yr^{-1}}$ after $\sim 12$\,kyr), whose
	      centrifugal radius stays at $\sim 0.7\times10^{3}$\,AU throughout, with a broad
	      spread. 
	\item \textbf{Fragmentation moves outward.}\label{concl:out} In our simulation the fragmentation sites move outward: the first burst
	      ($\sim 3.5\kyr$, $\sim 40$ sinks) forms at $\sim 500$\,AU and the second
	      ($\sim 8\kyr$, $\sim 11$ sinks) at $\sim 10^{3}$\,AU, outside where
	      most of the infall lands. The gas mass of the disk moves outward with them, the sinks empty the inner disk while the outer disk keeps gaining mass.
\end{enumerate}

The results presented here are for a single realization of collapse in a massive halo at $z\approx 21$, and are therefore subject to that caveat.  Pop III star formation is highly stochastic, so the fragment counts and
the shape of the protostellar mass function carry an uncertainty that a
single run cannot quantify. 
We leave a population study of Pop III collapse from a large-scale cosmological environment to future work.

\begin{acknowledgments}
We thank Shyam Menon, James Beattie, Bart Ripperda, Chris Thompson, Piyush Sharda, and several members of CITA for useful discussions, as well as SciNet maintenance and support teams for help with the usage of Trillium supercomputer.
Analysis was performed on the Trillium supercomputer at the SciNet HPC Consortium. SciNet is funded by Innovation, Science and Economic Development Canada; the Digital Research Alliance of Canada; the Ontario Research Fund: Research Excellence; and the University of Toronto.
VP thanks the CCA at the Flatiron Institute for hospitality while a portion of this research was carried out. The Center for Computational Astrophysics at the Flatiron Institute is supported by the Simons Foundation.
The Flatiron Institute is a division of the Simons foundation. The computations in this work were, in part, run at facilities supported by the Scientific Computing Core at the Flatiron Institute, a division of the Simons Foundation.
NM and VP were supported in part by the Natural Sciences \& Engineering Research Council of Canada (NSERC) [funding reference number RGPIN-2023-04901]. Cette recherche a \'et\'e financ\'ee par le Conseil de recherches en sciences naturelles et en g\'enie du Canada (CRSNG), [num\'eros de r\'ef\'erence RGPIN-2023-04901].
It was performed in part at Aspen Center for Physics, which is supported by National Science Foundation grant PHY-2210452.
CAFG was supported by NSF through grants AST-2108230, AST-2307327 and AST-2606015; by NASA through grants 80NSSC22K0809, 80NSSC22K1124 and 80NSSC24K1224; by STScI through grant JWST-AR-03252.001-A; and by BSF through grant \#2024262.{}
Support for P.F.H. was provided by a Simons Investigator grant.
\end{acknowledgments}

\software{GIZMO, NumPy, SciPy, Matplotlib, Astropy, h5py, yt, Meshoid}

\appendix

	\section{H$_2$ network coefficients}
	\label{app:chem}
	The rate equation for the H$_2$ abundance is given in \autoref{eq:nh2}.
	Here $\bar{n}_x$ is volume-averaged number density of $x$ inside of a cell.
	$C_m = \left< n^m \right> / \left< n \right>^m$ is the $m^{\text{th}}$-order
	clumping factor.
	$\alpha_\mathrm{GP}$ is the gas-phase formation rate coefficient via the
	H$^{-}$ and H$_2^+$ pathways \citep{Abel_1997,Galli_Palla_1998,Glover_Jappsen_2007}.
	The three-body formation coefficient $\alpha_{3\mathrm{B}} = 6\times10^{-32}\,T^{-0.25} + 2\times10^{-31}\,T^{-0.5}\,\mathrm{cm^{6}\,s^{-1}}$
	\citep{FIRE_3_Updated_Hopkin}, which is a single prescription that differs from the fits of \citet{abel02} and \citet{Palla_1983}, becomes the dominant H$_2$ formation channel
	for $n \gtrsim 10^8\,\mathrm{cm}^{-3}$.
	$\Gamma_{\mathrm{H_2}}$ is the coefficient associated with LW and ionizing photon
	dissociation.
	$\xi_{\mathrm{H_2}}$ accounts for cosmic-ray-induced ionization and dissociation
	(the code uses a fixed cosmic-ray energy density of $1\,\mathrm{eV\,cm^{-3}}$,
	giving $\zeta_\mathrm{CR} = 1.6\times10^{-17}$\,s$^{-1}$ and
	$\xi_{\mathrm{H_2}} = \zeta_\mathrm{CR}/2 = 8\times10^{-18}$\,s$^{-1}$ in gas denser than
	$\sim 10^3$ times the cosmic mean, a MW-like value;
	this rate is unconstrained at $z \sim 21$, where the physically
	expected value is far lower in the absence of supernovae. Besides destroying
	H$_2$, cosmic-ray ionization supplies free electrons that catalyse H$_2$
	(and HD) formation through the H$^-$ channel, so a rate this high makes
	the $\approx 200$\,K gas that contracts slowly before runaway collapse more
	molecular and colder (and can move it onto the colder HD-cooling track), while
	CR heating warms the lowest-density gas
	\citep{stacy_bromm07, nakauchi14, hummel16}. These differences are erased as
	the gas collapses: once runaway collapse begins, its thermal path converges
	to that without CRs \citep{hummel16};
	see \autoref{sec:caveats}).
	The last sum runs over collisional dissociation contributions from
	$e^-, \rm H^+, H_2, HI, HeI, HeII$, with rate coefficients $\beta_{\rm H_2, i}$ from \citet[][Table A1, reactions 7--11, 24--25 and 37--39]{Glover_Abel_2008}.

	\section{Disk identification}
	\label{app:diskid}
	To compute velocity dispersion, local Toomre $Q$, and other disk
	properties, we identify gas particles belonging to the disk at
	each snapshot using the following procedure:

	\begin{enumerate}
		\item We define a search radius
		      $r_\mathrm{search} = 2000\AU$, chosen because it roughly matches
		      the size of the disk that forms in our simulation. This radius
		      sets the region over which the bulk velocity and angular
		      momentum are measured, and we adopt it as the disk
		      scale throughout this work: the disk we identify below is
		      approximately equal to $r_\mathrm{search}$.
		\item Before any sinks form, the center is
		      placed at the densest gas particle within a refinement region.
		      After the first sink forms, we place the disk center at the most massive sink
		      particle. We find that this approach results in a center location that is more stable than using than the mass-weighted
		      center-of-mass (COM) of all sinks once secondary sinks form
		      at large separations.
		\item We compute a mass-weighted COM velocity $v_\mathrm{COM}$ from
		      gas particles within $r < r_\mathrm{search}$ and subtract it
		      from all particle velocities.
		\item We compute the disk normal $\Lhat$ from the
		      total angular momentum
		      $\mathbf{L} = \sum_i m_i\,(\mathbf{r}_i \times \mathbf{v}_i)$
		      summed over particles within $r_\mathrm{search}$, using the
		      COM-subtracted velocities from the previous step. The sign
		      of $\Lhat$ is fixed so that $L_z > 0$ to prevent
		      frame-to-frame flipping.
		\item We convert particle positions and velocities onto the
		      disk-plane coordinate system
		      $(r_\mathrm{cyl},\,\phi,\,z,\,v_r,\,v_\phi,\,v_z)$.
		\item We apply geometric and density cuts:
		      $r_\mathrm{cyl} < r_\mathrm{max} = 2000\AU$,
		      $f_\mathrm{kep} = |z|/r_\mathrm{cyl} < 0.3$ (aspect ratio), and
		      $\rho > \rho_\mathrm{thresh} = 10^{-15}\gcm$.
		\item We apply kinematic cuts:
		      $v_\phi > 0$ (co-rotating) and
		      $v_\phi / v_K > f_\mathrm{kep} = 0.3$, where
		      $v_K = \sqrt{G M_\mathrm{enc}(<r)/r}$ is the Keplerian speed
		      computed from the cumulative enclosed gas mass plus the total
		      sink mass placed at the origin.
		\item We extend the kinematically selected set to its bounding
		      cylinder (100th percentile in $r_\mathrm{cyl}$ and $|z|$) to
		      recover any disk material just outside the strict kinematic
		      boundary.
	\end{enumerate}
	Particles passing all cuts are labeled as disk members.
	The disk radius $\rdisk$ is taken as the 90th-percentile
	cylindrical radius of disk members.

	We tested this selection at nine epochs with eleven variants. The disk mass
	changes by $\lesssim 15\%$ for aspect ratios of $0.2$--$0.5$, for
	$f_\mathrm{kep} = 0$--$0.5$, for a center at the center of mass of gas and sinks, and
	for an iterated disk axis. It depends more strongly on the bounding-cylinder step and
	on the density threshold: dropping the former, or raising the latter to
	$10^{-14}\gcm$, lowers $M_\mathrm{disk}$ by up to a factor of $\sim 2.5$ at late
	times, when the disk is small. The star formation efficiency within the disk is
	robust to all of these choices ($SFE = 0.8$ at $9.5\kyr$ and $0.94$ at
	$15\kyr$).

	\section{Toomre-$Q$ estimator}
	\label{app:toomre}
	For each snapshot the gas is binned into 20 annular bins.
	In each annulus $b$ we compute mass-weighted face-on density:
	\begin{equation}
		\Sigma_b = \frac{\sum_{i\in b} m_i}{\pi(r_{\mathrm{out},b}^2 - r_{\mathrm{in},b}^2)}
	\end{equation}
	with $r_{\mathrm{in},b}$ and $r_{\mathrm{out},b}$ corresponding to the inner
	and outer radii of the annulus $b$ respectively.
	The radial velocity dispersion for each gas particle is measured as its deviation from the annulus average $\bar{v}_{r,b}$:
	\begin{equation}
		\sigma_{r,i} = \sqrt{v_{r,i}^2 - \bar{v}_{r,b}^2}
	\end{equation}
	The angular velocity is measured directly from the rotation profile:
	\begin{equation}
		\Omega_b = \frac{|\bar{v}_{\phi,b}|}{r_b}
	\end{equation}
	where $\bar{v}_{\phi,b}$ is the mass-weighted mean azimuthal velocity in annulus
	$b$ and $r_b$ is the bin center. The epicyclic frequency is computed numerically
	from the measured rotation curve \citep[e.g.][]{Toomre_1964},
	\begin{equation}
		\kappa_b^2 = \frac{2\,\Omega_b}{r_b}\,\frac{\mathrm{d}(r^2\,\Omega)}{\mathrm{d}r}\bigg|_{r_b}
	\end{equation}
	The Toomre parameter per annulus is then
	\begin{equation}
		Q_b = \frac{\zeta_b \,\kappa_b}{\pi\,G\,\Sigma_b}
	\end{equation}

	A pixel-by-pixel face-on $Q$ map is produced by computing
	local mass-weighted quantities for all of the relevant parameters:
	\begin{equation}
		Q(x,y) = \frac{\zeta_b (x,y)\;\kappa(r_{x,y})}{\pi\,G\,\Sigma(x,y)}
	\end{equation}
	where $r_{x,y} = \sqrt{x^2 + y^2}$, and $\zeta = \sqrt{\sigma_r^2 + c_s^2}$,
	which sums over thermal and turbulent contributions.

	\section{Star formation efficiency per free-fall time}
	\label{app:eff}
	We make the following assumptions. (i) The pre-stellar envelope is spherically symmetric, with a power-law density profile $\rho \propto r^{-2}$ out to the radius considered. The sound speed is not constant in the envelope: in \autoref{fig:kinematics_wide} it rises outward and stays close to a constant fraction of the Keplerian speed, $c_s \approx 0.25$--$0.3\,v_K$ ($\approx 1$--$2\,\mathrm{km\,s^{-1}}$ between $10^{4}$ and $10^{5}\AU$ before the first sink forms). For $\rho \propto r^{-2}$ the enclosed mass is $M \propto r$, so $v_K$ and hence $c_s$ are then constant with radius, and $A = 1/(2\beta^2)$ for $c_s = \beta v_K$; $\beta \approx 0.28$ gives the $A \approx 6$ found below. We use $c_s$ only through the product $A c_s^2$, which is set by the measured density normalization, so $\varepsilon_\mathrm{ff}$ does not depend on the value of $c_s$. (ii) The normalization $A$ of the profile is fixed by the gas mass inside our $0.1$\,pc sphere before the first sink forms, which gives $A \approx 6$ below, i.e.\ between the singular isothermal sphere ($A = 1$) and the Larson--Penston value. (iii) The gas mass in the sphere is taken to stay equal to this envelope mass, so that $\varepsilon_\mathrm{ff}$ follows from the stellar accretion rate alone. (iv) The free-fall time is that of the gas alone, whereas the measured $\varepsilon_\mathrm{ff}$ also includes the sinks in $\bar\rho$; the difference is quantified at the end of this appendix.

Approximate the envelope as a power law $\rho \propto r^{-2}$, written relative to the
	singular isothermal sphere (SIS) of sound speed $c_s$ as
	\begin{equation}
		\rho(r) = A\,\frac{c_s^{2}}{2\pi G r^{2}},
		\label{eq:envelope_A}
	\end{equation}
	where $A = 1$ for the SIS and $A \approx 4.4$ for the Larson--Penston envelope
	\citep{Shu_1977}. The mass and the mean density inside $r$ are
	\begin{equation}
		M(<r)        = \int_0^r 4\pi r'^{2}\rho\,\mathrm{d}r' = \frac{2A\,c_s^{2}\,r}{G}, \quad
		\bar\rho(<r) = \frac{3M}{4\pi r^{3}} = \frac{3A\,c_s^{2}}{2\pi G r^{2}},
	\end{equation}
	so the free-fall time at that mean density is
	\begin{equation}
		t_\mathrm{ff} = \sqrt{\frac{3\pi}{32 G\bar\rho}} = \frac{\pi r}{4\sqrt{A}\,c_s}.
	\end{equation}
	The mass inside $r$ divided by its own free-fall time, i.e.\ the rate at which this
	gas would turn into stars if all of it collapsed within one free-fall time, is then the
	same at every radius,
	\begin{equation}
		\frac{M}{t_\mathrm{ff}} = \frac{8}{\pi}\,A^{3/2}\,\frac{c_s^{3}}{G},
	\end{equation}
	and the efficiency per free-fall time of a sphere of radius $r$ in the envelope is
	\begin{equation}
		\varepsilon_\mathrm{ff} = \frac{\dot M_*\,t_\mathrm{ff}}{M_\mathrm{gas}}
		= \frac{\pi}{8}\,A^{-3/2}\,\frac{G\,\dot M_*}{c_s^{3}}.
	\end{equation}
	Before the first sink forms, the envelope holds $\approx 160\,\Msun$ inside our $0.1$\,pc sphere, where $c_s \approx 1\,\mathrm{km\,s^{-1}}$ (\autoref{fig:kinematics_wide}),
	i.e.\ $A \approx 6$, so that
	$M/t_\mathrm{ff} \approx 9\times10^{-3}\,\Msun\,\mathrm{yr^{-1}}$
	(this is $\approx 37\,c_s^{3}/G$ for $c_s = 1\,\mathrm{km\,s^{-1}}$; for the SIS it would be $\approx 2.5\,c_s^{3}/G$). Between $4$ and $10\kyr$ the stars
	accrete at $\dot M_* \approx 1.7\times10^{-2}\,\Msun\,\mathrm{yr^{-1}}$
	($\approx 70\,c_s^{3}/G$ for $c_s = 1\,\mathrm{km\,s^{-1}}$), which gives $\varepsilon_\mathrm{ff} \approx 1.9$ with the free-fall
	time of the gas alone. Including the sinks in $\bar\rho$, as in our definition, shortens
	$t_\mathrm{ff}$ and gives the measured median $\varepsilon_\mathrm{ff} \approx 1.4$
	over this period ($\lesssim 1$ before $4\kyr$ and after $10\kyr$).

\section{Magnetic field amplification channels and net toroidal flux}
\label{app:bamp}
The field does not remain wound up for long; by $\Delta t= 9.5$ kyr the field shows little order. Differential rotation,
$q = -\mathrm{d}\ln\Omega/\mathrm{d}\ln R$, winds a radial field into a toroidal one
that grows linearly in time, $B_\phi \approx -q\Omega t\,B_r$
\citep[][Sec. 3.3]{brandenburg_subramanian05}, so after a time $t$ the toroidal field
should exceed the radial one by a factor $q\Omega t$.

We measure the ratio of the two in midplane annuli at $R = 300$, $500$ and
$800\AU$ as the ratio of their mass-weighted rms values,
$B_{\phi,\mathrm{rms}}/B_{r,\mathrm{rms}} = (\langle B_\phi^2\rangle/\langle B_r^2\rangle)^{1/2}$,
which avoids the divergence of $B_\phi/B_r$ where $B_r$ changes sign. This ratio
stays at $\approx 2$, and the mass-weighted median pitch angle
$\arctan(|B_r|/|B_\phi|)$ at $\approx 38^\circ$, at all epochs, whereas winding
since $\Delta t = 0$ would give $q\Omega t \approx 15$ at $550\AU$ by
$2.8\kyr$ and $\sim 100$ later. Winding also brings field lines of
opposite polarity together \citep[][Sec. 6.5]{brandenburg_subramanian05}: the radial
wavenumber of the wound field grows as $k_R = q\Omega t/R$, so reversals are
$\pi R/(q\Omega t)$ apart.
This spacing stays larger than the local resolution until $\gtrsim 6\kyr$ at
$300\AU$, so numerical reconnection cannot explain the deficit at early times.
This suggests that the radial field is instead continually regenerated by the
turbulent, gravitationally driven motions of the disk, whose velocity dispersion
is comparable to the rotation speed, as in a gravitoturbulent dynamo, in which
vertical circulation around the spiral arms regenerates the poloidal field
\citep{deng20}, but we have not confirmed this.

The net sign of $B_\phi$ is not set by the winding, which
by symmetry leaves the azimuthal average near zero. It is instead set
by the uneven, one-sided mass distribution of the disk, which the star
formation traces. The denser, strongly $B_\phi<0$ arm
is the one in which most of the stars form (see the lower left panel at $\Delta t=3.6$ kyr of the field-morphology figure in \autoref{sec:bfield}); that arm is also where compression has made the field strongest. Because the field is frozen
into the gas, the inflow of this arm carries its negative toroidal flux
into the inner disk, while the positive flux of the lighter arm remains
at larger radii. Opposite-signed flux brought together reconnects, and
the net flux that remains in the inner disk is negative; the sign of the field is
simply that carried by the spiral arm that happened to become more massive.

We have verified that the dense gas
does carry most of the negative toroidal field: inside $2500\AU$ the mass-weighted
$\langle B_\phi\rangle$ exceeds the volume-weighted one by a factor of $\sim 20$--$800$.

The redistribution of magnetic flux, and the reduction of toroidal flux by reconnection, is, however, not the whole story. The net toroidal flux
through the meridional plane stays negative out to $1.6\times10^{4}\AU$ (not shown in the figure), with
$|\Phi_\phi^\mathrm{net}| / \Phi_\phi^\mathrm{unsigned} \approx 0.35$, instead of
being balanced by positive flux at larger radii; we conclude that a net ($m=0$) toroidal field
is also being generated, by winding of a net radial field, by the disk's
turbulent motions, or by some other mechanism.

\autoref{fig:phase_Bfield} shows the evolution of $|B(n)|$ as a function of $n$ for each of the 6 epochs. We fit a power law $|B(n)|\sim n^\alpha$ at each epoch.
The power-law exponent rises from $\alpha \approx 2/3$ at early times,
consistent with flux-freezing compression of spherically collapsing gas, toward
$\alpha \approx 1$ at late times.
The value $\alpha \approx 1$ is expected
when compression becomes anisotropic and the field is stretched in the plane
of the forming disk. However, kinematic-dynamo calculations give similar
slopes, $B \propto n^{0.8\text{--}0.9}$ \citep[see the summary in][]{hirano_machida22},
so the $B$--$n$ relation alone cannot separate anisotropic compression,
winding and dynamo amplification.
The magnitude of $|B|$ up to the time the first star forms is consistent with
flux freezing of the initial seed field.
Assuming flux freezing, the cosmological
expansion gives $\approx 8\times10^{-12}$\,G at the mean density at $z = 21$ (the redshift when the disk forms), which is what the low-density gas of the parent simulation has. Isotropic compression from the mean baryon density
($\bar n \approx 2\times10^{-3}\,\mathrm{cm^{-3}}$) to $n = 10^{10}\,\mathrm{cm^{-3}}$
then gives $\approx 2\times10^{-3}$\,G (\autoref{sec:mhd}), close to the $\approx
10^{-3}$\,G measured at that density before the first sink forms
(\autoref{fig:phase_Bfield}). Compression alone therefore can account for the
field strength at the onset of star formation.

The later steepening towards $\alpha \approx 1$ may come from anisotropic compression, winding or a dynamo.

The alignment of the in-plane flow and field lines does not allow us to determine whether a dynamo is acting. The median $|\cos\theta|$ rises from $0.69$ before the first sink
forms, which is the value for randomly oriented in-plane vectors ($\cos
45^\circ = 0.71$), to $0.94$ once the spiral arms are established. In a
rotating disk, winding aligns $\mathbf{B}$ with $\mathbf{v}$ whatever the
field strength. The ratio $E_\mathrm{mag}/E_\mathrm{kin}$, the plasma $\beta$
or the Alfv\'en Mach number would be the appropriate diagnostics; at late
times $E_\mathrm{mag}/E_\mathrm{therm} \approx 0.3$ near the center
(\autoref{fig:energy_evol}), so the field is not deeply kinematic there.
We cannot demonstrate that a small-scale dynamo is resolved, nor that it is absent (see \autoref{sec:caveats}), and neither the measured $\alpha$ nor the flow--field alignment can decide whether compression and shear alone account for the amplification.

This caution is reinforced by the scheme dependence of magnetic amplification
reported by \citet{ugalino26}. Evolving identical Milky-Way-like galaxy initial
conditions with Eulerian (ART, RAMSES) and (quasi-)Lagrangian (AREPO, SWIFT)
MHD codes, they found that the two families differ by orders of magnitude in
their amplification histories: the lower numerical diffusivity of the
Lagrangian schemes allows small-scale motions to persist and the field to grow
to near equipartition, whereas the grid-based codes remain close to the
shear-driven $E_\mathrm{mag} \propto t^2$ scaling. Because the meshless
finite-mass scheme used here is Lagrangian, the same tendency could inflate the
field strengths and lower the mass-to-flux ratios we measure relative to grid-based
calculations, although that comparison was made for galactic disks and did
not include GIZMO.
\bibliographystyle{aasjournalv7}
\bibliography{main}

\providecommand{\noopsort}[1]{}
\begin{thebibliography}{}
\expandafter\ifx\csname natexlab\endcsname\relax\def\natexlab#1{#1}\fi
\providecommand{\url}[1]{\href{#1}{#1}}
\providecommand{\dodoi}[1]{doi:~\href{http://doi.org/#1}{\nolinkurl{#1}}}
\providecommand{\doeprint}[1]{\href{http://ascl.net/#1}{\nolinkurl{http://ascl.net/#1}}}
\providecommand{\doarXiv}[1]{\href{https://arxiv.org/abs/#1}{\nolinkurl{https://arxiv.org/abs/#1}}}

\bibitem[{T. Abel {et~al.}(1997)Abel, Anninos, Zhang, \& Norman}]{Abel_1997}
Abel, T., Anninos, P., Zhang, Y., \& Norman, M.~L. 1997,
  \bibinfo{title}{{Modeling primordial gas in numerical cosmology},} New
  Astronomy, 2, 181, \dodoi{10.1016/S1384-1076(97)00010-9}

\bibitem[{T. Abel {et~al.}(2002)Abel, Bryan, \& Norman}]{abel02}
Abel, T., Bryan, G.~L., \& Norman, M.~L. 2002, \bibinfo{title}{{The Formation
  of the First Star in the Universe},} Science, 295, 93,
  \dodoi{10.1126/science.1063991}

\bibitem[{J.~S. {Bagla}(2002){Bagla}}]{2002JApA...23..185B}
{Bagla}, J.~S. 2002, \bibinfo{title}{{TreePM: A Code for Cosmological N-Body
  Simulations},} Journal of Astrophysics and Astronomy, 23, 185,
  \dodoi{10.1007/BF02702282}

\bibitem[{J. Barnes \& P. Hut(1986)Barnes \& Hut}]{Barnes_Hut_1986}
Barnes, J., \& Hut, P. 1986, \bibinfo{title}{A hierarchical O(N log N)
  force-calculation algorithm,} Nature, 324, 446–449,
  \dodoi{10.1038/324446a0}

\bibitem[{R. Beck(2015)Beck}]{beck15}
Beck, R. 2015, \bibinfo{title}{{Magnetic fields in spiral galaxies},} \aapr,
  24, 4, \dodoi{10.1007/s00159-015-0084-4}

\bibitem[{A.~C. Boley {et~al.}(2007)Boley, Hartquist, Durisen, \&
  Michael}]{boley07}
Boley, A.~C., Hartquist, T.~W., Durisen, R.~H., \& Michael, S. 2007,
  \bibinfo{title}{{The Internal Energy for Molecular Hydrogen in
  Gravitationally Unstable Protoplanetary Disks},} \apjl, 656, L89,
  \dodoi{10.1086/512235}

\bibitem[{I.~A. Bonnell {et~al.}(2001)Bonnell, Bate, Clarke, \&
  Pringle}]{bonnell01}
Bonnell, I.~A., Bate, M.~R., Clarke, C.~J., \& Pringle, J.~E. 2001,
  \bibinfo{title}{{Competitive accretion in embedded stellar clusters},}
  \mnras, 323, 785, \dodoi{10.1046/j.1365-8711.2001.04270.x}

\bibitem[{W.~B. Bonnor(1956)Bonnor}]{Bonnor_1956}
Bonnor, W.~B. 1956, \bibinfo{title}{Boyle’s Law and gravitational
  instability,} Monthly Notices of the Royal Astronomical Society, 116, 351,
  \dodoi{10.1093/mnras/116.3.351}

\bibitem[{A. Brandenburg \& E. Ntormousi(2023)Brandenburg \&
  Ntormousi}]{brandenburg_ntormousi23}
Brandenburg, A., \& Ntormousi, E. 2023, \bibinfo{title}{{Galactic Dynamos},}
  \araa, 61, 561, \dodoi{10.1146/annurev-astro-071221-052807}

\bibitem[{A. Brandenburg \& K. Subramanian(2005)Brandenburg \&
  Subramanian}]{brandenburg_subramanian05}
Brandenburg, A., \& Subramanian, K. 2005, \bibinfo{title}{Astrophysical
  magnetic fields and nonlinear dynamo theory,} Physics Reports, 417, 1

\bibitem[{V. Bromm {et~al.}(2002)Bromm, Coppi, \& Larson}]{bromm02}
Bromm, V., Coppi, P.~S., \& Larson, R.~B. 2002, \bibinfo{title}{{The Formation
  of the First Stars. I. The Primordial Star-forming Cloud},} \apj, 564, 23,
  \dodoi{10.1086/323947}

\bibitem[{G.~L. Bryan \& M.~L. Norman(1998)Bryan \& Norman}]{bryan98}
Bryan, G.~L., \& Norman, M.~L. 1998, \bibinfo{title}{Statistical Properties of
  X-Ray Clusters: Analytic and Numerical Comparisons,} The Astrophysical
  Journal, 495, 80, \dodoi{10.1086/305262}

\bibitem[{G. Chabrier(2003)Chabrier}]{chabrier03}
Chabrier, G. 2003, \bibinfo{title}{{Galactic Stellar and Substellar Initial
  Mass Function},} PASP, 115, 763, \dodoi{10.1086/376392}

\bibitem[{K.-J. Chen {et~al.}(2025)Chen, Ho, \& Tung}]{ho25}
Chen, K.-J., Ho, M.-Y., \& Tung, P.-C. 2025, \bibinfo{title}{Formation of
  Supersonic Turbulence in the Primordial Star-forming Cloud,} \apjl, 988, L67,
  \dodoi{10.3847/2041-8213/adf18d}

\bibitem[{P.~C. Clark {et~al.}(2011)Clark, Glover, Smith, Greif, Klessen, \&
  Bromm}]{clark11}
Clark, P.~C., Glover, S. C.~O., Smith, R.~J., {et~al.} 2011,
  \bibinfo{title}{{The Formation and Fragmentation of Disks Around Primordial
  Protostars},} Science, 331, 1040, \dodoi{10.1126/science.1198027}

\bibitem[{A.~C. Cummings {et~al.}(2016)Cummings, Stone, Heikkila, Lal, Webber,
  J{\'o}hannesson, Moskalenko, Orlando, \& Porter}]{cummings16}
Cummings, A.~C., Stone, E.~C., Heikkila, B.~C., {et~al.} 2016,
  \bibinfo{title}{Galactic Cosmic Rays in the Local Interstellar Medium:
  Voyager 1 Observations and Model Results,} The Astrophysical Journal, 831,
  18, \dodoi{10.3847/0004-637X/831/1/18}

\bibitem[{H. Deng {et~al.}(2020)Deng, Mayer, \& Latter}]{deng20}
Deng, H., Mayer, L., \& Latter, H. 2020, \bibinfo{title}{Global Simulations of
  Self-gravitating Magnetized Protoplanetary Disks,} The Astrophysical Journal,
  891, 154, \dodoi{10.3847/1538-4357/ab77b2}

\bibitem[{H. Deng {et~al.}(2019)Deng, Mayer, Latter, Hopkins, \& Bai}]{deng19}
Deng, H., Mayer, L., Latter, H., Hopkins, P.~F., \& Bai, X.-N. 2019,
  \bibinfo{title}{Local Simulations of MRI Turbulence with Meshless Methods,}
  The Astrophysical Journal Supplement Series, 241, 26,
  \dodoi{10.3847/1538-4365/ab0957}

\bibitem[{R. Ebert(1955)Ebert}]{Ebert_1955}
Ebert, R. 1955, \bibinfo{title}{Über die Verdichtung von H I-Gebieten. Mit 5
  Textabbildungen,} Zeitschrift fur Astrophysik, 37, 217

\bibitem[{C.-A.
  Faucher-Gigu{\`e}re(2020)Faucher-Gigu{\`e}re}]{Faucher-Giguere_2020}
Faucher-Gigu{\`e}re, C.-A. 2020, \bibinfo{title}{A cosmic UV/X-ray background
  model update,} Monthly Notices of the Royal Astronomical Society, 493, 1614,
  \dodoi{10.1093/mnras/staa302}

\bibitem[{C. Federrath {et~al.}(2011)Federrath, Sur, Schleicher, Banerjee, \&
  Klessen}]{Federrath2011}
Federrath, C., Sur, S., Schleicher, D. R.~G., Banerjee, R., \& Klessen, R.~S.
  2011, \bibinfo{title}{A NEW JEANS RESOLUTION CRITERION FOR (M)HD SIMULATIONS
  OF SELF-GRAVITATING GAS: APPLICATION TO MAGNETIC FIELD AMPLIFICATION BY
  GRAVITY-DRIVEN TURBULENCE,} The Astrophysical Journal, 731, 62,
  \dodoi{10.1088/0004-637X/731/1/62}

\bibitem[{D. Galli \& F. Palla(1998)Galli \& Palla}]{Galli_Palla_1998}
Galli, D., \& Palla, F. 1998, \bibinfo{title}{{The chemistry of the early
  Universe},} A\&A, 335, 403

\bibitem[{S.~C.~O. Glover \& T. Abel(2008)Glover \& Abel}]{Glover_Abel_2008}
Glover, S. C.~O., \& Abel, T. 2008, \bibinfo{title}{{Uncertainties in H$_{2}$
  and HD chemistry and cooling and their role in early structure formation},}
  \mnras, 388, 1627, \dodoi{10.1111/j.1365-2966.2008.13224.x}

\bibitem[{S.~C.~O. Glover \& A.-K. Jappsen(2007)Glover \&
  Jappsen}]{Glover_Jappsen_2007}
Glover, S. C.~O., \& Jappsen, A.-K. 2007, \bibinfo{title}{Star Formation at
  Very Low Metallicity. I. Chemistry and Cooling at Low Densities,} The
  Astrophysical Journal, 666, 1–19, \dodoi{10.1086/519445}

\bibitem[{N.~Y. Gnedin \& B.~T. Draine(2014)Gnedin \&
  Draine}]{Gnedin_Draine_2014}
Gnedin, N.~Y., \& Draine, B.~T. 2014, \bibinfo{title}{LINE OVERLAP AND
  SELF-SHIELDING OF MOLECULAR HYDROGEN IN GALAXIES,} The Astrophysical Journal,
  795, 37, \dodoi{10.1088/0004-637X/795/1/37}

\bibitem[{T.~H. Greif {et~al.}(2011)Greif, Springel, White, Glover, Clark,
  Smith, Klessen, \& Bromm}]{greif11}
Greif, T.~H., Springel, V., White, S. D.~M., {et~al.} 2011,
  \bibinfo{title}{{Simulations on a Moving Mesh: The Clustered Formation of
  Population III Protostars},} \apj, 737, 75,
  \dodoi{10.1088/0004-637X/737/2/75}

\bibitem[{M.~Y. Grudi{\'c} {et~al.}(2021)Grudi{\'c}, Guszejnov, Hopkins,
  Offner, \& Faucher-Gigu{\`e}re}]{STARFORGE_Towa_Grudic_2021}
Grudi{\'c}, M.~Y., Guszejnov, D., Hopkins, P.~F., Offner, S. S.~R., \&
  Faucher-Gigu{\`e}re, C.-A. 2021, \bibinfo{title}{STARFORGE: Towards a
  comprehensive numerical model of star cluster formation and feedback,}
  Monthly Notices of the Royal Astronomical Society, 506, 2199\textendash 2231,
  \dodoi{10.1093/mnras/stab1347}

\bibitem[{D. Guszejnov {et~al.}(2022)Guszejnov, Grudi{\'c}, Offner,
  Faucher-Gigu{\`e}re, Hopkins, \& Rosen}]{guszejnov22}
Guszejnov, D., Grudi{\'c}, M.~Y., Offner, S. S.~R., {et~al.} 2022,
  \bibinfo{title}{{Effects of the environment and feedback physics on the
  initial mass function of stars in the STARFORGE simulations},} \mnras, 515,
  4929, \dodoi{10.1093/mnras/stac2060}

\bibitem[{S. Higashi {et~al.}(2024)Higashi, Susa, Federrath, \&
  Chiaki}]{higashi24}
Higashi, S., Susa, H., Federrath, C., \& Chiaki, G. 2024,
  \bibinfo{title}{{Amplification and Saturation of Turbulent Magnetic Fields in
  Collapsing Primordial Gas Clouds},} \apj, 962, 158,
  \dodoi{10.3847/1538-4357/ad2066}

\bibitem[{S. Hirano \& M.~N. Machida(2022)Hirano \& Machida}]{hirano_machida22}
Hirano, S., \& Machida, M.~N. 2022, \bibinfo{title}{{Exponentially Amplified
  Magnetic Field Eliminates Disk Fragmentation around Population III
  Protostars},} \apjl, 935, L16, \dodoi{10.3847/2041-8213/ac85e0}

\bibitem[{S. Hirano {et~al.}(2023)Hirano, Shen, Nishijima, Sakai, \&
  Umeda}]{hirano_sakai23}
Hirano, S., Shen, Y., Nishijima, S., Sakai, Y., \& Umeda, H. 2023,
  \bibinfo{title}{{Formation of first star clusters under the supersonic gas
  flow -- I. Morphology of the massive metal-free gas cloud},} \mnras, 525,
  5737, \dodoi{10.1093/mnras/stad2693}

\bibitem[{M.-Y. {Ho} {et~al.}(2026){Ho}, {Chen}, \& {Tung}}]{ho25b}
{Ho}, M.-Y., {Chen}, K.-J., \& {Tung}, P.-C. 2026, \bibinfo{title}{{Turbulence
  in Primordial Dark Matter Halos and Its Impact on the First Star Formation},}
  \apj, 1004, 246, \dodoi{10.3847/1538-4357/ae731d}

\bibitem[{P.~F. Hopkins(2015)Hopkins}]{A_new_class_of_Hopkin_2015}
Hopkins, P.~F. 2015, \bibinfo{title}{A new class of accurate, mesh-free
  hydrodynamic simulation methods,} Monthly Notices of the Royal Astronomical
  Society, 450, 53\textendash 110, \dodoi{10.1093/mnras/stv195}

\bibitem[{P.~F. Hopkins(2016)Hopkins}]{Hopkins_2016_CG_MHD}
Hopkins, P.~F. 2016, \bibinfo{title}{{A constrained-gradient method to control
  divergence errors in numerical MHD},} \mnras, 462, 576,
  \dodoi{10.1093/mnras/stw1578}

\bibitem[{P.~F. Hopkins(2017)Hopkins}]{Hopkins_2017}
Hopkins, P.~F. 2017, \bibinfo{title}{Anisotropic diffusion in mesh-free
  numerical magnetohydrodynamics,} Monthly Notices of the Royal Astronomical
  Society, 466, 3387–3405, \dodoi{10.1093/mnras/stw3306}

\bibitem[{P.~F. Hopkins \& M.~Y. Grudić(2019)Hopkins \&
  Grudić}]{Hopkins_Grudic_2019}
Hopkins, P.~F., \& Grudić, M.~Y. 2019, \bibinfo{title}{Numerical problems in
  coupling photon momentum (radiation pressure) to gas,} Monthly Notices of the
  Royal Astronomical Society, 483, 4187–4196, \dodoi{10.1093/mnras/sty3089}

\bibitem[{P.~F. Hopkins {et~al.}(2020)Hopkins, Grudić, Wetzel, Kereš,
  Faucher-Giguère, Ma, Murray, \& Butcher}]{Hopkins2020}
Hopkins, P.~F., Grudić, M.~Y., Wetzel, A., {et~al.} 2020,
  \bibinfo{title}{Radiative stellar feedback in galaxy formation: Methods and
  physics,} Monthly Notices of the Royal Astronomical Society, 491,
  3702–3729, \dodoi{10.1093/mnras/stz3129}

\bibitem[{P.~F. Hopkins \& M.~J. Raives(2016)Hopkins \&
  Raives}]{Hopkins_Raives_2015}
Hopkins, P.~F., \& Raives, M.~J. 2016, \bibinfo{title}{Accurate, meshless
  methods for magnetohydrodynamics,} Monthly Notices of the Royal Astronomical
  Society, 455, 51–88, \dodoi{10.1093/mnras/stv2180}

\bibitem[{P.~F. Hopkins {et~al.}(2018)Hopkins, Wetzel, Kereš,
  Faucher-Giguère, Quataert, Boylan-Kolchin, Murray, Hayward, Garrison-Kimmel,
  Hummels, Feldmann, Torrey, Ma, Anglés-Alcázar, Su, Orr, Schmitz, Escala,
  Sanderson, Grudić, Hafen, Kim, Fitts, Bullock, Wheeler, Chan, Elbert, \&
  Narayanan}]{FIRE_2}
Hopkins, P.~F., Wetzel, A., Kereš, D., {et~al.} 2018, \bibinfo{title}{FIRE-2
  simulations: physics versus numerics in galaxy formation,} Monthly Notices of
  the Royal Astronomical Society, 480, 800–863, \dodoi{10.1093/mnras/sty1690}

\bibitem[{P.~F. Hopkins {et~al.}(2022)Hopkins, Wetzel, Wheeler, Sanderson,
  Grudić, Sameie, Boylan-Kolchin, Orr, Ma, Faucher-Giguère, Kereš, Quataert,
  Su, Moreno, Feldmann, Bullock, Loebman, Anglés-Alcázar, Stern, Necib,
  Choban, \& Hayward}]{FIRE_3_Updated_Hopkin}
Hopkins, P.~F., Wetzel, A., Wheeler, C., {et~al.} 2022, \bibinfo{title}{FIRE-3:
  updated stellar evolution models, yields, and microphysics and fitting
  functions for applications in galaxy simulations,} Monthly Notices of the
  Royal Astronomical Society, 519, 3154–3181, \dodoi{10.1093/mnras/stac3489}

\bibitem[{P.~F. Hopkins {et~al.}(2024)Hopkins, Grudic, Su, Wellons,
  Angles-Alcazar, Steinwandel, Guszejnov, Murray, Faucher-Giguere, Quataert, \&
  Keres}]{FORGE_d_in_FIRE_Hopkin_2023}
Hopkins, P.~F., Grudic, M.~Y., Su, K.-Y., {et~al.} 2024,
  \bibinfo{title}{FORGE\textquotesingle d in FIRE: Resolving the End of Star
  Formation and Structure of AGN Accretion Disks from Cosmological Initial
  Conditions,} The Open Journal of Astrophysics, 7,
  \dodoi{10.21105/astro.2309.13115}

\bibitem[{T. Hosokawa {et~al.}(2016)Hosokawa, Hirano, Kuiper, Yorke, Omukai, \&
  Yoshida}]{hosokawa16}
Hosokawa, T., Hirano, S., Kuiper, R., {et~al.} 2016, \bibinfo{title}{{Formation
  of Massive Primordial Stars: Intermittent UV Feedback with Episodic Mass
  Accretion},} \apj, 824, 119, \dodoi{10.3847/0004-637X/824/2/119}

\bibitem[{T. Hosokawa {et~al.}(2011)Hosokawa, Omukai, Yoshida, \&
  Yorke}]{hosokawa11}
Hosokawa, T., Omukai, K., Yoshida, N., \& Yorke, H.~W. 2011,
  \bibinfo{title}{{Protostellar Feedback Halts the Growth of the First Stars in
  the Universe},} Science, 334, 1250, \dodoi{10.1126/science.1207433}

\bibitem[{J.~A. Hummel {et~al.}(2016)Hummel, Stacy, \& Bromm}]{hummel16}
Hummel, J.~A., Stacy, A., \& Bromm, V. 2016, \bibinfo{title}{The first stars:
  formation under cosmic ray feedback,} Monthly Notices of the Royal
  Astronomical Society, 460, 2432, \dodoi{10.1093/mnras/stw1127}

\bibitem[{O. Jaura {et~al.}(2022)Jaura, Glover, Wollenberg, Klessen, Geen, \&
  Haemmerl{\'e}}]{jaura22}
Jaura, O., Glover, S. C.~O., Wollenberg, K. M.~J., {et~al.} 2022,
  \bibinfo{title}{Trapping of H II regions in Population III star formation,}
  Monthly Notices of the Royal Astronomical Society, 512, 116,
  \dodoi{10.1093/mnras/stac487}

\bibitem[{S. {Khullar} {et~al.}(2026){Khullar}, {Grudi{\'c}}, {Matzner},
  {Hopkins}, {Murray}, \& {Faucher-Gigu{\`e}re}}]{khullar2026}
{Khullar}, S., {Grudi{\'c}}, M.~Y., {Matzner}, C.~D., {et~al.} 2026,
  \bibinfo{title}{{Star Formation Mitosis: The Effects of Particle Splitting on
  the Star Formation Outcome},} arXiv e-prints, arXiv:2610.00513,
  \dodoi{10.48550/arXiv.2610.00513}

\bibitem[{J.-h. Kim {et~al.}(2013)Kim, Abel, Agertz, Bryan, Ceverino,
  Christensen, Conroy, Dekel, Gnedin, Goldbaum, Guedes, Hahn, Hobbs, Hopkins,
  Hummels, Iannuzzi, Keres, Klypin, Kravtsov, Krumholz, Kuhlen, Leitner, Madau,
  Mayer, Moody, Nagamine, Norman, Onorbe, O’Shea, Pillepich, Primack, Quinn,
  Read, Robertson, Rocha, Rudd, Shen, Smith, Szalay, Teyssier, Thompson,
  Todoroki, Turk, Wadsley, Wise, Zolotov, Adi, \& Collaboration}]{AGORA2014}
Kim, J.-h., Abel, T., Agertz, O., {et~al.} 2013, \bibinfo{title}{THE AGORA
  HIGH-RESOLUTION GALAXY SIMULATIONS COMPARISON PROJECT,} The Astrophysical
  Journal Supplement Series, 210, 14, \dodoi{10.1088/0067-0049/210/1/14}

\bibitem[{W.-T. Kim \& E.~C. Ostriker(2001)Kim \& Ostriker}]{kim_ostriker01}
Kim, W.-T., \& Ostriker, E.~C. 2001, \bibinfo{title}{{Amplification,
  Saturation, and Q Thresholds for Runaway: Growth of Self-Gravitating
  Structures in Models of Magnetized Galactic Gas Disks},} \apj, 559, 70,
  \dodoi{10.1086/322330}

\bibitem[{R.~S. Klessen \& S.~C.~O. Glover(2023)Klessen \&
  Glover}]{klessen_glover23}
Klessen, R.~S., \& Glover, S. C.~O. 2023, \bibinfo{title}{{The First Stars:
  Formation, Properties, and Impact},} ARA\&A, 61, 65,
  \dodoi{10.1146/annurev-astro-071221-053453}

\bibitem[{D. Koh {et~al.}(2021)Koh, Abel, \& Jedamzik}]{koh21}
Koh, D., Abel, T., \& Jedamzik, K. 2021, \bibinfo{title}{{First Star Formation
  in the Presence of Primordial Magnetic Fields},} \apjl, 909, L21,
  \dodoi{10.3847/2041-8213/abe8dd}

\bibitem[{E. Komatsu {et~al.}(2011)Komatsu {et~al.}}]{komatsu11}
Komatsu, E., {et~al.} 2011, \bibinfo{title}{{Seven-year Wilkinson Microwave
  Anisotropy Probe (WMAP) Observations: Cosmological Interpretation},} \apjs,
  192, 18, \dodoi{10.1088/0067-0049/192/2/18}

\bibitem[{P. Kroupa(2001)Kroupa}]{kroupa01}
Kroupa, P. 2001, \bibinfo{title}{{On the variation of the initial mass
  function},} \mnras, 322, 231, \dodoi{10.1046/j.1365-8711.2001.04022.x}

\bibitem[{M.~R. Krumholz {et~al.}(2012)Krumholz, Dekel, \& McKee}]{krumholz12}
Krumholz, M.~R., Dekel, A., \& McKee, C.~F. 2012, \bibinfo{title}{{A Universal,
  Local Star Formation Law in Galactic Clouds, Nearby Galaxies, High-redshift
  Disks, and Starbursts},} \apj, 745, 69, \dodoi{10.1088/0004-637X/745/1/69}

\bibitem[{M. Kulkarni {et~al.}(2021)Kulkarni, Visbal, \& Bryan}]{kulkarni21}
Kulkarni, M., Visbal, E., \& Bryan, G.~L. 2021, \bibinfo{title}{The Critical
  Dark Matter Halo Mass for Population III Star Formation: Dependence on
  Lyman-Werner Radiation, Baryon-dark Matter Streaming Velocity, and Redshift,}
  The Astrophysical Journal, 917, 40, \dodoi{10.3847/1538-4357/ac08a3}

\bibitem[{R.~B. Larson(1969)Larson}]{Larson_1969}
Larson, R.~B. 1969, \bibinfo{title}{{Numerical calculations of the dynamics of
  a collapsing proto-star},} MNRAS, 145, 271, \dodoi{10.1093/mnras/145.3.271}

\bibitem[{M.~A. Latif {et~al.}(2022)Latif, Whalen, \&
  Khochfar}]{latif_khochfar22}
Latif, M.~A., Whalen, D., \& Khochfar, S. 2022, \bibinfo{title}{{The birth mass
  function of Population III stars},} \apj, 925, 28,
  \dodoi{10.3847/1538-4357/ac3916}

\bibitem[{E.~J. Lee {et~al.}(2015)Lee, Chang, \& Murray}]{lee15}
Lee, E.~J., Chang, P., \& Murray, N.~W. 2015, \bibinfo{title}{{Time Varying
  Dynamical Star Formation Rate},} \apj, 800, 49,
  \dodoi{10.1088/0004-637X/800/1/49}

\bibitem[{C.~D. Levermore(1984)Levermore}]{Levermore_1984}
Levermore, C.~D. 1984, \bibinfo{title}{Relating Eddington factors to flux
  limiters,} Journal of Quantitative Spectroscopy and Radiative Transfer, 31,
  149–160, \dodoi{10.1016/0022-4073(84)90112-2}

\bibitem[{P.~S. Li {et~al.}(2018)Li, Klein, \& McKee}]{li_klein_mckee18}
Li, P.~S., Klein, R.~I., \& McKee, C.~F. 2018, \bibinfo{title}{Formation of
  stellar clusters in magnetized, filamentary infrared dark clouds,} Monthly
  Notices of the Royal Astronomical Society, 473, 4220–4241,
  \dodoi{10.1093/mnras/stx2611}

\bibitem[{M.~N. Machida {et~al.}(2025)Machida, Hirano, \& Basu}]{machida25}
Machida, M.~N., Hirano, S., \& Basu, S. 2025, \bibinfo{title}{{Effect of
  Magnetic Field on the Accretion Phase of Population III Star Formation},}
  \apj, 988, 6, \dodoi{10.3847/1538-4357/addc56}

\bibitem[{C.~D. Matzner \& P.~H. Jumper(2015)Matzner \&
  Jumper}]{matzner_jumper15}
Matzner, C.~D., \& Jumper, P.~H. 2015, \bibinfo{title}{Star Cluster Formation
  with Stellar Feedback and Large-scale Inflow,} The Astrophysical Journal,
  815, 68, \dodoi{10.1088/0004-637X/815/1/68}

\bibitem[{C.~F. McKee \& E.~C. Ostriker(2007)McKee \&
  Ostriker}]{mckee_ostriker07}
McKee, C.~F., \& Ostriker, E.~C. 2007, \bibinfo{title}{Theory of Star
  Formation,} Annual Review of Astronomy and Astrophysics, 45, 565

\bibitem[{C.~F. McKee \& J.~C. Tan(2003)McKee \& Tan}]{mckee03}
McKee, C.~F., \& Tan, J.~C. 2003, \bibinfo{title}{{The Formation of Massive
  Stars from Turbulent Cores},} \apj, 585, 850, \dodoi{10.1086/346149}

\bibitem[{Y.~J. Meziani {et~al.}(2026{\natexlab{a}})Meziani, Hopkins,
  Grudi{\'c}, Khullar, Faucher-Gigu{\`e}re, \& Gandhi}]{meziani26}
Meziani, Y.~J., Hopkins, P.~F., Grudi{\'c}, M.~Y., {et~al.} 2026{\natexlab{a}},
  \bibinfo{title}{{FORGE'd in the Early Universe: The Effect of Protostellar
  Outflows on Pop III Accretion},} arXiv e-prints.
\newblock \doarXiv{2602.02953}

\bibitem[{Y.~J. Meziani {et~al.}(2026{\natexlab{b}})Meziani, Hopkins,
  {\noopsort{Z}}Gandhi, Faucher-Gigu{\`e}re, Grudi{\'c}, Khullar, \&
  Offner}]{meziani26b}
Meziani, Y.~J., Hopkins, P.~F., {\noopsort{Z}}Gandhi, P.~J., {et~al.}
  2026{\natexlab{b}}, \bibinfo{title}{{FORGE'd in the Early Universe II:
  Constraining the Pop III Initial Mass Function},}

\bibitem[{D.~W. Murray {et~al.}(2017)Murray, Chang, Murray, \&
  Pittman}]{murray16}
Murray, D.~W., Chang, P., Murray, N.~W., \& Pittman, J. 2017,
  \bibinfo{title}{{Collapse in Self-gravitating Turbulent Fluids},} \mnras,
  465, 1316, \dodoi{10.1093/mnras/stw2796}

\bibitem[{N.~W. Murray \& P. Chang(2015)Murray \& Chang}]{murray15}
Murray, N.~W., \& Chang, P. 2015, \bibinfo{title}{{Star Formation in
  Self-gravitating Turbulent Fluids},} \apj, 804, 44,
  \dodoi{10.1088/0004-637X/804/1/44}

\bibitem[{D. Nakauchi {et~al.}(2014)Nakauchi, Inayoshi, \& Omukai}]{nakauchi14}
Nakauchi, D., Inayoshi, K., \& Omukai, K. 2014, \bibinfo{title}{Conditions for
  HD cooling in the first galaxies revisited: interplay between far-ultraviolet
  and cosmic ray feedback in Population III star formation,} Monthly Notices of
  the Royal Astronomical Society, 442, 2667, \dodoi{10.1093/mnras/stu1042}

\bibitem[{S.~S.~R. Offner {et~al.}(2009)Offner, Klein, McKee, \&
  Krumholz}]{Offner_2009}
Offner, S. S.~R., Klein, R.~I., McKee, C.~F., \& Krumholz, M.~R. 2009,
  \bibinfo{title}{The Effects of Radiative Transfer on Low-Mass Star
  Formation,} The Astrophysical Journal, 703, 131–149,
  \dodoi{10.1088/0004-637X/703/1/131}

\bibitem[{S.~P. Oh \& Z. Haiman(2002)Oh \& Haiman}]{oh_haiman02}
Oh, S.~P., \& Haiman, Z. 2002, \bibinfo{title}{Second-Generation Objects in the
  Universe: Radiative Cooling and Collapse of Halos with Virial Temperatures
  above $10^4$ K,} The Astrophysical Journal, 569, 558, \dodoi{10.1086/339393}

\bibitem[{K. Omukai(2000)Omukai}]{omukai00}
Omukai, K. 2000, \bibinfo{title}{Protostellar Collapse with Various
  Metallicities,} The Astrophysical Journal, 534, 809, \dodoi{10.1086/308776}

\bibitem[{K. Omukai \& F. Palla(2003)Omukai \& Palla}]{omukai_palla03}
Omukai, K., \& Palla, F. 2003, \bibinfo{title}{Formation of the First Stars by
  Accretion,} The Astrophysical Journal, 589, 677

\bibitem[{L.~S. Oswald {et~al.}(2025)Oswald, Weltevrede, Posselt, Johnston,
  Karastergiou, \& Lower}]{oswald25}
Oswald, L.~S., Weltevrede, P., Posselt, B., {et~al.} 2025, \bibinfo{title}{{The
  Thousand-Pulsar-Array programme on MeerKAT -- XVI. Mapping the Galactic
  magnetic field with pulsar observations},} \mnras,
  \dodoi{10.1093/mnras/staf645}

\bibitem[{P. Padoan {et~al.}(2020)Padoan, Pan, Juvela, Haugb{\o}lle, \&
  Nordlund}]{padoan20}
Padoan, P., Pan, L., Juvela, M., Haugb{\o}lle, T., \& Nordlund, {\AA}. 2020,
  \bibinfo{title}{The Origin of Massive Stars: The Inertial-inflow Model,} The
  Astrophysical Journal, 900, 82

\bibitem[{F. Palla {et~al.}(1983)Palla, Salpeter, \& Stahler}]{Palla_1983}
Palla, F., Salpeter, E.~E., \& Stahler, S.~W. 1983, \bibinfo{title}{{Primordial
  star formation: the role of molecular hydrogen},} \apj, 271, 632,
  \dodoi{10.1086/161231}

\bibitem[{J. Park {et~al.}(2022)Park, Ricotti, \& Sugimura}]{Park_2022}
Park, J., Ricotti, M., \& Sugimura, K. 2022, Population {III} star formation in
  an {X}-ray background: {III}. Periodic radiative feedback and luminosity
  induced by elliptical orbits, \doarXiv{2212.04564}

\bibitem[{M.~V. Penston(1969)Penston}]{Penston_1969}
Penston, M.~V. 1969, \bibinfo{title}{{Dynamics of self-gravitating gaseous
  spheres -- III. Analytical results in the free-fall of isothermal cases},}
  MNRAS, 144, 425, \dodoi{10.1093/mnras/144.4.425}

\bibitem[{T. Peters {et~al.}(2010)Peters, Klessen, Mac~Low, \&
  Banerjee}]{peters10}
Peters, T., Klessen, R.~S., Mac~Low, M.-M., \& Banerjee, R. 2010,
  \bibinfo{title}{Limiting Accretion onto Massive Stars by
  Fragmentation-induced Starvation,} The Astrophysical Journal, 725, 134

\bibitem[{J. Prieto {et~al.}(2011)Prieto, Padoan, Jim{\'e}nez, \&
  Infante}]{prieto11}
Prieto, J., Padoan, P., Jim{\'e}nez, R., \& Infante, L. 2011,
  \bibinfo{title}{{Population III Stars from Turbulent Fragmentation at
  Redshift $\sim$11},} \apjl, 731, L38, \dodoi{10.1088/2041-8205/731/2/L38}

\bibitem[{L.~R. Prole {et~al.}(2022{\natexlab{a}})Prole, Clark, Klessen, \&
  Glover}]{prole22a}
Prole, L.~R., Clark, P.~C., Klessen, R.~S., \& Glover, S. C.~O.
  2022{\natexlab{a}}, \bibinfo{title}{Fragmentation-induced starvation in
  Population III star formation: a resolution study,} Monthly Notices of the
  Royal Astronomical Society, 510, 4019, \dodoi{10.1093/mnras/stab3697}

\bibitem[{L.~R. Prole {et~al.}(2022{\natexlab{b}})Prole, Clark, Klessen,
  Glover, \& Pakmor}]{prole22}
Prole, L.~R., Clark, P.~C., Klessen, R.~S., Glover, S. C.~O., \& Pakmor, R.
  2022{\natexlab{b}}, \bibinfo{title}{{Primordial magnetic fields in Population
  III star formation: a magnetized resolution study},} \mnras, 516, 2223,
  \dodoi{10.1093/mnras/stac2327}

\bibitem[{E. Ripamonti \& T. Abel(2004)Ripamonti \&
  Abel}]{ripamonti_fragmentation_2004}
Ripamonti, E., \& Abel, T. 2004, \bibinfo{title}{Fragmentation and the
  formation of primordial protostars: the possible role of collision-induced
  emission,} Monthly Notices of the Royal Astronomical Society, 348, 1019,
  \dodoi{10.1111/j.1365-2966.2004.07422.x}

\bibitem[{W.~W. Roberts \& C. Yuan(1970)Roberts \& Yuan}]{roberts_yuan70}
Roberts, W.~W., \& Yuan, C. 1970, \bibinfo{title}{{Application of the
  Density-Wave Theory to the Spiral Structure of the Milky Way System. III.
  Magnetic Field: Large-Scale Hydromagnetic Shock Formation},} \apj, 161, 887,
  \dodoi{10.1086/150592}

\bibitem[{B. Robertson \& P. Goldreich(2012)Robertson \&
  Goldreich}]{robertson_goldreich12}
Robertson, B., \& Goldreich, P. 2012, \bibinfo{title}{{Adiabatic Heating of
  Contracting Turbulent Fluids},} \apjl, 750, L31,
  \dodoi{10.1088/2041-8205/750/2/L31}

\bibitem[{K.~E. Sadanari {et~al.}(2023)Sadanari, Omukai, Sugimura, Matsumoto,
  \& Tomida}]{sadanari23}
Sadanari, K.~E., Omukai, K., Sugimura, K., Matsumoto, T., \& Tomida, K. 2023,
  \bibinfo{title}{{Non-ideal magnetohydrodynamic simulations of the first star
  formation: the effect of ambipolar diffusion},} \mnras, 519, 3076,
  \dodoi{10.1093/mnras/stac3724}

\bibitem[{K.~E. Sadanari {et~al.}(2024)Sadanari, Omukai, Sugimura, Matsumoto,
  \& Tomida}]{tanaka24}
Sadanari, K.~E., Omukai, K., Sugimura, K., Matsumoto, T., \& Tomida, K. 2024,
  \bibinfo{title}{{Impact of Turbulent Magnetic Fields on Disk Formation and
  Fragmentation in First Star Formation},} PASJ, 76, 823,
  \dodoi{10.1093/pasj/psae051}

\bibitem[{C. Shang {et~al.}(2010)Shang, Bryan, \& Haiman}]{shang10}
Shang, C., Bryan, G.~L., \& Haiman, Z. 2010, \bibinfo{title}{{Supermassive
  black hole formation by direct collapse: keeping protogalactic gas H$_2$ free
  in dark matter haloes with virial temperatures T$_{vir}$ > 10$^4$ K},}
  \mnras, 402, 1249, \dodoi{10.1111/j.1365-2966.2009.15960.x}

\bibitem[{P. Sharda {et~al.}(2021)Sharda, Federrath, Krumholz, \&
  Schleicher}]{sharda21}
Sharda, P., Federrath, C., Krumholz, M.~R., \& Schleicher, D. R.~G. 2021,
  \bibinfo{title}{{Magnetic field amplification in accretion discs around the
  first stars: implications for the primordial IMF},} \mnras, 503, 2014,
  \dodoi{10.1093/mnras/stab531}

\bibitem[{P. Sharda \& S.~H. Menon(2025)Sharda \& Menon}]{sharda_menon25}
Sharda, P., \& Menon, S.~H. 2025, \bibinfo{title}{{Population III star
  formation in the presence of turbulence, magnetic fields, and ionizing
  radiation feedback},} \mnras, 540, 1745, \dodoi{10.1093/mnras/staf803}

\bibitem[{P. Sharda {et~al.}(2025)Sharda, Menon, Gerasimov, Bromm, Burkhart,
  Haemmerl{\'e}, van Veenen, \& Wibking}]{prole25}
Sharda, P., Menon, S.~H., Gerasimov, R., {et~al.} 2025,
  \bibinfo{title}{{Magnetic fields limit the mass of Population III stars even
  before the onset of protostellar radiation feedback},} \mnras, 541, L1,
  \dodoi{10.1093/mnrasl/slaf043}

\bibitem[{F.~H. Shu(1977)Shu}]{Shu_1977}
Shu, F.~H. 1977, \bibinfo{title}{Self-similar collapse of isothermal spheres
  and star formation.,} The Astrophysical Journal, 214, 488–497,
  \dodoi{10.1086/155274}

\bibitem[{Y. Sofue \& M. Fujimoto(1983)Sofue \& Fujimoto}]{sofue_fujimoto83}
Sofue, Y., \& Fujimoto, M. 1983, \bibinfo{title}{{A bisymmetric spiral magnetic
  field and the spiral arms in our Galaxy},} \apj, 265, 722,
  \dodoi{10.1086/160718}

\bibitem[{Y. Sofue {et~al.}(1986)Sofue, Fujimoto, \& Wielebinski}]{sofue86}
Sofue, Y., Fujimoto, M., \& Wielebinski, R. 1986, \bibinfo{title}{Global
  structure of magnetic fields in spiral galaxies,} Annual Review of Astronomy
  and Astrophysics, 24, 459

\bibitem[{V. Springel(2005)Springel}]{Springel_2005}
Springel, V. 2005, \bibinfo{title}{The cosmological simulation code GADGET-2,}
  Monthly Notices of the Royal Astronomical Society, 364, 1105,
  \dodoi{10.1111/j.1365-2966.2005.09655.x}

\bibitem[{V. Springel(2010)Springel}]{Springel_2010b}
Springel, V. 2010, \bibinfo{title}{E pur si muove: Galilean-invariant
  cosmological hydrodynamical simulations on a moving mesh,} Monthly Notices of
  the Royal Astronomical Society, 401, 791–851,
  \dodoi{10.1111/j.1365-2966.2009.15715.x}

\bibitem[{A. Stacy \& V. Bromm(2007)Stacy \& Bromm}]{stacy_bromm07}
Stacy, A., \& Bromm, V. 2007, \bibinfo{title}{Impact of cosmic rays on
  Population III star formation,} Monthly Notices of the Royal Astronomical
  Society, 382, 229, \dodoi{10.1111/j.1365-2966.2007.12247.x}

\bibitem[{A. Stacy \& V. Bromm(2013)Stacy \& Bromm}]{Stacy_Bromm_2013}
Stacy, A., \& Bromm, V. 2013, \bibinfo{title}{Constraining the statistics of
  Population III binaries,} Monthly Notices of the Royal Astronomical Society,
  433, 1094–1107, \dodoi{10.1093/mnras/stt789}

\bibitem[{A. Stacy {et~al.}(2016)Stacy, Bromm, \& Lee}]{stacy16}
Stacy, A., Bromm, V., \& Lee, A.~T. 2016, \bibinfo{title}{{Building Up the
  Population III Initial Mass Function from Cosmological Initial Conditions},}
  \mnras, 462, 1307, \dodoi{10.1093/mnras/stw1728}

\bibitem[{A. Stacy {et~al.}(2010)Stacy, Greif, \& Bromm}]{stacy10}
Stacy, A., Greif, T.~H., \& Bromm, V. 2010, \bibinfo{title}{{The First Stars:
  Formation of Binaries and Small Multiple Systems},} \mnras, 403, 45,
  \dodoi{10.1111/j.1365-2966.2009.16113.x}

\bibitem[{A. Stacy {et~al.}(2012)Stacy, Greif, \& Bromm}]{stacy12}
Stacy, A., Greif, T.~H., \& Bromm, V. 2012, \bibinfo{title}{{The First Stars:
  Mass Growth Under Protostellar Feedback},} \mnras, 422, 290,
  \dodoi{10.1111/j.1365-2966.2012.20605.x}

\bibitem[{A. Stacy {et~al.}(2022)Stacy, McKee, Lee, Klein, \& Li}]{stacy22}
Stacy, A., McKee, C.~F., Lee, A.~T., Klein, R.~I., \& Li, P.~S. 2022,
  \bibinfo{title}{{Magnetic fields in the formation of the first stars -- II.
  Results},} \mnras, 511, 5042, \dodoi{10.1093/mnras/stac372}

\bibitem[{S.~W. Stahler {et~al.}(1986)Stahler, Palla, \& Salpeter}]{stahler86}
Stahler, S.~W., Palla, F., \& Salpeter, E.~E. 1986, \bibinfo{title}{Primordial
  stellar evolution: the protostar phase,} The Astrophysical Journal, 302, 590

\bibitem[{A. Storck {et~al.}(2026)Storck, Katz, Devriendt, Slyz, Cadiou,
  Choustikov, Rey, Saxena, Agertz, \& Kimm}]{storck26}
Storck, A., Katz, H., Devriendt, J., {et~al.} 2026, \bibinfo{title}{{MEGATRON:
  The environments of Population III stars at Cosmic Dawn and their connection
  to present-day galaxies},} \mnras, 548, stag529,
  \dodoi{10.1093/mnras/stag529}

\bibitem[{K. Sugimura {et~al.}(2020)Sugimura, Matsumoto, Hosokawa, Hirano, \&
  Omukai}]{sugimura20}
Sugimura, K., Matsumoto, T., Hosokawa, T., Hirano, S., \& Omukai, K. 2020,
  \bibinfo{title}{{The Birth of a Massive First-star Binary},} \apjl, 892, L14,
  \dodoi{10.3847/2041-8213/ab7d37}

\bibitem[{K. Sugimura {et~al.}(2023)Sugimura, Matsumoto, Hosokawa, Hirano, \&
  Omukai}]{sugimura23}
Sugimura, K., Matsumoto, T., Hosokawa, T., Hirano, S., \& Omukai, K. 2023,
  \bibinfo{title}{{Formation of Massive and Wide First-star Binaries in
  Radiation Hydrodynamic Simulations},} \apj, 959, 17,
  \dodoi{10.3847/1538-4357/ad02fc}

\bibitem[{K. Sugimura {et~al.}(2014)Sugimura, Omukai, \& Inoue}]{sugimura14}
Sugimura, K., Omukai, K., \& Inoue, A.~K. 2014, \bibinfo{title}{{The critical
  radiation intensity for direct collapse black hole formation: dependence on
  the radiation spectral shape},} \mnras, 445, 544,
  \dodoi{10.1093/mnras/stu1778}

\bibitem[{H. Susa {et~al.}(2014)Susa, Hasegawa, \& Tominaga}]{susa14}
Susa, H., Hasegawa, K., \& Tominaga, N. 2014, \bibinfo{title}{{The Mass
  Spectrum of the First Stars},} \apj, 792, 32,
  \dodoi{10.1088/0004-637X/792/1/32}

\bibitem[{A. Toomre(1964)Toomre}]{Toomre_1964}
Toomre, A. 1964, \bibinfo{title}{On the gravitational stability of a disk of
  stars.,} The Astrophysical Journal, 139, 1217–1238, \dodoi{10.1086/147861}

\bibitem[{J.~K. Truelove {et~al.}(1997)Truelove, Klein, McKee, Holliman,
  Howell, \& Greenough}]{truelove97}
Truelove, J.~K., Klein, R.~I., McKee, C.~F., {et~al.} 1997,
  \bibinfo{title}{{The Jeans Condition: A New Constraint on Spatial Resolution
  in Simulations of Isothermal Self-gravitational Hydrodynamics},} \apjl, 489,
  L179, \dodoi{10.1086/310975}

\bibitem[{M.~J. Turk {et~al.}(2011{\natexlab{a}})Turk, Clark, Glover, Greif,
  Abel, Klessen, \& Bromm}]{turk11}
Turk, M.~J., Clark, P., Glover, S. C.~O., {et~al.} 2011{\natexlab{a}},
  \bibinfo{title}{Effects of Varying the Three-body Molecular Hydrogen
  Formation Rate in Primordial Star Formation,} The Astrophysical Journal, 726,
  55

\bibitem[{M.~J. Turk {et~al.}(2011{\natexlab{b}})Turk, Clark, Glover, Greif,
  Abel, Klessen, \& Bromm}]{turk11b}
Turk, M.~J., Clark, P., Glover, S. C.~O., {et~al.} 2011{\natexlab{b}},
  \bibinfo{title}{{Effects of Varying the Three-Body Molecular Hydrogen
  Formation Rate in Primordial Star Formation},} \apj, 726, 55,
  \dodoi{10.1088/0004-637X/726/1/55}

\bibitem[{M.~I. Ugalino {et~al.}(2026)Ugalino, Diemer, Semenov, Kim, Marinacci,
  Pfrommer, Schaller, Shchutskyi, Springel, Teyssier, \& Zier}]{ugalino26}
Ugalino, M.~I., Diemer, B., Semenov, V.~A., {et~al.} 2026,
  \bibinfo{title}{{Towards robust simulations of the galactic dynamo: a
  cross-code comparison between Eulerian and Lagrangian schemes},} arXiv
  e-prints.
\newblock \doarXiv{2610.02308}

\bibitem[{A. Wetzel {et~al.}(2023)Wetzel, Hayward, Sanderson, Garrison-Kimmel,
  Faucher-Gigu{\`e}re, Kere{\v{s}}, \& Hopkins}]{wetzel23}
Wetzel, A., Hayward, C.~C., Sanderson, R.~E., {et~al.} 2023,
  \bibinfo{title}{{Public Data Release of the FIRE-2 Cosmological Zoom-in
  Simulations of Galaxy Formation},} \apjs, 265, 44,
  \dodoi{10.3847/1538-4365/acb99a}

\bibitem[{J.~H. Wise \& T. Abel(2007)Wise \& Abel}]{wise_abel07}
Wise, J.~H., \& Abel, T. 2007, \bibinfo{title}{Resolving the Formation of
  Protogalaxies. I. Virialization,} The Astrophysical Journal, 665, 899

\bibitem[{K.~M.~J. Wollenberg {et~al.}(2020)Wollenberg, Glover, Clark, \&
  Klessen}]{wollenberg20}
Wollenberg, K. M.~J., Glover, S. C.~O., Clark, P.~C., \& Klessen, R.~S. 2020,
  \bibinfo{title}{{Formation sites of Population III star formation: The
  effects of different levels of rotation and turbulence on the fragmentation
  behaviour of primordial gas},} \mnras, 494, 1871,
  \dodoi{10.1093/mnras/staa289}

\bibitem[{N. Yoshida {et~al.}(2003)Yoshida, Abel, Hernquist, \&
  Sugiyama}]{yoshida03}
Yoshida, N., Abel, T., Hernquist, L., \& Sugiyama, N. 2003,
  \bibinfo{title}{Simulations of Early Structure Formation: Primordial Gas
  Clouds,} The Astrophysical Journal, 592, 645

\bibitem[{N. Yoshida {et~al.}(2008)Yoshida, Omukai, \& Hernquist}]{yoshida08}
Yoshida, N., Omukai, K., \& Hernquist, L. 2008, \bibinfo{title}{{Protostar
  Formation in the Early Universe},} Science, 321, 669,
  \dodoi{10.1126/science.1160259}

\end{thebibliography}

\end{document}